\documentclass[iop,apj]{emulateapj}

\usepackage{amssymb,subfigure,rotating}

\slugcomment{To appear in ApJ}

\shorttitle{SED Models for LLAGNs in LINERs}
\shortauthors{Nemmen et al.}

\begin{document}

\title{Spectral Energy Distribution Models for Low-luminosity Active Galactic Nuclei in LINERs}

\author{Rodrigo S. Nemmen,\altaffilmark{1,2,3} Thaisa Storchi-Bergmann\altaffilmark{2} and Michael Eracleous\altaffilmark{3}}

\altaffiltext{1}{NASA Goddard Space Flight Center, Greenbelt, MD 20771, USA}
\altaffiltext{2}{Instituto de F\'isica, Universidade Federal do Rio Grande do Sul, Campus do Vale, Porto Alegre, RS, Brazil.}
\altaffiltext{3}{Department of Astronomy and Astrophysics, Pennsylvania State University, 525 Davey Lab, University Park, PA 16802.}

\begin{abstract}
Low-luminosity active galactic nuclei (LLAGNs) represent the bulk of the AGN population in the present-day universe and they trace the low-level accreting supermassive black holes.
In order to probe the accretion and jet physical properties in LLAGNs as a class, we model the broadband radio to X-rays spectral energy distributions (SEDs) of 21 LLAGNs in low-ionization nuclear emission-line regions (LINERs) with a coupled accretion-jet model. The accretion flow is modeled as an inner ADAF outside of which there is a truncated standard thin disk. 
We find that the radio emission is severely underpredicted by ADAF models and is explained by the relativistic jet. 
The origin of the X-ray radiation in most sources can be explained by three distinct scenarios: the X-rays can be dominated by emission from the ADAF, or the jet, or the X-rays can arise from a jet-ADAF combination in which both components contribute to the emission with similar importance. For 3 objects both the jet and ADAF fit equally well the X-ray spectrum and can be the dominant source of X-rays whereas for 11 LLAGNs a jet-dominated model accounts better than the ADAF-dominated model for the data.
The individual and average SED models that we computed can be useful for different studies of the nuclear emission of LLAGNs.
From the model fits, we estimate important parameters of the central engine powering LLAGNs in LINERs, such as the mass accretion rate and the mass-loss rate in the jet and the jet power - relevant for studies of the kinetic feedback from jets.
\end{abstract}


\section{Introduction}

One of the central paradigms in extragalactic astronomy is that today's galaxies host supermassive black holes (SMBHs) at their centers and have a symbiotic evolution with them \citep{magorrian98, ferrarese05}. The central black holes seeded in proto-galaxies at high redshift grow in mass during cosmic history through a sequence of mergers of massive black holes and accretion episodes (e.g., \citealt{volonteri05,fani11}). The massive black holes grow in such a way that they co-evolve with the galactic bulges \citep{kor95, ferrarese00, geb00, ferrarese05} but do not seem to do so with the galaxy disks \citep{kor11a} or dark matter haloes \citep{kor11b}.

The growth of SMBHs is dominated by the quasar phase \citep{marconi04,mclure04,sijacki07,merloni08} during which they accrete from their large gas reservoir via geometrically thin accretion disks which are radiatively efficient \citep{shakura73,novikov73} as suggested by different lines of evidence \citep{shields78,koratkar99,shang05}.
The quasar population has a strong cosmological evolution with its density decreasing by a factor of $\sim 100$ from $z \sim 2$ -- where the density reaches its peak -- to the present time (e.g., \citealt{osmer04,silverman05}). 

In the present-day universe, SMBHs are underfed compared to the ones at high $z$ and are ``sleeping''. Most of SMBH activity at low $z$ is dominated by the weak end of the AGN luminosity function in the form of low-luminosity AGNs (LLAGNs; \citealt{ho95,ho97,nagar05,ho08}). 
Though the LLAGN phase dominates the time evolution of SMBHs, it contributes little to their mass growth \citep{hopkins06b,sijacki07,merloni08,xu10}.

The bulk of the LLAGN population ($\approx 2/3$; \citealt{ho08,ho09}) reside in low-ionization nuclear emission-line regions (LINERs; \citealt{heckman80,ho97apjs}). LLAGNs are extremely sub-Eddington systems which are many orders of magnitude less luminous than quasars, with average bolometric luminosities $\left \langle L_{\rm bol} \right \rangle \sim 10^{40}-10^{41} \ {\rm erg \ s}^{-1}$ and an average Eddington ratio of $L_{\rm bol}/L_{\rm Edd} \sim 10^{-5}$ \citep{ho09} where $L_{\rm Edd}$ is the Eddington luminosity. 
The observational properties of LLAGNs are quite different from those of more luminous AGNs. Regarding the broadband spectral energy distributions (SEDs), LLAGNs seem not to have the thermal continuum prominence in the ultraviolet (UV) -- the ``big blue bump'' -- which is one of the signatures of the presence of an optically thick, geometrically thin accretion disk \citep{ho99, nemmen06, wu07, ho08, erac10}. Regarding the emission-lines, LLAGNs typically have weak and narrow Fe K$\alpha$ emission \citep{terashima02} and a handful of LINERs display broad double-peaked H$\alpha$ lines (e.g., \citealt{sb03}); these properties of the emission-line spectrum are consistent with the absence of a thin accretion disk, or a thin accretion disk whose inner radius is truncated at $\gtrsim 100 GM/c^2$ \citep{chen89a, chen89b}. Last but not least, with the typical fuel supply of hot diffuse gas (via Bondi accretion) and cold dense gas (via stellar mass loss) available in nearby galaxies, LLAGNs would be expected to produce much higher luminosities than observed on the assumption of standard thin disks with a $10\%$ radiative efficiency \citep{ho09}.
Taken together, this set of observational properties favors the scenario in which the accretion flow in LLAGNs is advection-dominated or radiatively inefficient.

Advection-dominated accretion flows (ADAFs\footnote{In this paper, we consider ADAFs and radiatively inefficient accretion flows (RIAFs) to be the same kind of accretion flow solution. For a clarification regarding the terminology, see \citealt{yu11}.}; 
for reviews see \citealt{nar98,yuan07,nar08}) are very hot, geometrically thick, optically thin flows which are typified by low radiative efficiencies ($L \ll 0.1 \dot{M} c^2$) and occur at low accretion rates ($\dot{M} \lesssim 0.01 \dot{M}_{\rm Edd}$). SMBHs are thought to spend most of their lives in the ADAF state \citep{hopkins06b,xu10}, the best studied individual case being Sgr A* (e.g., \citealt{yuan07}). 

In many LLAGNs, another component in the accretion flow besides the ADAF is required in order to account for different observations, including a ``red bump'' in the SEDs \citep{lasota96,quat99,nemmen06,yu11} and as mentioned before the double-peaked Balmer emission lines (e.g., \citealt{sb03,erac09}): the emission from a thin accretion disk whose inner radius is truncated at the outer radius of the ADAF. In many LLAGNs the accretion flow may begin as a standard thin disk but somehow at a certain transition radius the accretion flow abruptly switches from a cold to a hot ADAF mode. The details of how this transition might happen are still not well understood \citep{manmoto00ssd,yuan04,nar08}, but it seems to be analogous to the transition between the different spectral states in black hole binary systems \citep{remi06,done07}.

\citet{maoz07} challenged the scenario of the central engines of LLAGNs consisting of ADAFs and truncated thin disks. Maoz argues that LLAGNs in LINERs have UV/X-ray luminosity ratios similar on average to those of brighter Seyfert 1s and based on that observation he posits that thin disks extending all the way down to the radius of the innermost stable circular orbits (ISCO) persist even for LLAGNs, despite their smaller accretion rates. \citet{yu11} showed that the SEDs compiled by \citet{maoz07} are naturally fitted by ADAF models with the addition of a truncated thin disk and also discussed on theoretical grounds why the ADAF model has a superior explanation capability than the pure thin disk model. The ADAF model is the only model that can naturally account for the set of observational properties of LLAGNs within a self-contained theoretical framework. Hence, it is the physical scenario adopted in this work.

ADAFs are relevant to the understanding of AGN feedback since they are quite efficient at producing powerful outflows and jets, as suggested by theoretical studies including analytical theory \citep{ny94,meier01,nemmen07} and numerical simulations \citep{mckinney04,hawley06,sasha10}). This is in line with the different observational studies which demonstrate that LLAGNs are generally radio-loud \citep{ho02, ho08} and accompanied by powerful jets (e.g., \citealt{heinz07, merloni08}). 
In fact, the so-called ``radio mode'' of AGN feedback invoked in semi-analytic and hydrodynamic simulations of galaxy formation (e.g., \citealt{bower06,croton06,sijacki07,fani11}) would presumably correspond to the ADAF accretion state actively producing jets as explicitly incorporated in some works \citep{sijacki07, oka08}. 

It is clear that an understanding of the physical nature of LLAGNs in LINERs will shed light on the nature of black hole accretion, outflows and consequently black hole growth and AGN feedback in present-day galaxies. One of the best ways of exploring the astrophysics of black hole accretion and outflows is by using multiwavelength observations of black hole systems and comparing the spectra predicted by specific models with the data. The goal of this work is therefore to probe the physics of accretion and ejection in the LLAGN population, by modeling their nuclear, broadband, radio to X-rays SEDs which provide constraints on physical models for the emission of the accretion flow and the jet.
Furthermore, with a large enough sample of systems, we can derive from the fits to the data the parameters that characterize the central engines and build a census of the ``astrophysical diet'' of low-state AGNs.

In this work, we will present how the currently favored model for the central engines of weakly accreting black holes is able to account for the observed SEDs of LLAGNs. We will also present the typical accretion/ejection -- or feeding/feedback -- properties of LLAGNs in LINERs as inferred from our detailed SED fits. With these parameters, we will then be able to draw a connection between the SMBHs in nearby active galaxies and the one in the Milky Way, Sgr A*. We will also present the inferred production site of the continuum emission and other properties inferred from the fits. 

The structure of this paper is as follows. In Section \ref{sec:sample} we describe the sample of 21 LLAGNs in LINERs that we used, the data and classifications. Section \ref{sec:models} describes the physical model that we adopted in order to interpret and fit the SEDs and derive the central engine parameters. In Section \ref{sec:seds} we describe the model fits to the SEDs, explaining in more detail the SED models for 5 objects which we consider the ``reference'' or fiducial LLAGN examples. Section \ref{ap:seds} presents the model fits to the other 16 SEDs in our sample. In Section \ref{sec:pars} we describe the distributions of central engine parameters resulting from our detailed model fits. We present the average SED resulting from our fits -- including the predicted emission in wavebands which can be observed with future facilities such as ALMA and the James Webb Space Telescope -- and the inferred nature of the broadband emission in each wavelength in \textsection \ref{sec:avgsed}. We discuss in \ref{sec:disc} which model is favored to explain our LLAGN sample, in particular which component in the flow dominates the X-ray emission. We conclude by presenting a summary of our results in \ref{sec:end} and comparing the SMBH ``diet'' of massive black holes in LINER AGNs and Sgr A*. 


\section{Sample} \label{sec:sample}

\citet{erac10} (hereafter EHF10) compiled a sample of 35 SEDs of LLAGNs found in LINERs which include high spatial resolution optical and UV observations with the \textit{Hubble Space Telescope} (HST), as well as X-ray observations with \emph{Chandra}. Most of the SEDs studied by EHF10 have also high-resolution radio (observed with the Very Large Array -- VLA -- or VLBA/VLBI) and infrared observations available. In order to select the best SEDs to be modelled from the sample of EHF10, we applied the following selection criteria: we selected only those objects for which their X-ray flux is not an upper limit. We also demand that there is a determination of the black hole mass. These selection criteria leave us with 24 LINERs. In three of these objects (NGC 404, NGC 5055 and NGC 6500) the emission is likely to be dominated by a stellar population (EHF10). Hence we discard these three LLAGNs and we are left with 21 objects which are listed in Table \ref{tab:sample}. This table also lists their corresponding Hubble and LINER Types, black hole masses, bolometric luminosities and Eddington ratios.

\begin{deluxetable*}{llrcccccc}
\tabletypesize{\scriptsize}
\tablecaption{Sample of galaxies and their basic properties\tablenotemark{a} 
\label{tab:sample}}
\tablewidth{0pt}
\tablehead{
\colhead{Galaxy} & \colhead{Hubble} & \colhead{Distance\tablenotemark{b}} & \colhead{$\log$}             & \colhead{LINER} & \colhead{$L_{\rm X}$} & \colhead{$L_{\rm bol}$} & \colhead{$L_{\rm bol}/L_{\rm Edd}$} & SED \\
\colhead{}       & \colhead{Type}   & \colhead{(Mpc)}    & \colhead{$(M_{\rm BH}/M_\odot)$} & \colhead{Type}  & \colhead{(erg s$^{-1}$)\tablenotemark{c}} & \colhead{(erg s$^{-1}$)\tablenotemark{d}} & & type\tablenotemark{e} 
}
\startdata
NGC  266                 &  SB(rs)ab   & 62.4 (1) & 7.6\tablenotemark{f} & L1 & $7.4 \times 10^{40}$ & $2.2 \times 10^{42}$ & $4 \times 10^{-4}$ & B \\
NGC 1097                 &  SB(rl)b    & 14.5 (1) & 8.1 & L1      & $4.3 \times 10^{40}$ & $8.5 \times 10^{41}$ & $5 \times 10^{-5}$ & A \\
NGC 1553                 &  SA(r1)0    & 17.2 (2) & 7.9 & L2/T2   & $1.3 \times 10^{40}$ & $3.8 \times 10^{41}$ & $4 \times 10^{-5}$ & B \\
NGC 2681                 &  SBA(rs)0/a & 16.0 (2) & 7.1 & L1      & $6.1 \times 10^{38}$ & $1.8 \times 10^{40}$ & $1 \times 10^{-5}$ & B \\
NGC 3031 (M81)           &  SA(s)ab    &  3.6 (3) & 7.8 & S1.5/L1 & $1.9 \times 10^{40}$ & $2.1 \times 10^{41}$ & $3 \times 10^{-5}$ & A \\
NGC 3169                 &  SA(s)a     & 19.7 (1) & 7.8 & L2      & $1.1 \times 10^{41}$ & $3.3 \times 10^{42}$ & $4 \times 10^{-4}$ & B \\
NGC 3226                 &  E2         & 21.9 (2) & 8.1 & L1      & $5.0 \times 10^{40}$ & $1.5 \times 10^{42}$ & $1 \times 10^{-4}$ & B \\
NGC 3379  (M105)         &  E1         &  9.8 (2) & 8.2 & L2/T2   & $1.7 \times 10^{37}$ & $5.1 \times 10^{38}$ & $3 \times 10^{-8}$ & B \\
NGC 3998                 &  SA(r)0     & 13.1 (2) & 8.9 & L1      & $2.6 \times 10^{41}$ & $1.4 \times 10^{43}$ & $1 \times 10^{-4}$ & A \\
NGC 4143                 &  SAB(s)0    & 14.8 (2) & 8.3 & L1      & $1.1 \times 10^{40}$ & $3.2 \times 10^{41}$ & $1 \times 10^{-5}$ & A \\
NGC 4261                 &  E2-3       & 31.6 (2) & 8.7 & L2      & $1.0 \times 10^{41}$ & $6.8 \times 10^{41}$ & $1 \times 10^{-5}$ & B \\
NGC 4278                 &  E1-2       & 14.9 (2) & 8.6 & L1      & $9.1 \times 10^{39}$ & $2.7 \times 10^{41}$ & $5 \times 10^{-6}$ & A \\
NGC 4374 (M84, 3C 272.1) &  E1         & 17.1 (2) & 8.9 & L2      & $3.5 \times 10^{39}$ & $5.0 \times 10^{41}$ & $5 \times 10^{-6}$ & A \\
NGC 4457                 &  SAB(s)0/a  & 10.7 (4) & 6.9 & L2      & $1.0 \times 10^{39}$ & $3.0 \times 10^{40}$ & $3 \times 10^{-5}$ & B \\
NGC 4486 (M87, 3C 274)   &  E0-1       & 14.9 (2) & 9.8 & L2      & $1.6 \times 10^{40}$ & $9.8 \times 10^{41}$ & $7 \times 10^{-6}$ & A \\
NGC 4494                 &  E1-2       & 15.8 (2) & 7.6 & L2      & $9.2 \times 10^{38}$ & $2.8 \times 10^{40}$ & $6 \times 10^{-6}$ & B \\
NGC 4548 (M91)           &  SBb(rs)    & 15.0 (3) & 7.6 & L2      & $5.4 \times 10^{39}$ & $1.6 \times 10^{41}$ & $3 \times 10^{-5}$ & B \\
NGC 4552 (M89)           &  E          & 14.3 (2) & 8.2 & T2      & $2.6 \times 10^{39}$ & $7.8 \times 10^{40}$ & $4 \times 10^{-6}$ & A \\
NGC 4579 (M58)           &  SAB(rs)b   & 21.0 (4) & 7.8 & L2      & $1.8 \times 10^{41}$ & $1.0 \times 10^{42}$ & $1 \times 10^{-4}$ & A \\
NGC 4594 (M104)          &  SA(s)a     &  9.1 (2) & 8.5 & L2      & $1.6 \times 10^{40}$ & $4.8 \times 10^{41}$ & $1 \times 10^{-5}$ & A \\
NGC 4736 (M94)           &  (R)SA(r)ab &  4.8 (2) & 7.1 & L2      & $5.9 \times 10^{38}$ & $1.8 \times 10^{40}$ & $1 \times 10^{-5}$ & A \\
\enddata
\tablenotetext{a}{The information in this table was obtained from
  EHF10, see text.}
\tablenotetext{b}{The number in parenthesis gives the source and
  method of the distance measurement, as follows: (1) From the catalog
  of \citet{tully88}, determined from a model for peculiar velocities and
  assuming $H_0=75{\rm\; km\;s^{-1}\;Mpc^{-1}}$; (2) From \citet{tonry01}, who used the surface brightness fluctation  method. Following \citet{jensen03}, the distance modulus
  reported by \citet{tonry01} was corrected by subtracting 0.16
  mag; (3) From \citet{freedman94,freedman01}, who used Cepheid
  variables; (4) From \citet{gavazzi09} who used the Tully-Fisher
  method.}
\tablenotetext{c}{$L_{\rm X}$ is the X-ray luminosity in the $2-10$
  keV range.}
\tablenotetext{d}{The bolometric luminosities of NGC 1097, NGC 3031,
  NGC 3998, NGC 4374, NGC 4486, NGC 4579 and NGC 4594 were estimated
  by integrating the SEDs. For all other galaxies $L_{\rm bol}$ was
  determined by scaling $L_{\rm X}$ as described in \textsection \ref{sec:sample}.}
\tablenotetext{e}{Classification of the SED depending on the quality
  of the data sampling (see Section \ref{sec:sample}).}
\tablenotetext{f}{The mass estimate for NGC 266 is subject to a large
  uncertainty ($\approx 1$ dex) since it is based on using the
  fundamental plane of \citet{merloni03} (see text).}
\end{deluxetable*}

According to the quality of the available data, we can further classify our sample of LINERs in two groups. Group A contains the objects that have the data with the best quality available. For instance, the radio, optical and UV bands have each at least one data point which is not an upper limit to the luminosity. Obviously, the objects in this group provide better constraints to the accretion-jet models. Group B comprises the sources that provide not as good contraints to the models, because they lack observational constraints in the radio to UV region of the spectrum. Some objects in this group have only upper limits to the flux in the UV, others have no data points or only upper limits in the radio band. 
All the objects in both groups have observations in the near-IR band which are treated as upper limits. We take this approach because the observations in the near-IR were taken with lower spatial resolutions, i.e. larger apertures that include considerable contamination from the emission of the host galaxy. Therefore, near-IR observations can only be considered as upper limits to the nuclear emission.

The masses of the SMBHs were estimated from the stellar velocity dispersions using the $M_{\rm BH}-\sigma$ relationship \citep{ferrarese00, geb00, tremaine02}, with the exception of NGC 266, whose black hole mass was estimated from the measured 5 GHz radio and the 2-10 keV X-ray luminosities using the correlation of \citet{merloni03} (the ``fundamental plane of black hole activity''), and NGC 3031 (M81) and NGC 4486 (M87) whose black hole masses were estimated from the stellar and/or gas kinematics (\citealt{bower00,devereux03,geb09}; see Table \ref{tab:sample}). 
As EHF10 note, for the objects with multiple mass determinations using different methods, the estimated masses are consistent with each other within a factor of 2. Due to the intrinsic scatter in the fundamental plane, the mass estimated for NGC 266 is subject to an uncertainty by a factor of $\approx 10$ \citep{merloni03}.

The optical-UV data of all objects were corrected for extinction as discussed in EHF10. 
In order to compute the bolometric luminosities from the SEDs, EHF10 used two methods. For the objects with the best sampled SEDs, they computed $L_{\rm bol}$ by integrating the SEDs directly, ignoring upper limit data points. The objects with well-sampled SEDs are NGC 3031, NGC 3998, NGC 4374, NGC 4486, NGC 4579 and NGC 4594. For these objects they assumed that pairs of neighboring points in the SEDs defined a power law, integrated each segment individually and summed the segments to obtain $L_{\rm bol}$. From this set of best sampled SEDs they estimated the average ``bolometric correction'' from the 2-10 keV luminosity to the bolometric luminosity ($L_{\rm bol}=50 L_X$), which they used to obtain $L_{\rm bol}$ for the remaining objects, for which the SEDs are not as well sampled. These bolometric luminosities are listed in Table \ref{tab:sample}.

\section{Models for the accretion flow and jet} \label{sec:models}

\begin{figure}[!ht]
\centering
\includegraphics[scale=0.25]{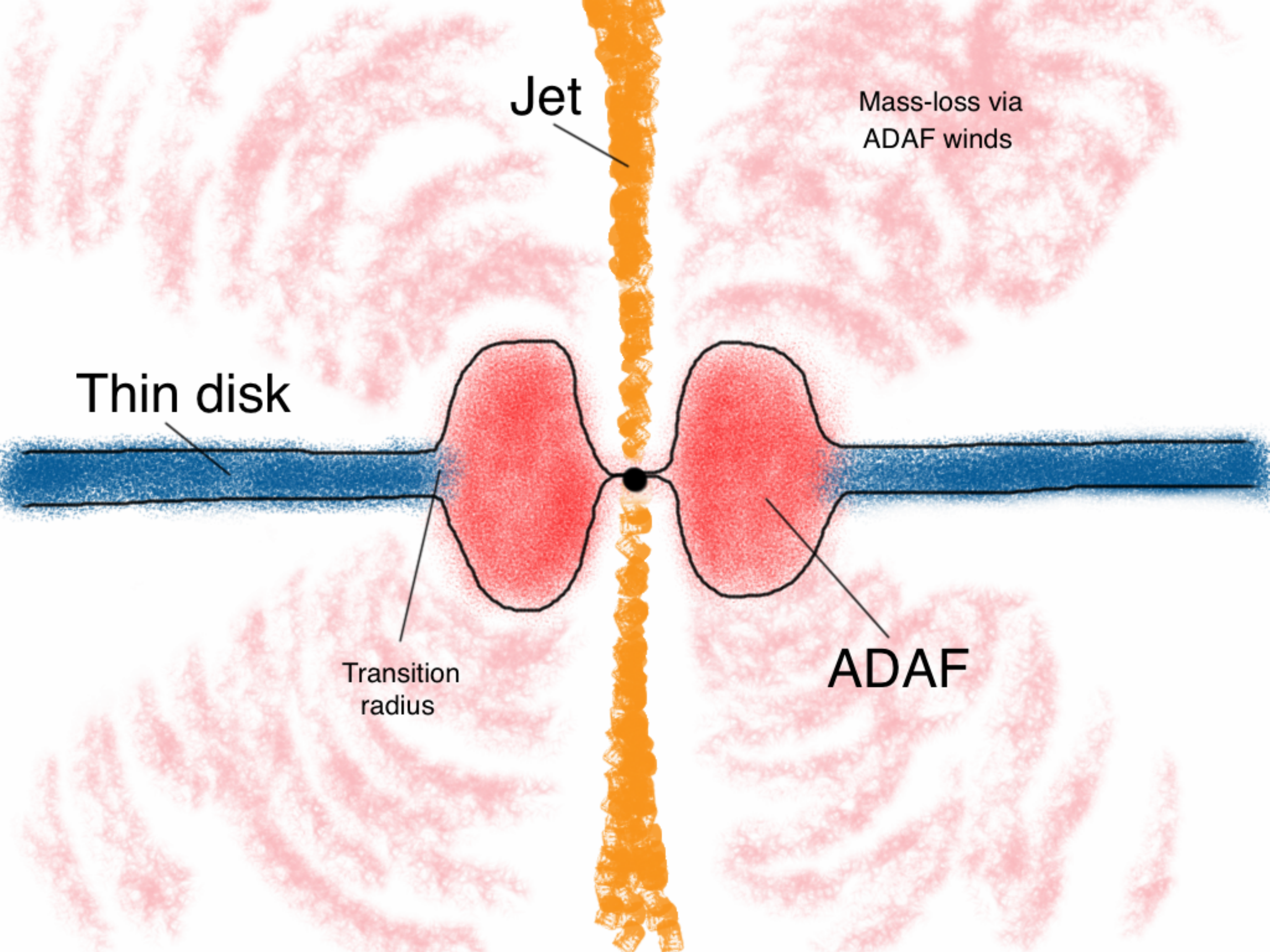}
\caption{Cartoon illustrating the model for the central engines of LLAGNs. It consists of three components: an inner ADAF, an outer truncated thin disk and a relativistic jet.}
\label{fig:cartoon}
\end{figure}

We fit the observed broadband SEDs of the LLAGNs in our sample using a model which consists of three components: an inner ADAF, an outer truncated thin accretion disk and a jet. The components of the model are illustrated in Figure \ref{fig:cartoon}. We describe here the main features of this model. 

\subsection{ADAF component}

The inner part of the accretion flow is in the form of an ADAF which is a hot, geometrically thick, optically thin two-temperature accretion flow, which has low radiative efficiency (e.g., \citealt{nar98, kato98}). ADAFs are characterized by the presence of outflows or winds, which prevent a considerable fraction of the gas that is available at large radii from being accreted onto the black hole. This has been suggested by numerical simulations \citep{stone99,hawley01,igu03,devilliers03,mckinney04} and analytical work \citep{ny94,bb99,nar00,quataert00}. In addition, observational studies suggest that winds coming from the inner parts of accretion flows are ubiquitous in AGNs (e.g., \citealt{tombesi10a,tombesi10b}) and it is natural to take them into account when studying the LLAGN population.

In order to take this mass-loss into account, we follow \citet{bb99} and introduce the parameter $s$ by 
\begin{equation}	\label{eq:massloss}
\dot{M}=\dot{M}_{\rm out} \left( \frac{R}{R_{\rm out}} \right)^{s}, 
\end{equation}
to describe the radial variation of the accretion rate $\dot{M}_{\rm out}$ measured at the outer radius of the ADAF, $R_{\rm out}$. Following the results of numerical simulations of the dynamics of ADAFs, in our models we allow $s$ to vary over the range $0 < s < 1$, taking into account the current uncertainty in the determination of the amount of gas that is lost via winds in the accretion flow. 

The other parameters that describe the ADAF solution are the black hole mass $M$; the viscosity parameter $\alpha$; the modified plasma $\beta$ parameter, defined as the ratio between the gas and total pressures, $\beta=P_g/P_{\rm tot}$; $\delta$, the fraction of energy dissipated via turbulence that directly heats electrons; and the adiabatic index $\gamma$. In our calculations, we adopt $\alpha=0.3$, $\beta=0.9$ and $\gamma=1.5$. Traditional ADAF models adopted $\delta$ to be small ($\delta \lesssim 0.01$; e.g., \citealt{ny95}). On the other hand, it has been argued that the value of $\delta$ can be potentially increased due to different physical processes - such as magnetic reconnection - that affect the heating of protons and electrons in hot plasmas (e.g., \citealt{quatgru99,sharma07}). Given the theoretical uncertainty related to the value of $\delta$, we allow it to vary over the range $0.01 \leq \delta \leq 0.5$. 

The case in which $\delta = 0.5$ and $s = 0.3$ deserves special attention, since these values were obtained from fitting current ADAF models to the SED of Sgr A* \citep{yuan03,yuan06}. 
We note that while \citet{yuan09} fix the values of $\delta$ and $s$ in their fits, we leave these parameters free taking into account the theoretical uncertainties.
There are degeneracies between $s$ and $\delta$ when fitting the data with ADAF models \citep{quat99}, such that different combinations of the values of these parameters may result in similar SEDs. 

The cooling mechanisms incorporated in the calculations are synchrotron emission, bremsstrahlung and inverse Comptonization of the seed photons produced by the first two radiative processes. Given the values of the parameters of the ADAF, in order to compute its spectrum we first numerically solve for the global structure and dynamics of the flow, as outlined in \citet{yuan00,yuan03}. Obtaining the global solution of the differential equations for the structure of the accretion flow is a two-point boundary value problem. This problem is solved numerically using the shooting method, by varying the eigenvalue $j$ (the specific angular momentum of the flow at the horizon) until the sonic point condition at the sonic radius $R_s$ is satisfied, in addition to the outer boundary conditions. 

There are three outer boundary conditions that the ADAF solution must satisfy, specified in terms of the three variables of the problem: the ion temperature $T_{i}$, the electron temperature $T_{e}$ and the radial velocity $v$ (or equivalently the angular velocity $\Omega$). Following \citet{yuan08}, when the outer boundary of the ADAF is at the radius $R_{\rm out}=10^4 R_S$ (where $R_S$ is the Schwarzschild radius) we adopt the outer boundary conditions $T_{\rm out,i}=0.2 T_{\rm vir}$, $T_{\rm out,e}=0.19 T_{\rm vir}$ and $\lambda_{\rm out}=0.2$, where the virial temperature is given by $T_{\rm vir}=3.6 \times 10^{12} (R_S/R) \ \rm K$, $\lambda \equiv v/c_s$ is the Mach number and $c_s$ is the adiabatic sound speed. When the outer boundary is at $R_{\rm out} \sim 10^2 R_S$ we adopt the boundary conditions $T_{\rm out,i}=0.6 T_{\rm vir}$, $T_{\rm out,e}=0.08 T_{\rm vir}$ and $\lambda_{\rm out}=0.5$. After the global solution is calculated, the spectrum of the accretion flow is obtained (see e.g., \citealt{yuan03} for more details). We verified that if these boundary conditions are varied by a factor of a few, the resulting spectrum does not change much.

\subsection{Thin disk component}

Outside the ADAF there is an outer thin accretion disk with an inner radius truncated at $R_{\rm tr}=R_{\rm out}$ and extending up to $10^5 R_S$. Therefore, in practice the outer radius of the ADAF corresponds to the transition radius to the thin disk.
The other parameters that describe the thin disk solution are the inclination angle $i$, the black hole mass and the accretion rate $\dot{M}_{\rm out}$ (the same as the accretion rate at the outer boundary of the ADAF). In the cases where $R_{\rm out} \sim 10^4 R_S$ we simply ignore the contribution of the thin disk spectrum.

The thin disk emits locally as a blackbody and we take into account the reprocessing of the X-ray radiation from the ADAF. This reprocessing effect has only a little impact on the spectrum of the thin disk though, with the resulting SED being almost identical to that of a standard thin disk (e.g., \citealt{frank02}). 

\subsection{Jet component}

The SEDs of LLAGNs are generally radio-loud (\citealt{ho99, ho01, ho02, tera03}; EHF10); but see \citealt{maoz07,sikora07}). The radio prominence of these SEDs is usually explained by invoking the synchrotron emission of relativistic jets, since the the accretion flow does not produce enough radio emission to account for the observed radio luminosity (e.g., \citealt{quat99, ulvestad01, wu05, nemmen06, wu07, yuan09}). Some authors even argue that the entire SED of LLAGNs may be explained by the jet component (e.g., \citealt{falcke04,markoff08}). We therefore include in our modelling the contribution from the emission from a relativistic jet.

We adopt a jet model based on the internal shock scenario which is used to interpret gamma-ray burst afterglows (e.g., \citealt{piran99, spada01}; see \citealt{yuan05} for more details). This model has been adopted in previous works to understand the broadband SEDs of X-ray binaries and AGNs \citep{spada01, yuan05, nemmen06, wu07}. According to this jet model, some fraction of the material in the innermost regions of the accretion flow is transferred to the jet producing an outflow rate $\dot{M}_{\rm jet}$ and a standing shock wave in the region of the jet closest to the black hole is formed. This shock wave is created from the bending of the supersonic accretion flow near the black hole in the direction of the jet. We calculate the shock jump (Rankine-Hugoniot) conditions to find the electron and ion temperatures of the plasma ($T_e$ and $T_i$). We find that the jet spectrum is not very sensitive to changes in $T_e$ and $T_i$, since the emission is completely dominated by non-thermal electrons (see below). We therefore adopt $T_i=6.3 \times 10^{11} \ {\rm K}$ and $T_e=10^{9} \ {\rm K}$ in our calculations of the jet emission. 

The jet is modelled as having a conical geometry with half-opening angle $\phi$ and a bulk Lorentz factor $\Gamma_j$ which are independent of the distance from the black hole. The jet is along the axis of the ADAF and makes an angle $i$ with the line of sight. The internal shocks in the jet are presumably created by the collisions of shells of plasma with different velocities. These shocks accelerate a small fraction $\xi_e$ of the electrons into a power-law energy distribution with index $p$. The energy density of accelerated electrons and the amplified magnetic field are described by two free parameters, $\epsilon_e$ and $\epsilon_B$. Following \citet{nemmen06,wu07}, we adopt in our calculations of the jet emission the values $\phi=0.1$ radians, $\xi_e=10 \%$ and $\Gamma_j=2.3$, which corresponds to $v/c \approx 0.9$ (except in the case of M87, which has an independent estimate of $\Gamma_j$ available). Unless otherwise noted, we adopt $i=30^\circ$. Therefore, there are four free parameters in the jet model: $\dot{M}_{\rm jet}$, $p$, $\epsilon_e$ and $\epsilon_B$.
In our calculations we consider the synchrotron emission of the jet, with the optically thick part of the synchrotron spectrum contributing mainly in the radio and the optically thin part contributing mainly in the X-rays. 


\subsection{Free parameters and fit procedure}

When fitting the observed SEDs with our coupled accretion-jet model, we have eight free parameters. Four of these parameters describe the emission of the accretion flow: the accretion rate $\dot{M}_{\rm out}$, the transition radius between the inner ADAF and the outer truncated thin disk $R_{\rm tr}$, the fraction of viscously dissipated energy that directly heats the electrons $\delta$ and the strength of the wind from the ADAF $s$. The other four parameters characterize the jet emission: the mass-loss rate in the jet $\dot{M}_{\rm jet}$, the electron energy spectral index $p$, and the electron and magnetic energy parameters $\epsilon_e$ and $\epsilon_B$. $\dot{M}_{\rm out}$ and $\dot{M}_{\rm jet}$ are obviously expected to be correlated, but given our ignorance about the mechanism of jet formation we vary these parameters independently in our fits. 
We require for consistency that $\dot{M}_{\rm jet} / \dot{M}_{\rm out} < 1$. Since the ADAF loses mass via winds (equation \ref{eq:massloss}), it is also of interest to check the value of $\dot{M}_{\rm jet} / \dot{M}(3 R_S)$, in which the accretion rate is calculated at the radius of the innermost stable circular orbit for a Schwarzschild black hole. 

Throughout this paper we will use the dimensionless mass accretion rates $\dot{m}=\dot{M}/\dot{M}_{\rm Edd}$, noting that the Eddington accretion rate is defined as
$\dot{M}_{\rm Edd} \equiv 22 M /(10^{9} M_\odot) \; M_\odot \, {\rm yr}^{-1}$ 
assuming a $10\%$ radiative efficiency. We will also express the black hole mass in terms of the mass of the sun, $m=M/M_{\odot}$, and the radius in terms of the Schwarzschild radius, $r=R/R_S$.

Our fitting procedure can be summarized as follows. We vary the free jet parameters in order to fit the radio observations, since the radio is emitted predominantly by the jet. Once we fit the radio band with the jet model, we try to fit the truncated thin disk model to the IR-optical data, in order to constrain the transition radius and $\dot{M}_{\rm out}$ (keeping in mind that the theoretical spectrum should not exceed the observed luminosities). 
We then use the X-ray data to constrain the ADAF component in order to refine the estimates of $R_{\rm tr}$ and $\dot{M}_{\rm out}$, and to estimate $s$ and $\delta$. Depending on the combination of parameters that we use, the jet can have an important contribution to the emission not only in the radio, but also in the X-ray band. We sum the emission of the inner ADAF, outer thin disk and jet, and compare this sum with the observed SEDs. 

We explore the parameter space of the models considering the many possible fits to the data. 
We search the literature for independent constraints from other works on the values of the model parameters, such as the mass accretion rate, inclination angle and the transition radius, as well as the jet power which is obtained from the jet model.

\section{Model fits to well sampled SEDs} \label{sec:seds}

We describe in this section the results obtained from fitting our coupled accretion-jet model to the SEDs that were selected using our selection criteria in Section \ref{sec:sample}. We describe in more detail the results for NGC 4594 (\textsection \ref{sec:n4594}), NGC 4374 (\textsection \ref{sec:m84}), NGC 4486 (\textsection \ref{sec:m87}), NGC 3031 (\textsection \ref{sec:m81}) and NGC 3998 (\textsection \ref{sec:n3998}). These five objects will serve as the ``reference'' examples that will illustrate the general results for the whole sample. These objects were chosen for a more detailed discussion because of their good multiwavelength sampling. We show the model fits to the other 16 SEDs and the resulting model parameters in Section \ref{ap:seds}.

In the SED plots that follow below, the error bars represent the uncertainty in the extinction corrections applied to the the optical-UV data by EHF10. The lower bars correspond to the measurements without any extinction correction, while the upper bars represent the same observations after a maximal extinction correction (see EHF10).
The arrows represent upper limits to the nuclear luminosity. These upper limits are caused by either non-detections, or because the corresponding observations were taken at lower spatial resolutions. In the latter case, the upper limits include a potentially significant contamination by the emission from the host galaxy. This happens particularly for the observations in the near-IR band.

Table \ref{tab:models} lists the model parameters that result from the SED fits for the AD and JD type of models. We list whenever available the Bondi accretion rate and the corresponding reference from which $\dot{m}_{\rm Bondi}$ was taken. $P_{\rm jet}^{\rm obs}$ corresponds to the jet power estimated either from observations or using the correlation between jet power and radio luminosity of \citet{merloni07}. $P_{\rm jet}^{\rm mod}$ represents the jet power resulting from our jet model, calculated as $P_{\rm jet} = \dot{M}_{\rm jet} \Gamma_j^2 c^2$. 

\subsection{The case of NGC 4594} \label{sec:n4594}

\citet{pelle05} studied \textit{Chandra} X-ray observations of NGC 4594 and derived the Bondi accretion rate $\dot{m}_{\rm Bondi}=2^{+6.8}_{-1.3} \times 10^{-3}$.

\begin{figure*}
\centerline{
\includegraphics[scale=0.5]{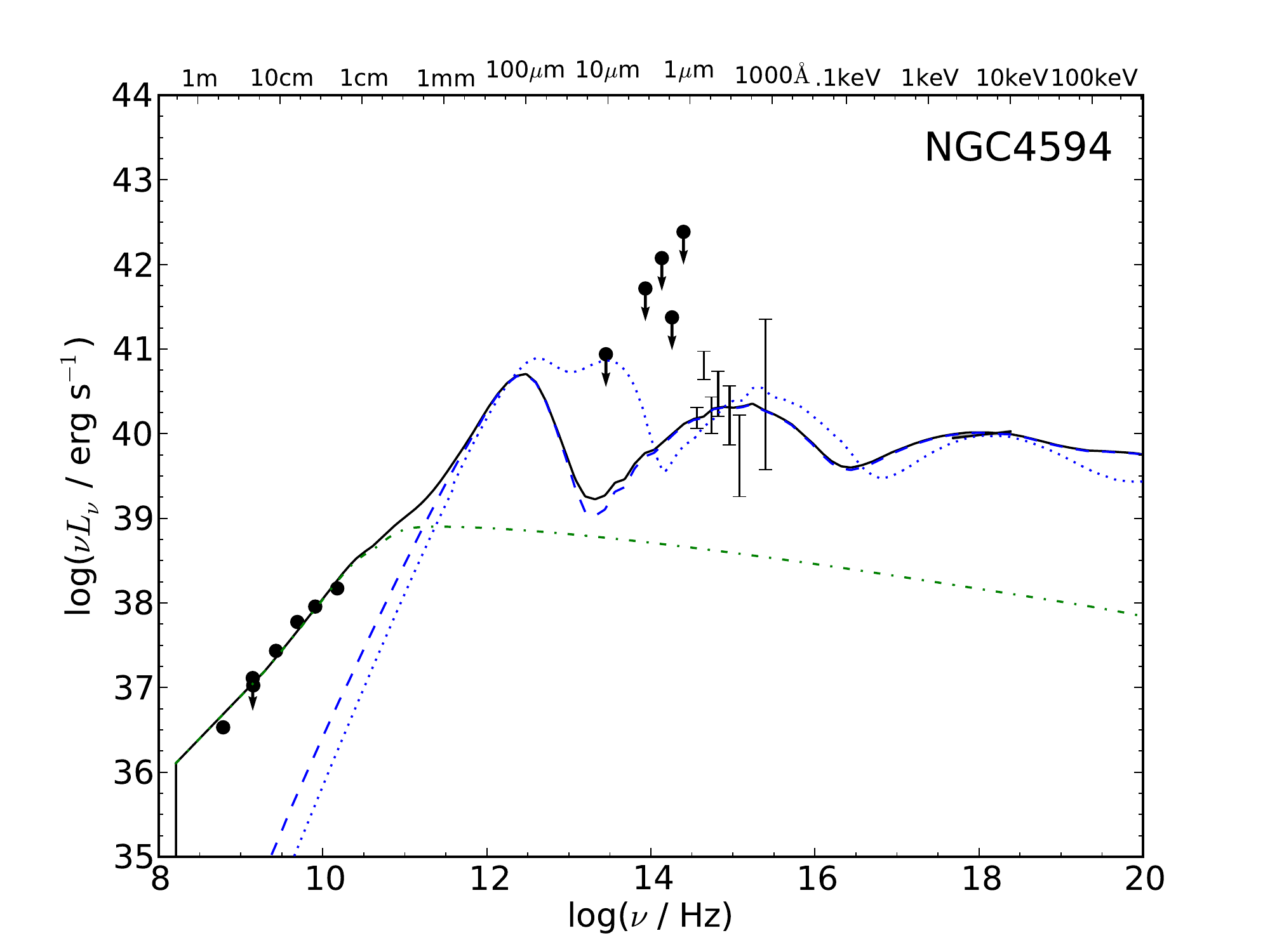}
\hskip -0.3truein
\includegraphics[scale=0.5]{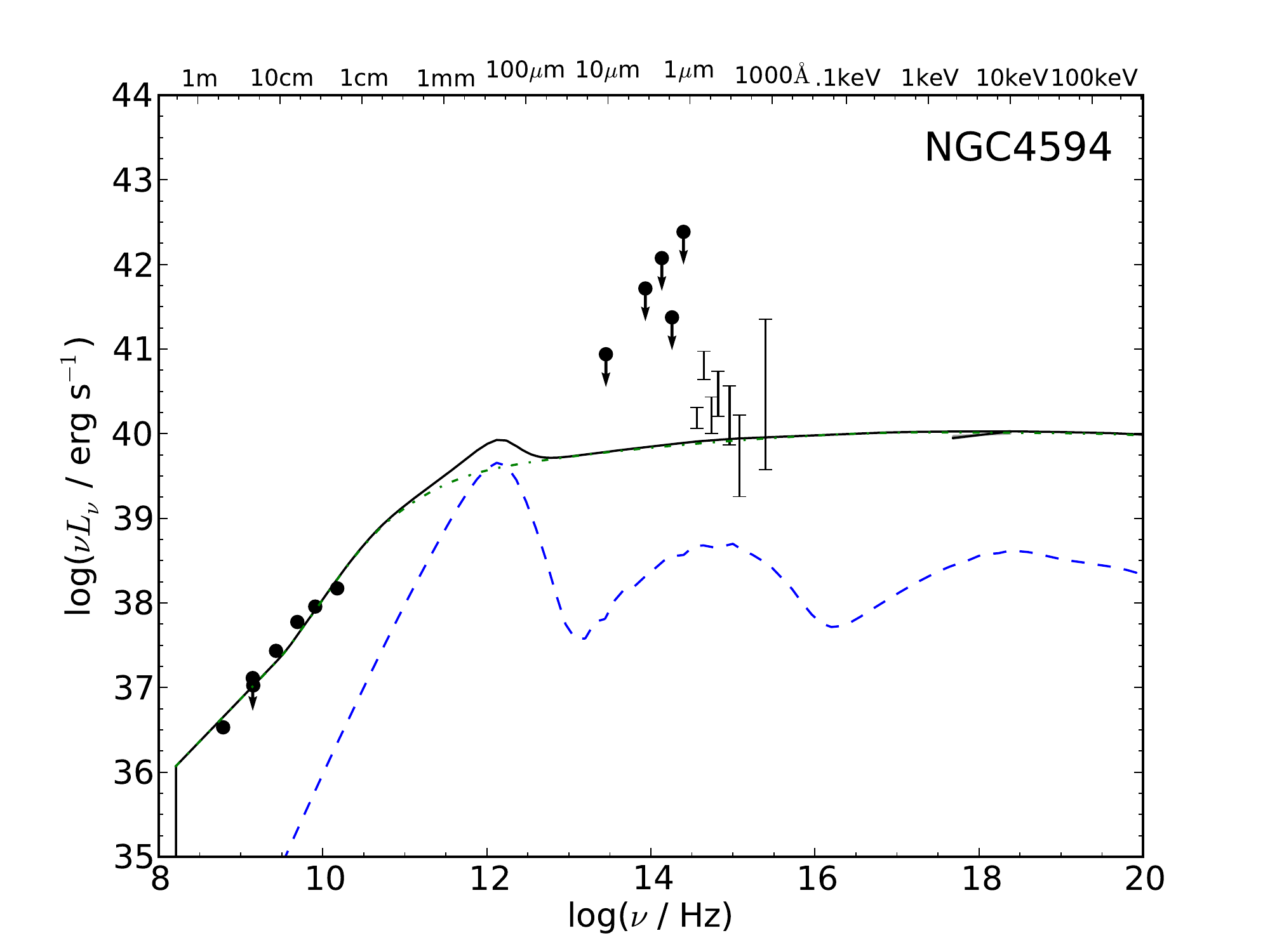}
}
\caption{Models for the SED of NGC 4594 (Sombrero). The dashed and dot-dashed lines show respectively the emission from the ADAF and jet, while the solid line shows the sum of the radiation from these components. \textbf{Left:} model in which the ADAF dominates the observed X-ray emission. The dotted line shows for comparison an ADAF + thin disk model in which the transition radius is $300 R_S$.  \textbf{Right:} model in which the jet dominates the X-ray output. See the text for the parameters.}
\label{fig:n4594}
\end{figure*}

The SED of NGC 4594 is plotted in Figure \ref{fig:n4594}, together with two different spectral fits. The left panel of Figure \ref{fig:n4594} shows an ADAF-jet model in which the \emph{ADAF component (dashed line) completely dominates the emission from the IR to the X-ray bands} and is consistent with the available optical-UV and X-ray data. In this model, the jet (\emph{dot-dashed line}) dominates the radio emission and contributes  only weakly to the X-ray flux. The solid line corresponds to the sum of the ADAF and jet emission. 

The parameters of the ADAF model are $\dot{m}_{\rm out}=6.3 \times 10^{-3}$ (consistent with the Bondi accretion rate estimated by \citealt{pelle05}), $r_{\rm out}=10^4$, $\delta=0.01$ and $s=0.3$.
The jet parameters are $\dot{m}_{\rm jet}=9 \times 10^{-7}$, $p=2.3$, $\epsilon_e=5 \times 10^{-3}$ and $\epsilon_B=0.03$. The jet kinetic power is $P_{\rm jet}/L_{\rm bol} \approx 4$. In other words, the kinetic power carried by particles in the jet exceeds the overall radiative power. The ratio of the mass-loss rate in the jet to the accretion rate estimated at $3 R_S$  is  $\dot{m}_{\rm jet} / \dot{m}(3 R_S) \approx 1.6 \times 10^{-3}$. In other words, only $\approx 0.2 \%$ of the mass that is ultimately accreted by the black hole is channeled into the jet.

The dotted line in the left panel of Figure \ref{fig:n4594} shows an ADAF + truncated thin disk model for which $r_{\rm tr}=300$, $\dot{m}_{\rm out}=9.1 \times 10^{-4}$, $\delta=0.01$ and $s=0.1$. The thin disk emission corresponds to the ``bump'' at $10 \mu$m. This model demonstrates that accretion flows with a much smaller transition radius are also consistent with the data. 
Since we have only upper limit points in the mid to near-IR for NGC 4594, the available data are not sufficient to constrain the presence of a truncated thin disk in this source.

The right panel of Fig. \ref{fig:n4594} shows a model for the SED of NGC 4594 in which the jet dominates the radio and X-ray emission. In fact, the right panel shows that \emph{the relativistic jet can account well for the entire observed SED}. The jet parameters are $\dot{m}_{\rm jet}=4.5 \times 10^{-7}$, $p=2.01$, $\epsilon_e=0.8$ and $\epsilon_B=3 \times 10^{-2}$, with $P_{\rm jet}=8.4 \times 10^{42} \ {\rm erg \ s}^{-1}$, $P_{\rm jet}/L_{\rm bol} \approx 4$ and $\dot{m}_{\rm jet} / \dot{m}(3 R_S) \approx 0.009$.

We display in the right panel of Figure \ref{fig:n4594} one illustrative ADAF model which is consistent with the possibility that  the jet emission dominates in this source, i.e. the ADAF contribution in all bands (except at 1 mm) is very small compared to the jet. The ADAF parameters are $\dot{m}_{\rm out} = \dot{m}_{\rm Bondi}$, $r_{\rm out}=10^4$, $\delta=0.01$ and $s=0.3$. 
We can see that the ADAF contribution in this case is only significant in the wavelength ranges $\sim 1 \ {\rm mm} - 100 \ \mu {\rm m}$. The importance of the ADAF emission in the other bands is small compared to the jet. 



By exploring the parameter space of the accretion/jet models within the range of plausible values allowed by theory, we demonstrate that there are basically two possible types of models which can accommodate the observed SED of NGC 4594 and the other LINERs in this section. In the first type, the emission from the ADAF dominates the observed X-rays; in the second type of model, the jet emission dominates the X-rays. We will hereafter use the abbreviations AD (as in \emph{ADAF-dominated}) and JD (as in \emph{jet-dominated}) when referring to the former and latter types of models, respectively. A third type of model is also possible in which the jet and the ADAF contribute with similar intensities to the high energy emission. These results apply not only to the five LINERs discussed in more detail in this section but also to the other sources in our sample with a few exceptions.

\subsection{NGC 4374}	\label{sec:m84}


A prominent jet resolved with the VLA is observed to create cavities in the X-ray emitting gas \citep{allen06, fino08}. The Bondi accretion rate is $\dot{m}_{\rm Bondi}=4 \times 10^{-4}$ \citep{pelle05, allen06}.

\begin{figure*}
\centerline{
\includegraphics[scale=0.5]{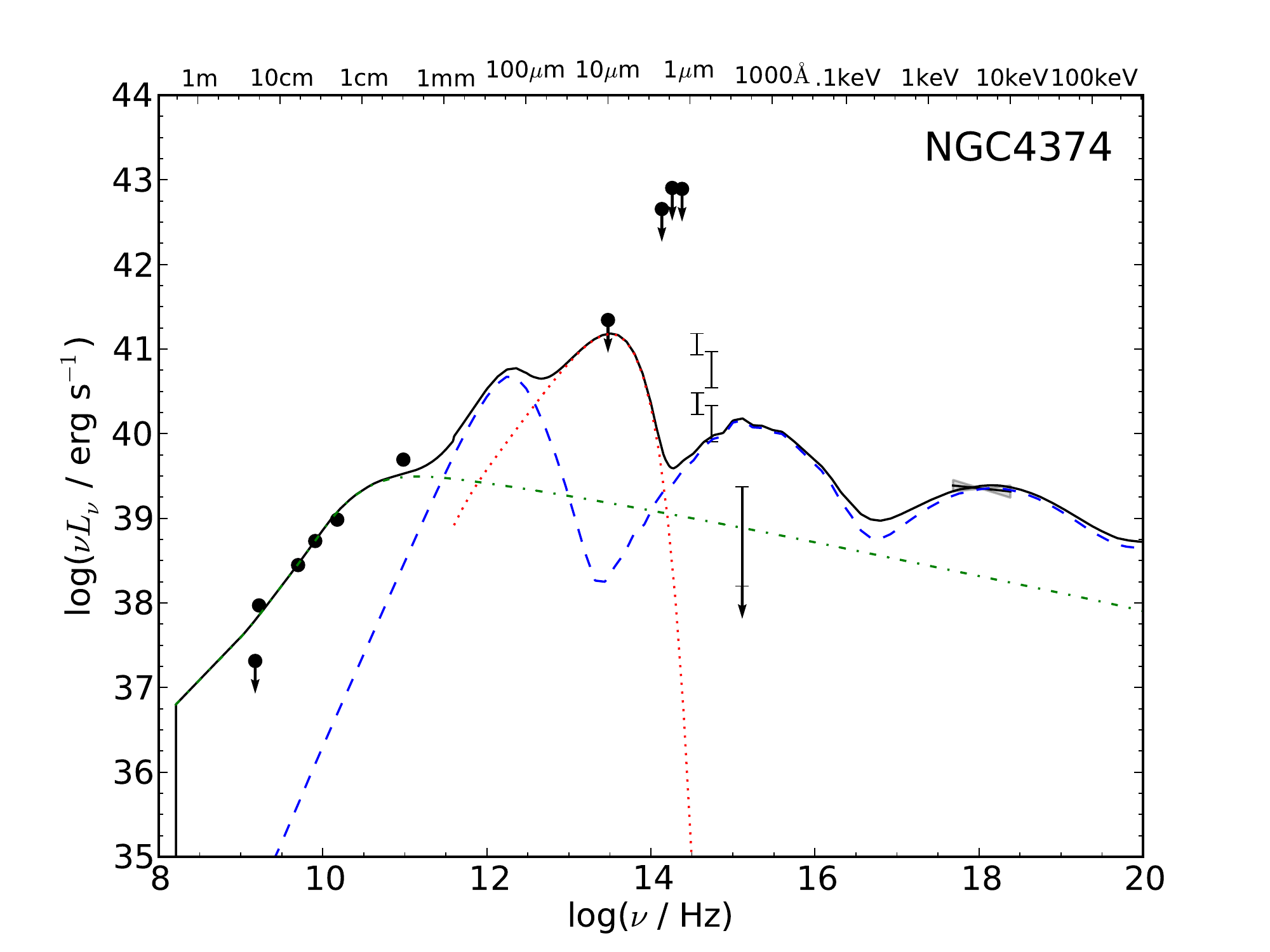}
\hskip -0.3truein
\includegraphics[scale=0.5]{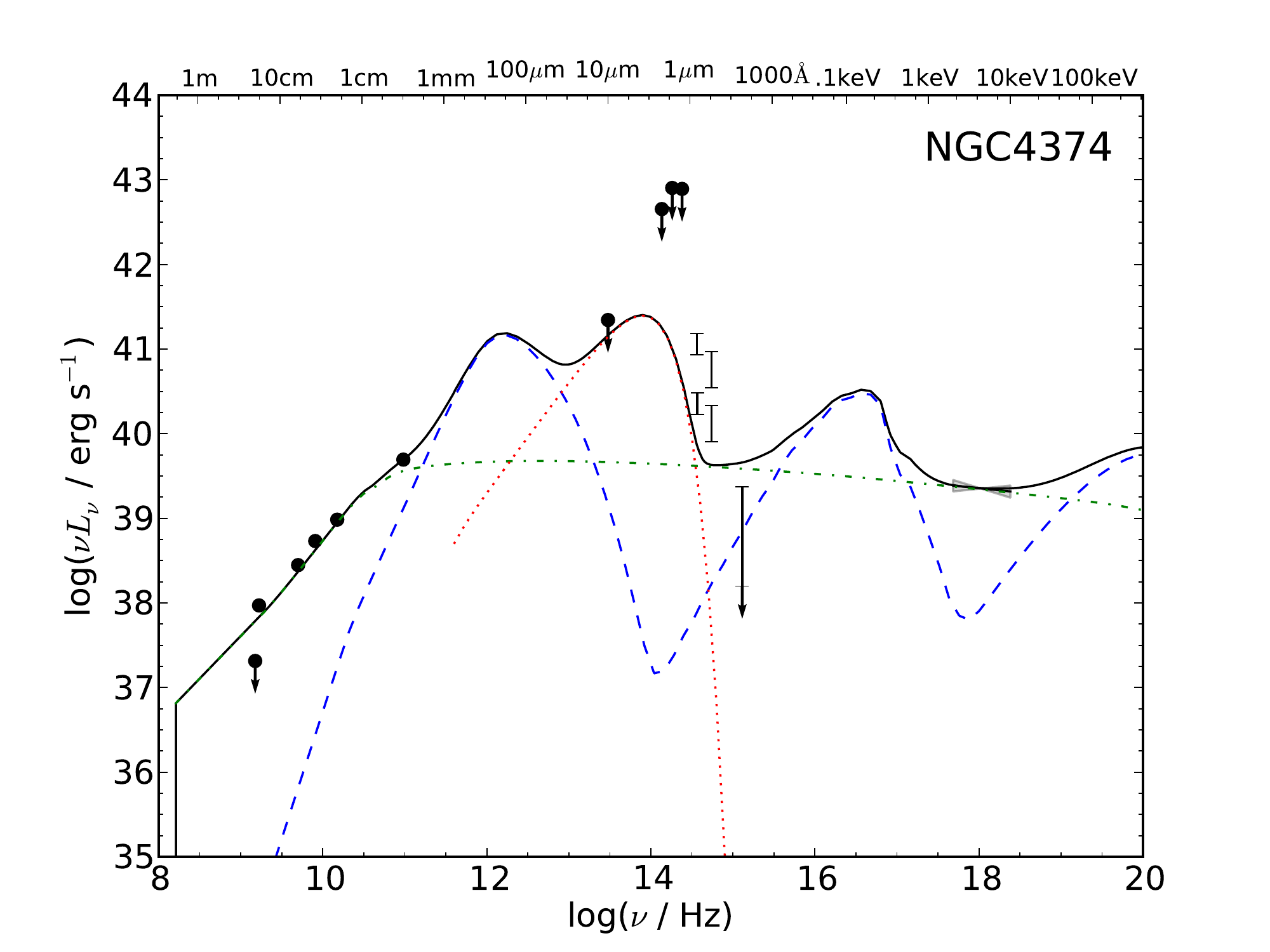}
}
\caption{Models for the SED of NGC 4374/M84 showing the emission of the ADAF (\emph{dashed}), jet (\emph{dot-dashed}), truncated thin disk (\emph{dotted}) and the total emission (\emph{solid}). \textbf{Left:} model in which the ADAF dominates the observed X-ray emission (``AD model''). \textbf{Right:} model in which the jet dominates the X-ray output (``JD model'').}
\label{fig:n4374}
\end{figure*}

The left panel of Figure \ref{fig:n4374} shows an ADAF-jet model (\textit{dashed line}) in which the ADAF completely dominates the emission from the near-IR to the X-ray bands and explains well the available optical-UV and X-ray data, as was the case for NGC 4594. The synchrotron peak of the ADAF also dominates the emission in the range $\sim 1 \ {\rm mm} - 100 \ \mu {\rm m}$. The jet (\textit{dot-dashed line}) dominates the radio emission, but contributes weakly to the X-ray flux. The thin disk dominates the IR emission and its contribution peaks at $\lambda \sim 10 \ \mu$m, creating a ``red bump'' in the SED as opposed to the ``big blue bump'' observed in quasars (e.g., \citealt{koratkar99, nemmen10}). 
The parameters of the accretion flow are $\dot{m}_{\rm out}=3.95 \times 10^{-4}$, $r_{\rm tr}=150$, $\delta=0.01$ and $s=0.1$. 
As was the case for NGC 4594, the mid to near-IR data are insufficient to make the case of a truncated thin disk compelling and similar models with much larger values of $r_{\rm tr}$ are not ruled out by the available IR data. 

The jet parameters are $\dot{m}_{\rm jet}=4 \times 10^{-7}$, $p=2.4$, $\epsilon_e=0.01$ and $\epsilon_B=0.1$. The resulting jet power is $P_{\rm jet}=2.1 \times 10^{42} \ {\rm erg \ s}^{-1} \approx 4 L_{\rm bol}$ which is in excellent agreement with the jet power estimated by \citet{merloni07} based on the calorimetry of the X-ray cavities observed with \emph{Chandra} ($3.9 \times 10^{42} \ {\rm erg \ s}^{-1}$). The ratio $\dot{m}_{\rm jet} / \dot{m}(3 R_S)$ is $1.5 \times 10^{-3}$. 

The right panel of Fig \ref{fig:n4374} shows a model in which the jet dominates the radio and the 1-100 keV X-ray band. The emission in the region between 1000 $\rm \AA$ and 1 keV is dominated by the Compton peak of the ADAF spectrum. As was the case for the previous model, the IR emission is dominated by a truncated thin disk with a small transition radius ($r_{\rm tr}=30$) which is even smaller than the AD fit. At $\nu > 100$ keV the bremmstrahlung emission of the ADAF is the dominant radiation process. 
The parameters of ADAF are $\dot{m}_{\rm out}=1.5 \times 10^{-4}$ (in agreement with the Bondi rate) and $\delta=s=0.3$. The jet parameters are $\dot{m}_{\rm jet}=1.6  \times 10^{-6}$, $p=2.2$, $\epsilon_e=9  \times 10^{-3}$ and $\epsilon_B=8 \times 10^{-3}$, with $P_{\rm jet}=8.4 \times 10^{42} \ {\rm erg \ s}^{-1} \approx 17 L_{\rm bol}$. The ratio $\dot{m}_{\rm jet} / \dot{m}(3 R_S)$ is $\dot{m}_{\rm jet} / \dot{m}(3 R_S) \approx 0.02$. There are no data available to constrain the predicted ADAF bumps at soft and hard X-rays in this model. Notice that, similarly to the JD fit to the SED of NGC 4594, the shape of X-ray spectrum from the ADAF with $\delta=0.3$ (dashed line) does not agree with the data.

\subsection{NGC 4486} \label{sec:m87}

\citet{dmt03} previously calculated the Bondi accretion rate using Chandra X-ray data and found it to be $\dot{M}_{\rm Bondi} \sim 0.1 \ M_\odot \ \rm{yr}^{-1}$ ($\dot{m}_{\rm Bondi} \approx 7 \times 10^{-4}$). They also fitted the multiwavelength SED of M87 with an ADAF model for different values of $\delta$ (but not including mass-loss in the accretion flow, i.e. $s=0$) and found that the ADAF emission roughly explains the SED. 
\citet{biretta99} analysed HST observations of the jet in M87 and estimated that $\Gamma_j \geq 6$ and $10^\circ<i<19^\circ$. Based on the results of \citet{biretta99}, we adopt in our jet modelling the parameters $\Gamma_j=6$ and $i=10^\circ$. 
The kinetic power carried by the jet is estimated to be in the range $10^{43} - 10^{44} \ {\rm erg \ s}^{-1}$ (e.g., \citealt{bicknell99, owen00, allen06, merloni07}), therefore the jet power resulting from fitting our jet model must be within this range.

\begin{figure*}
\centerline{
\includegraphics[scale=0.5]{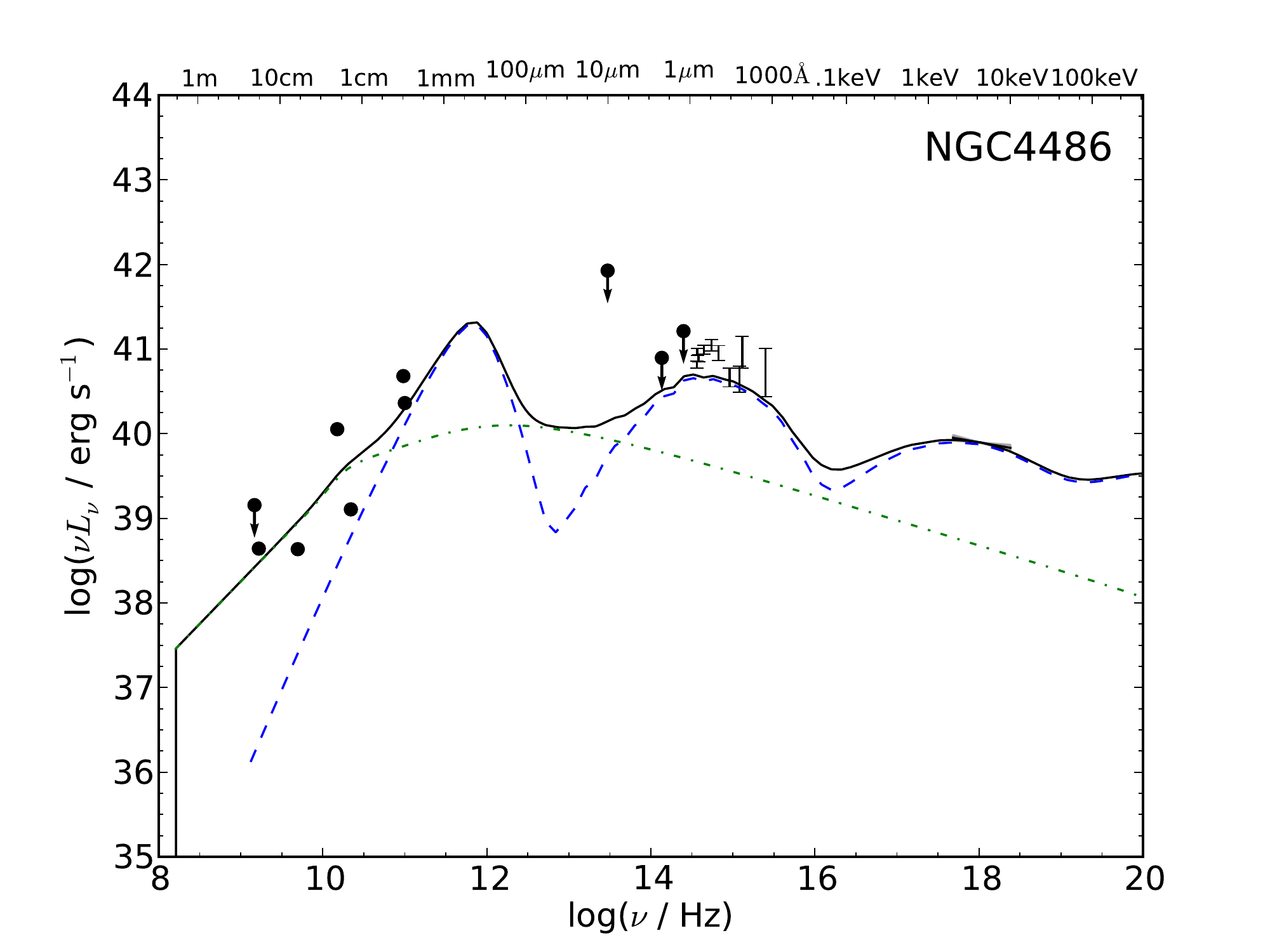}
\hskip -0.3truein
\includegraphics[scale=0.5]{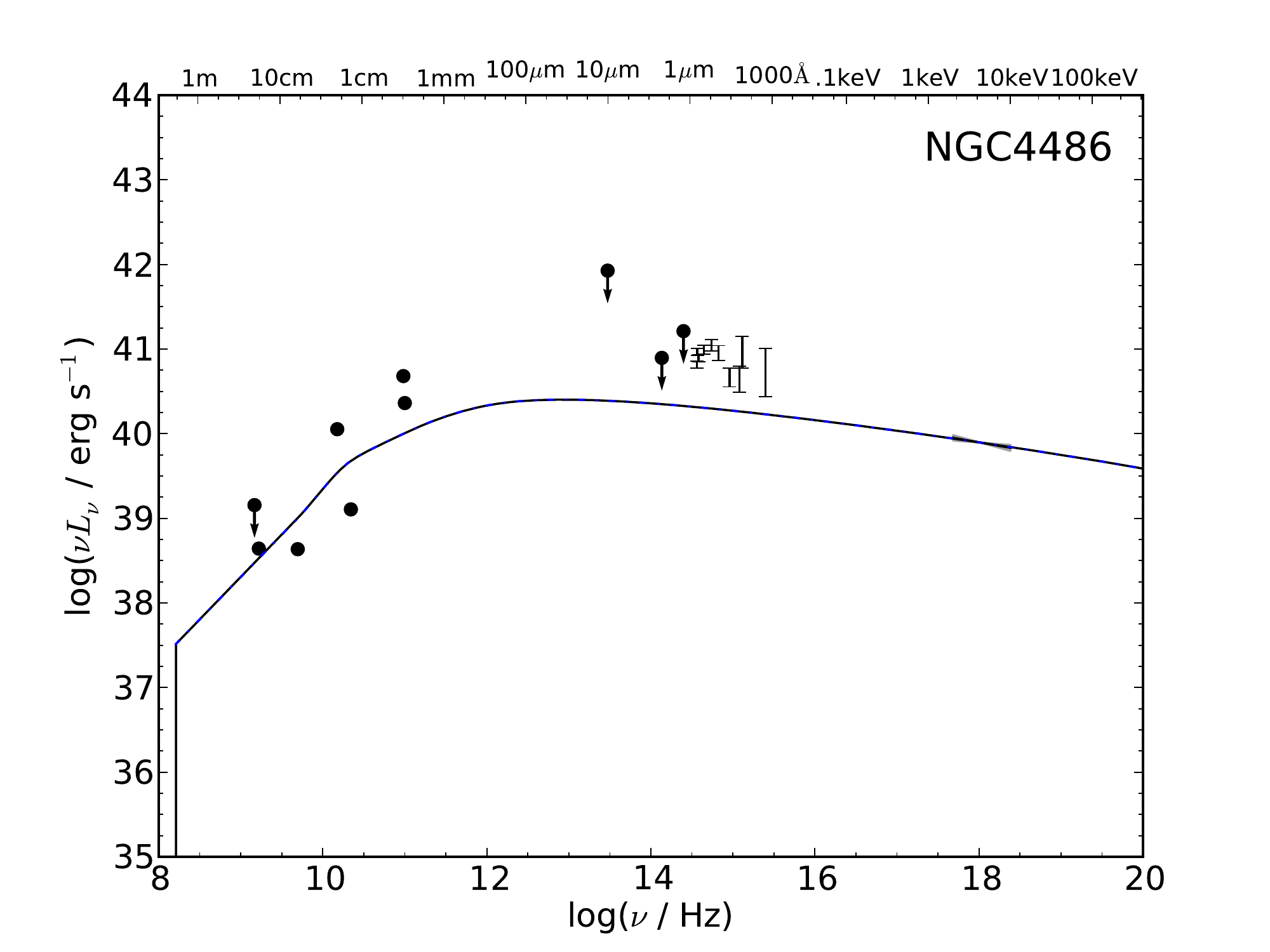}
}
\caption{Same as Figure \ref{fig:n4374} for M87/NGC 4486.}
\label{fig:m87}
\end{figure*}

The left panel of Figure \ref{fig:m87} shows an ADAF-jet model in which the ADAF dominates the optical-UV (OUV) and X-ray emission (i.e. AD-type model), while the jet dominates the radio band. The parameters of the accretion flow are $\dot{m}_{\rm out}=5.5 \times 10^{-4}$ (in agreement with the Bondi rate), $r_{\rm out}=10^4$, $\delta=0.01$ and $s=0.1$. The jet parameters are $\dot{m}_{\rm jet}=5 \times 10^{-8}$, $p=2.6$, $\epsilon_e=10^{-3}$ and $\epsilon_B=8 \times 10^{-3}$. The jet power is given by $P_{\rm jet}=6 \times 10^{42} \ {\rm erg \ s}^{-1}$, with $P_{\rm jet}/L_{\rm bol} \approx 6$ and $\dot{m}_{\rm jet} / \dot{m}(3 R_S) \approx 9 \times 10^{-5}$. 

The right panel in Fig \ref{fig:m87} shows a jet model which explains quite well the radio and X-ray observations (JD-type model), although it underpredicts the OUV data by a factor of a few. The jet parameters are $\dot{m}_{\rm jet}=3  \times 10^{-8}$, $p=2.3$, $\epsilon_e=10^{-3}$ and $\epsilon_B=8 \times 10^{-3}$, with $P_{\rm jet}=9 \times 10^{42} \ {\rm erg \ s}^{-1}$ and $P_{\rm jet}/L_{\rm bol} \approx 9$. 



The SED of M87 was previously fitted using ADAF and/or jet models \citep{dmt03, yuan09, li09}. \citet{dmt03} modelled the SED of M87 with an ADAF model using different values of $\delta$ but not including mass-loss (i.e., $s=0$). The model adopted by \citet{li09} is quite similar to that of \citet{dmt03} although the former incorporate general relativistic corrections. \citet{dmt03, li09} obtained that the ADAF emission with no mass-loss approximately reproduces the SED and results in accretion rates consistent with the Bondi rate. Our AD model is similar to that of \citet{dmt03} but it also incorporates the effect of winds, as suggested by numerical simulations. 

\citet{yuan09} tried to model the SED using an ADAF model with $\delta=0.5$ but failed to fit to data with an AD model. The reason is that for high values of $\delta$ ($\delta > 0.1$) and small accretion rates ($\dot{m}_{\rm out} \ll 0.01$) the ADAF X-ray spectrum is harder than the data. \citet{yuan09} instead successfully fit the data with a JD model using $p=0.5$.

\subsection{NGC 3031}	\label{sec:m81}

This source is among the brightest LLAGNs known since it is the nearest AGN besides Centaurus A and has been the subject of a broadband multiwavelength monitoring campaign \citep{markoff08, miller10}. Due to the many available observational constraints, accretion-jet models can be well constrained for this LLAGN. For illustrative purposes, here we will report our modelling results using the data compiled by EHF10.

\begin{figure*}
\centerline{
\includegraphics[scale=0.5]{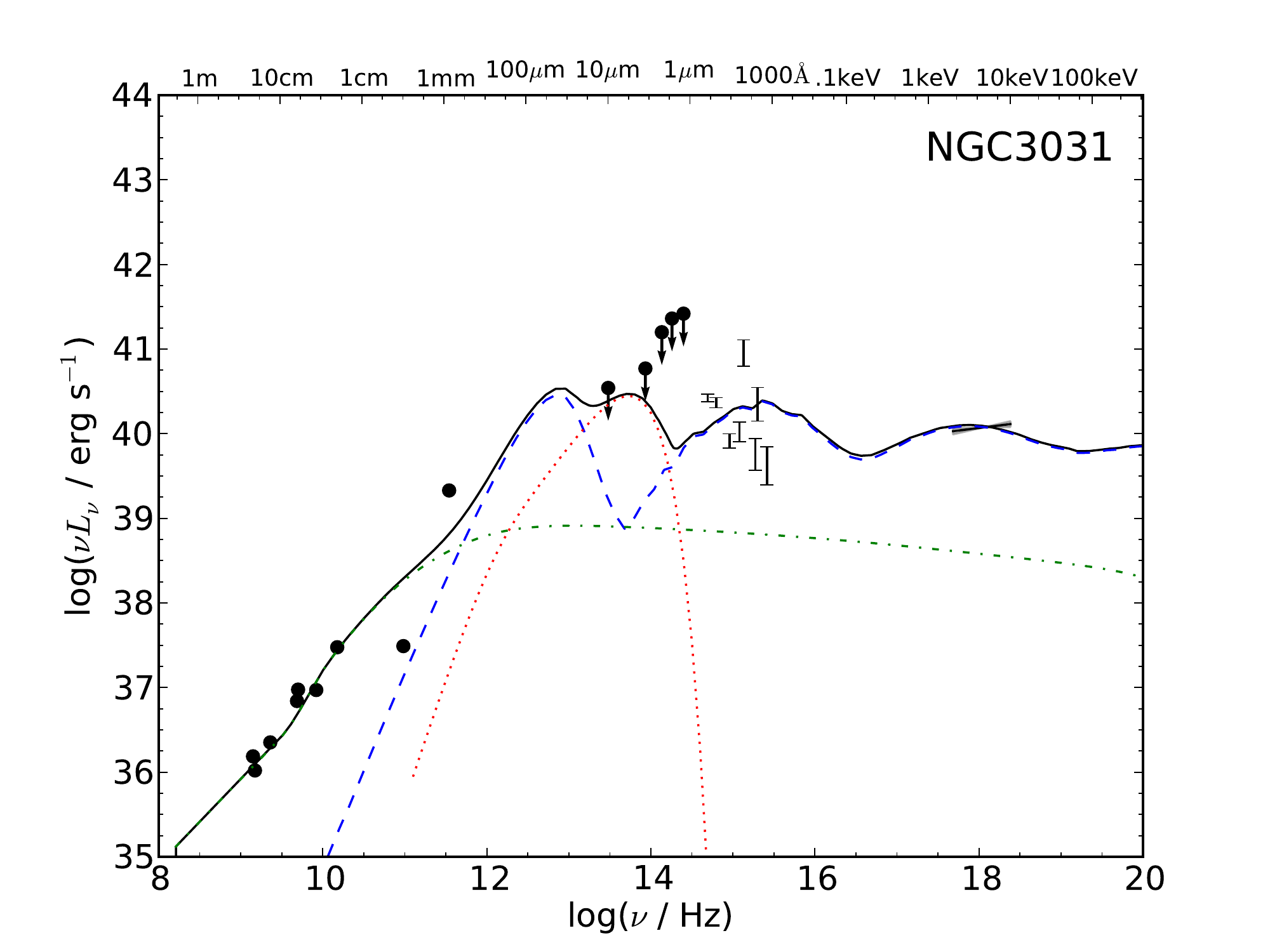}
\hskip -0.3truein
\includegraphics[scale=0.5]{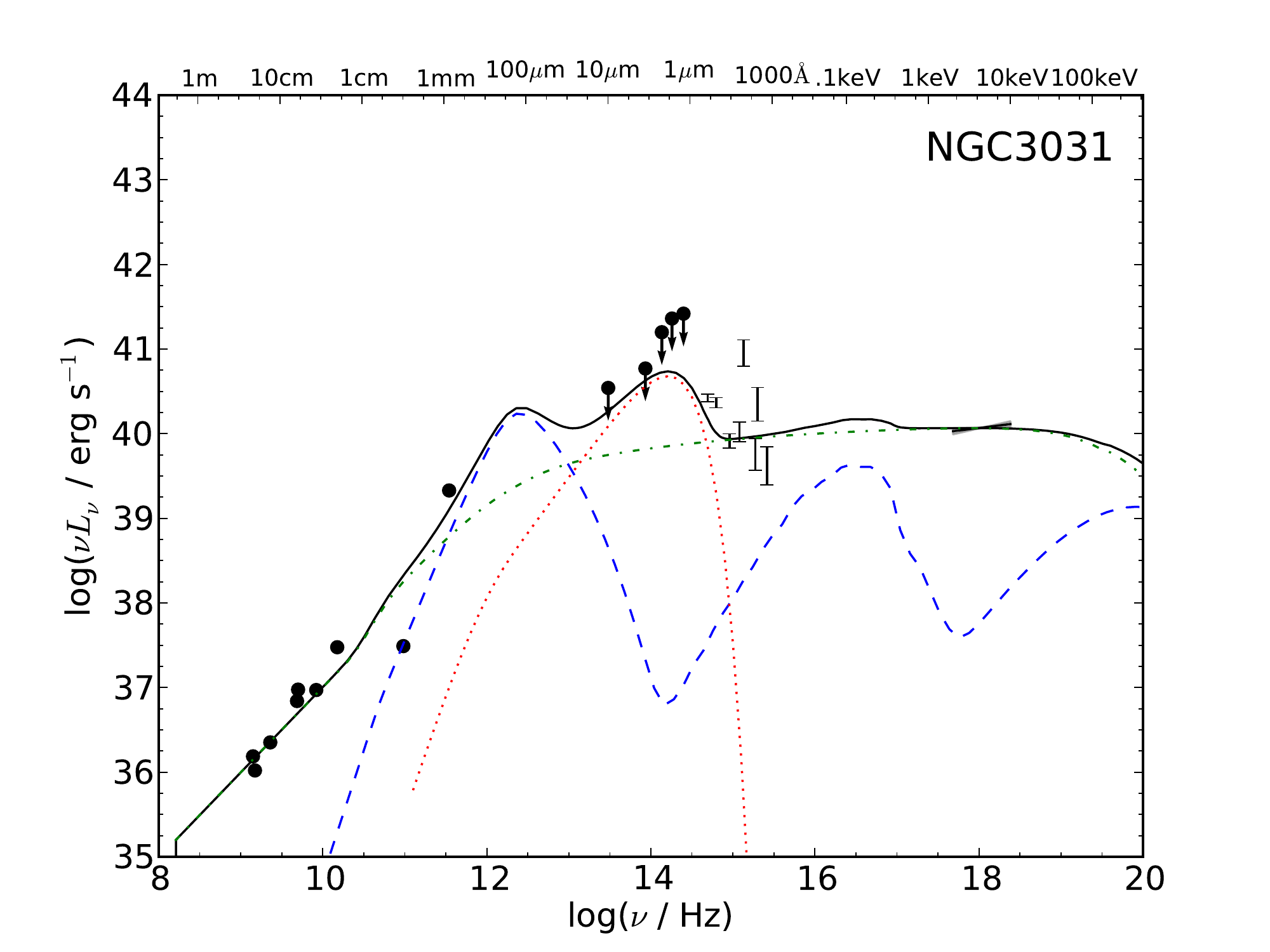}
}
\caption{Same as Figure \ref{fig:n4374} for M81.}
\label{fig:m81}
\end{figure*}

\citet{devereux07} modeled the profile of the broad double-peaked H$\alpha$ line with a relativistic thin disk model. They were able to explain the profile with an inclination angle of $50^\circ$. Such a high inclination angle is supported by radio observations of the jet \citep{bieten00}. The H$\alpha$ line profile of M81 is consistent with an inner radius of $\approx 280 - 360 R_S$ for the line-emitting thin disk \citep{devereux07}. In our models for the SED, we therefore adopt an inclination angle $i=50^\circ$ and we impose the constraint on the models that $r_{\rm tr} \lesssim 360$.

The left panel of Figure \ref{fig:m81} shows the AD fit model for the SED of M81. In this model, the synchrotron radiation from the jet is the dominant mechanism of radio emission. In the wavelength range between $\approx 10 \ \mu$m and $\approx 1 \ \mu$m a ``red bump'' can be seen which corresponds to the emission of the truncated thin disk. The thin disk dominates the emission in this band. In the interval between 1 mm and $\approx 100 \ \mu$m the ADAF is the dominant component. The ADAF is also the flow component that dominates the emission for wavelengths less than $1 \ \mu$ until X-rays. The sum of the radiation of all components is plotted as the solid line. 
The accretion flow parameters are $\dot{m}_{\rm out}=3 \times 10^{-3}$, $r_{\rm tr}=360$ (consistent with the H$\alpha$ modeling results), $\delta=0.01$ and $s=0.16$. The jet parameters are $\dot{m}_{\rm jet}=2 \times 10^{-6}$, $p=2.2$, $\epsilon_e=0.1$ and $\epsilon_B=0.01$. The jet power is given by $P_{\rm jet}=8 \times 10^{41} \ {\rm erg \ s}^{-1}$ with $P_{\rm jet}/L_{\rm bol} \approx 4$ and $\dot{m}_{\rm jet} / \dot{m}(3 R_S) \approx 10^{-3}$.

In this AD model, a low value of $\delta=0.01$ is required in order to reproduce the X-ray spectral shape characterized by $\Gamma=1.88$. As was the case for M87, if we increase the value to $\delta=0.3$ keeping the other parameters fixed except $s=0.4$ the predicted value of the photon index is reduced and the ADAF spectrum becomes harder than the data. 

The right panel of Fig. \ref{fig:m81} shows a JD model in which the jet component dominates the radio, optical-UV and X-ray bands of the SED. In the wavelength range between 1 mm and $\approx 100 \ \mu$m the ADAF is the dominant component, although there are no observational constraints to the model in this region. Between $\approx 10 \ \mu$m and $6000 \ {\rm \AA}$ the truncated thin disk dominates the emission as is the case for the AD model. The parameters of the accretion flow are $\dot{m}_{\rm out}=8 \times 10^{-4}$, $r_{\rm tr}=50$, $\delta=0.3$ and $s=0.6$. The parameters of the jet are $\dot{m}_{\rm jet}=1.2 \times 10^{-5}$, $p=2.05$, $\epsilon_e=0.6$ and $\epsilon_B=10^{-4}$ ($P_{\rm jet}=4.8 \times 10^{42} \ {\rm erg \ s}^{-1}$, $P_{\rm jet}/L_{\rm bol} \approx 23$ and 
$\dot{m}_{\rm jet} / \dot{m}(3 R_S) \approx 3 \times 10^{-5}$).

We adopted two different transition radii in the fits shown in figure \ref{fig:m81}. The accretion flow model with $r_{\rm tr}=360$ 
overpredicts the UV data (left panel of fig. \ref{fig:m81}). A smaller transition radius of $r_{\rm tr}=50$ on the other hand does a better overall job of describing the optical-UV data. Such small transition radius seems also to be favored by the modelling of the Fe K$\alpha$ emission line \citep{young07}. We note however that with the available near-IR and optical data it is not possible to set a strong constraint on the properties of the truncated thin disk.

M81 has been the subject of several modeling efforts \citep{quat99,markoff08,yuan09}. Our goal here is not to discuss extensively this particular LINER since this was done elsewhere \citep{miller10}, but rather put the general features of our AD and JD models for M81 in context with the results found by other authors. 

\citet{quat99} did not consider the jet contribution and favored an AD scenario for the origin of the SED (they used an ``old'' ADAF model assuming $\delta=0.01$ and $s=0$, i.e. no winds).  \citet{markoff08,yuan09} on the other hand were also able to explain the observations, but in the context of a JD model. \citet{yuan09} also considered the simultaneous contribution of the ADAF and jet but restricted themselves to $\delta=0.5$ in their fits. Mainly for this reason, \citet{yuan09} were unable to find an AD model for the SED, since as we mentioned before, ADAF X-ray spectra with such values of $\delta$ are quite hard. 

It should be noted that \citet{quat99} estimated a transition radius which is twice ($r_{\rm tr}=100$) the value favored in the present work and a much higher accretion rate. The reason for this is that these authors previously fitted the SED of M81 using an outdated value of the black hole mass, which was fifteen times smaller and more uncertain than the value we use (estimated via spatially resolved gas and stellar kinematics, see EHF10). As a consequence, the multicolor blackbody radiated by their thin disk for the same accretion rate and transition radius that we use is considerably hotter and fainter. As a consequence, they needed to increase $r_{\rm tr}$ and $\dot{m}$.

\subsection{NGC 3998}	\label{sec:n3998}


The shortest wavelength (1750 \AA) UV data point in the SED of NGC 3998 presented in EHF10 is anomalously high because of variability \citep{devereux11}. As discussed by Devereux this data point was obtained many years before all the other observations, when the source was much brighter. For this reason, we exclude this UV measurement from our analysis and plots.

\begin{figure*}
\centerline{
\includegraphics[scale=0.5]{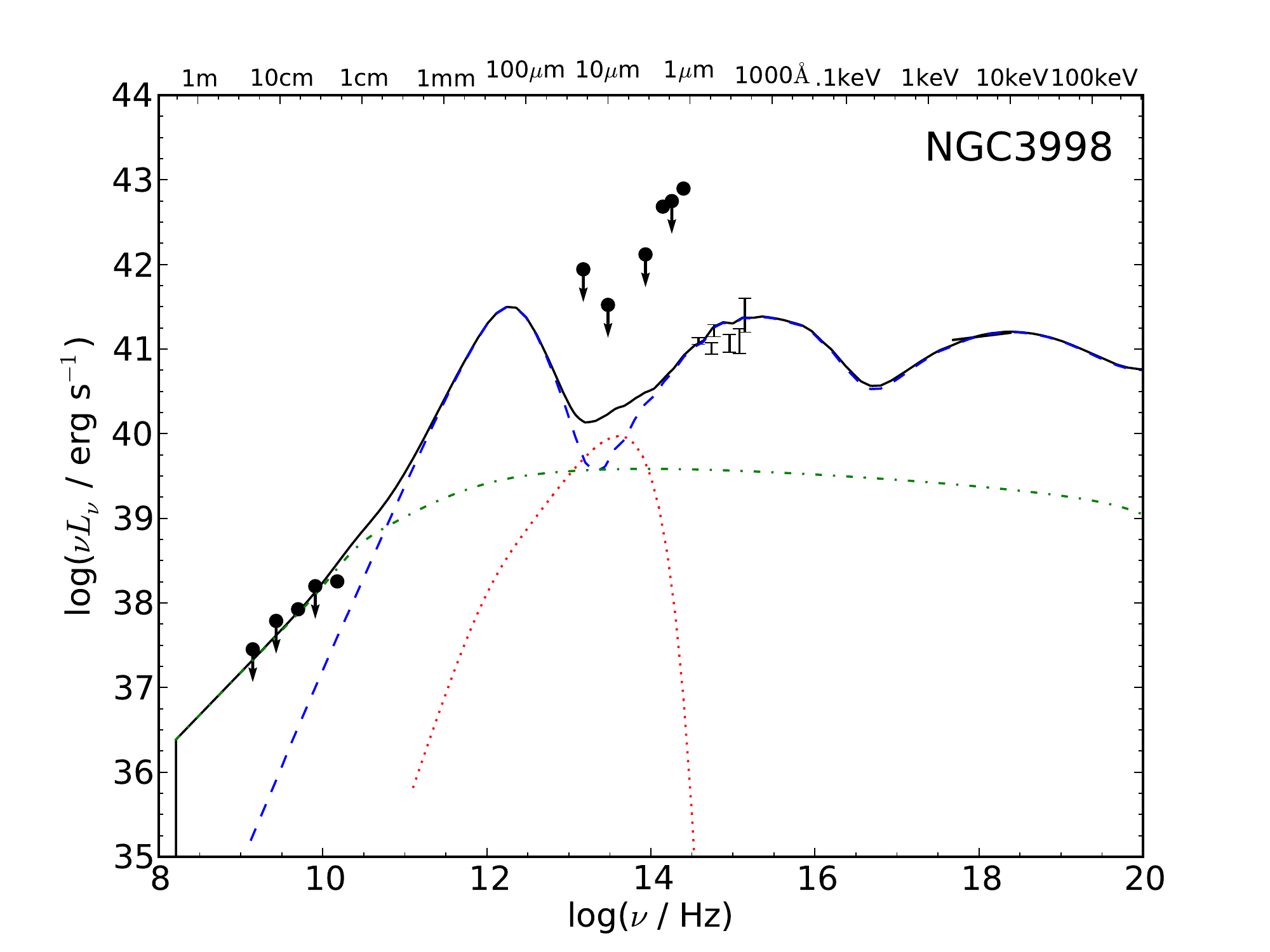}
\hskip -0.3truein
\includegraphics[scale=0.5]{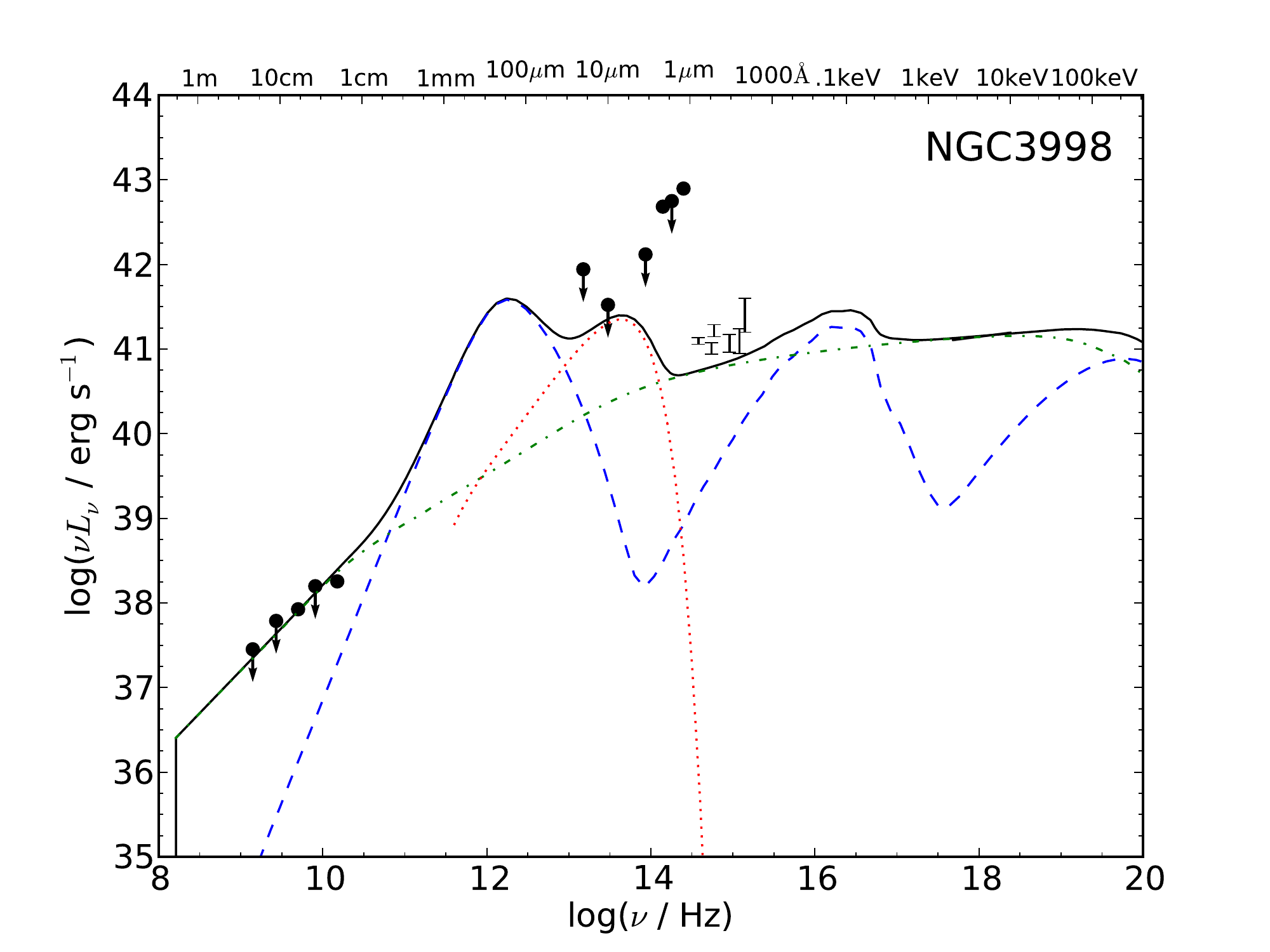}
}
\caption{Same as Figure \ref{fig:n4374} for NGC 3998.}
\label{fig:n3998}
\end{figure*}

The left panel of Figure \ref{fig:n3998} shows an AD model in which the ADAF dominates the emission from the IR to the X-ray bands and explains well the available optical-UV and X-ray data. As usual, the jet dominates the radio emission. The ADAF parameters are $\dot{m}_{\rm out}=7.2 \times 10^{-3}$, $r_{\rm out}=10^4$, $\delta=0.1$ and $s=0.4$. The jet parameters are $\dot{m}_{\rm jet}=1.7 \times 10^{-6}$, $p=2.2$, $\epsilon_e=0.01$ and $\epsilon_B=10^{-3}$. For the resulting jet power from this model ($P_{\rm jet}=9 \times 10^{42} \ {\rm erg \ s}^{-1}$, $\dot{m}_{\rm jet} / \dot{m}(3 R_S) \approx 6 \times 10^{-3}$), the jet kinetic energy is quite modest compared to the radiative output ($P_{\rm jet}/L_{\rm bol} \approx 0.6$).

In the AD model above we adopted a large outer radius for the ADAF, but smaller radii are not ruled out by the data. For instance, we are able to obtain a reasonable fit to the SED with $r_{\rm tr}=500$, $\dot{m}_{\rm out}=10^{-3}$  and $s=\delta=0.01$. This model is consistent with the IR upper limits and accounts for the X-ray data. The transition radius cannot be much smaller than this value, otherwise the emission of the truncated thin disk would exceed the IR upper limits. Furthermore, as noted by \citet{ptak04}, the lack of Fe K$\alpha$ line emission also suggests that the value of $r_{\rm tr}$ is not so small. 

The right panel of Fig. \ref{fig:n3998} shows a JD model in which the jet dominates the radio, optical-UV and the X-ray emission. Note that although the jet accounts quite well for the X-rays, it somewhat underestimates the optical-UV data. The parameters of the jet are $\dot{m}_{\rm jet}=3.5 \times 10^{-6}$, $p=2.01$, $\epsilon_e=0.75$ and $\epsilon_B=3 \times 10^{-5}$ ($P_{\rm jet}=1.8 \times 10^{43} \ {\rm erg \ s}^{-1}$, $P_{\rm jet}/L_{\rm bol} \approx 1.3$ and $\dot{m}_{\rm jet} / \dot{m}(3 R_S) \approx 0.025$). For illustration, we also show in the right panel an ADAF model computed with parameters such that its contribution in X-rays is weak ($\dot{m}_{\rm out}=4 \times 10^{-4}$, $r_{\rm tr}=100$, $\delta=0.3$ and $s=0.3$). Note that, as for the cases discussed in the previous section, for values of $\delta \gtrsim 0.3$ the ADAF X-ray spectrum is harder than the data. 

The SED of this NGC 3998 was modelled before by \citet{ptak04}. Here we confirm the main results of their work, namely that both AD and JD models are able to account well for the broadband SED.

\section{Fits to sparsely sampled SEDs} \label{ap:seds}

We present in this section the model fits to the 16 SEDs that were not discussed in the previous section. 

The SED fits are shown in Figures \ref{fig:seds01}-\ref{fig:seds06}. In the subsections below, we briefly describe the details of the fits to the SED of each object. The model parameters are summarized in Table \ref{tab:models}.

\subsection{NGC 0266}


The AD model accounts slightly better for the best-fit X-ray slope, but note the pronnounced uncertainty on the value of observed photon index. The JD model requires $p<2$ which is below the usual range $2<p<3$ suggested by relativistic shock theory \citep{bednarz98, kirk00}.

\subsection{NGC 1097}

Note that even though the SED of NGC 1097 is as well sampled as many of the LLAGNs in Section \ref{sec:seds}, we chose to include it in this section rather than discussing it in more detail in \textsection \ref{sec:seds} because its SED was already well-studied by \citet{nemmen06}.

We take into account here the HST UV spectrum which was not included in EHF10 because it contains contributions from an obscured nuclear starburst in addition to the AGN. This spectrum was obtained from \citet{sb05} who studied and modeled it in detail.

Our models do not take into account the contribution of the nuclear starburst. Therefore, the AD and JD models do not reproduce the ``UV bump'' between $\approx 1000$ \AA\ and $\approx 7000$ \AA\ (see Fig. 3 in \citealt{sb05}).

\begin{figure*}
\centerline{
\includegraphics[scale=0.5]{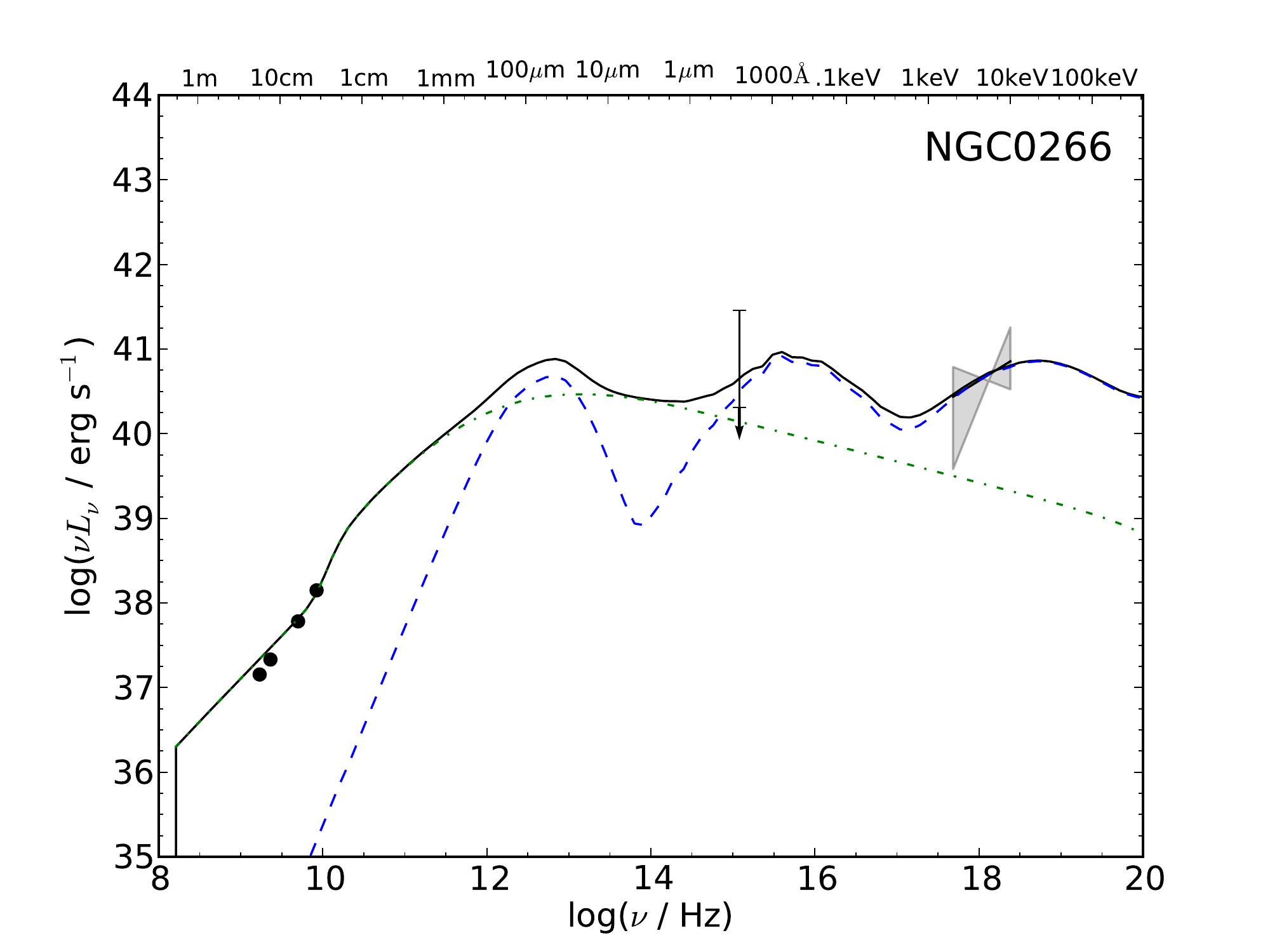}
\hskip -0.3truein
\includegraphics[scale=0.5]{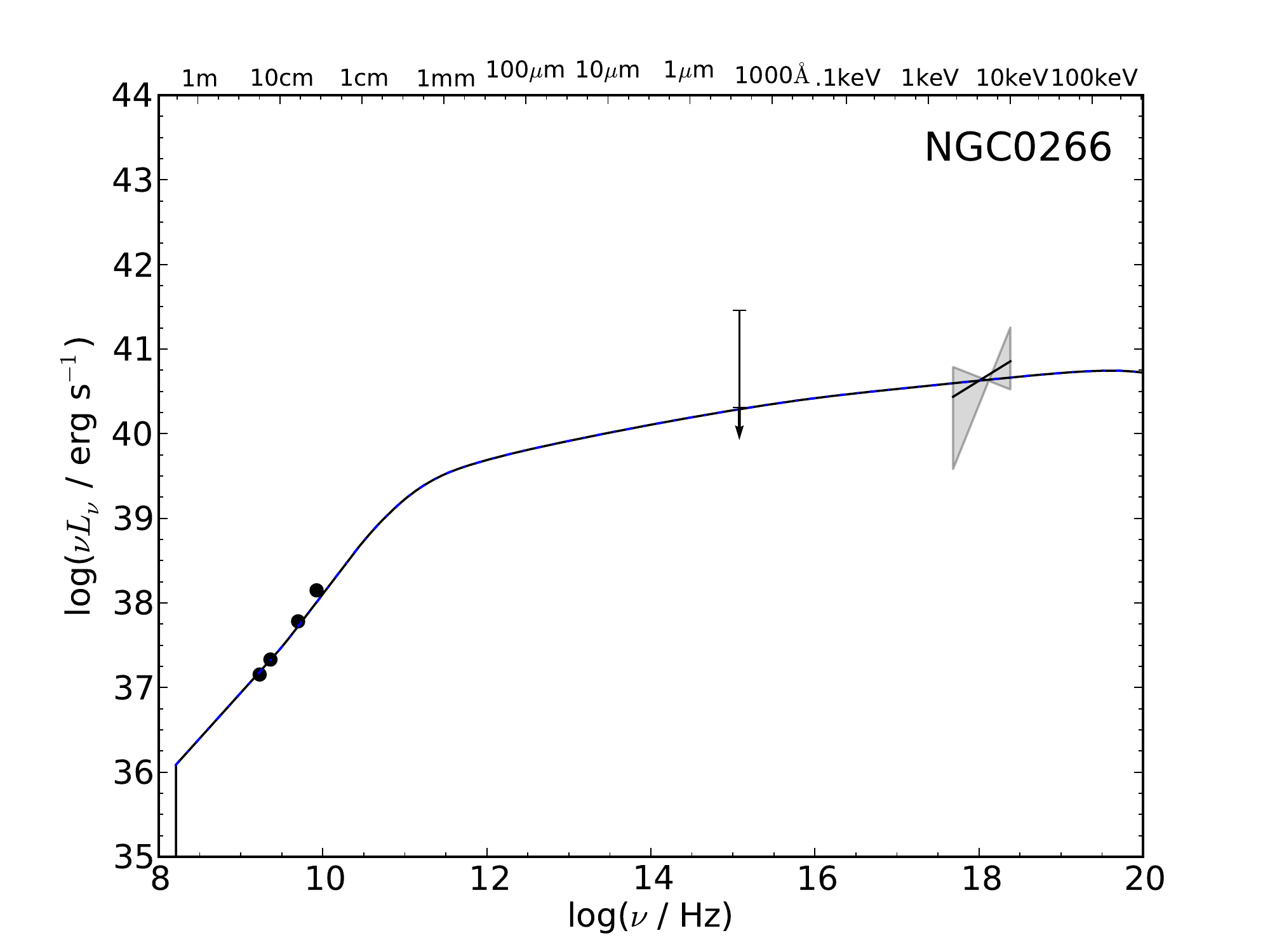}
}
\centerline{
\includegraphics[scale=0.5]{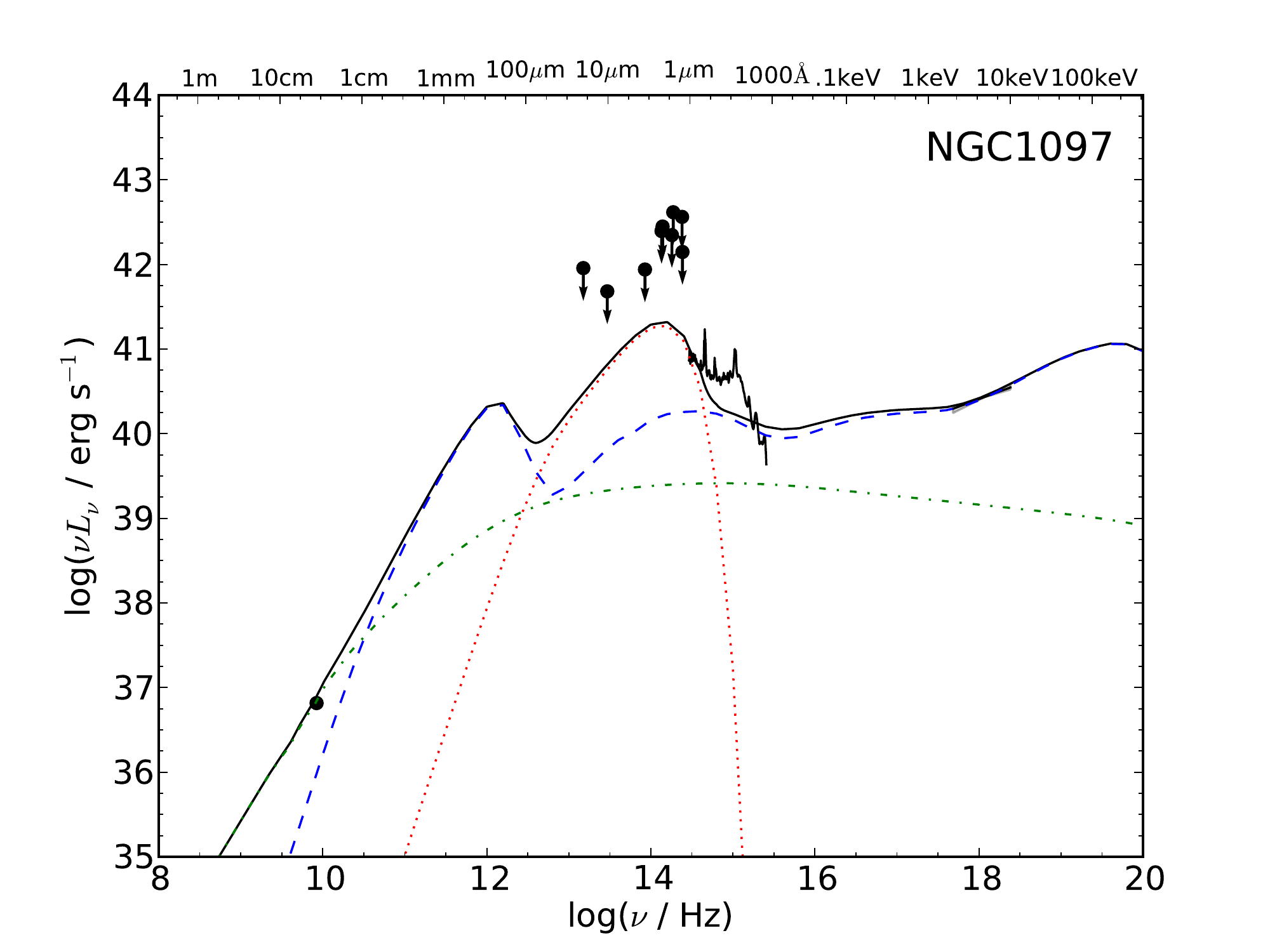}
\hskip -0.3truein
\includegraphics[scale=0.5]{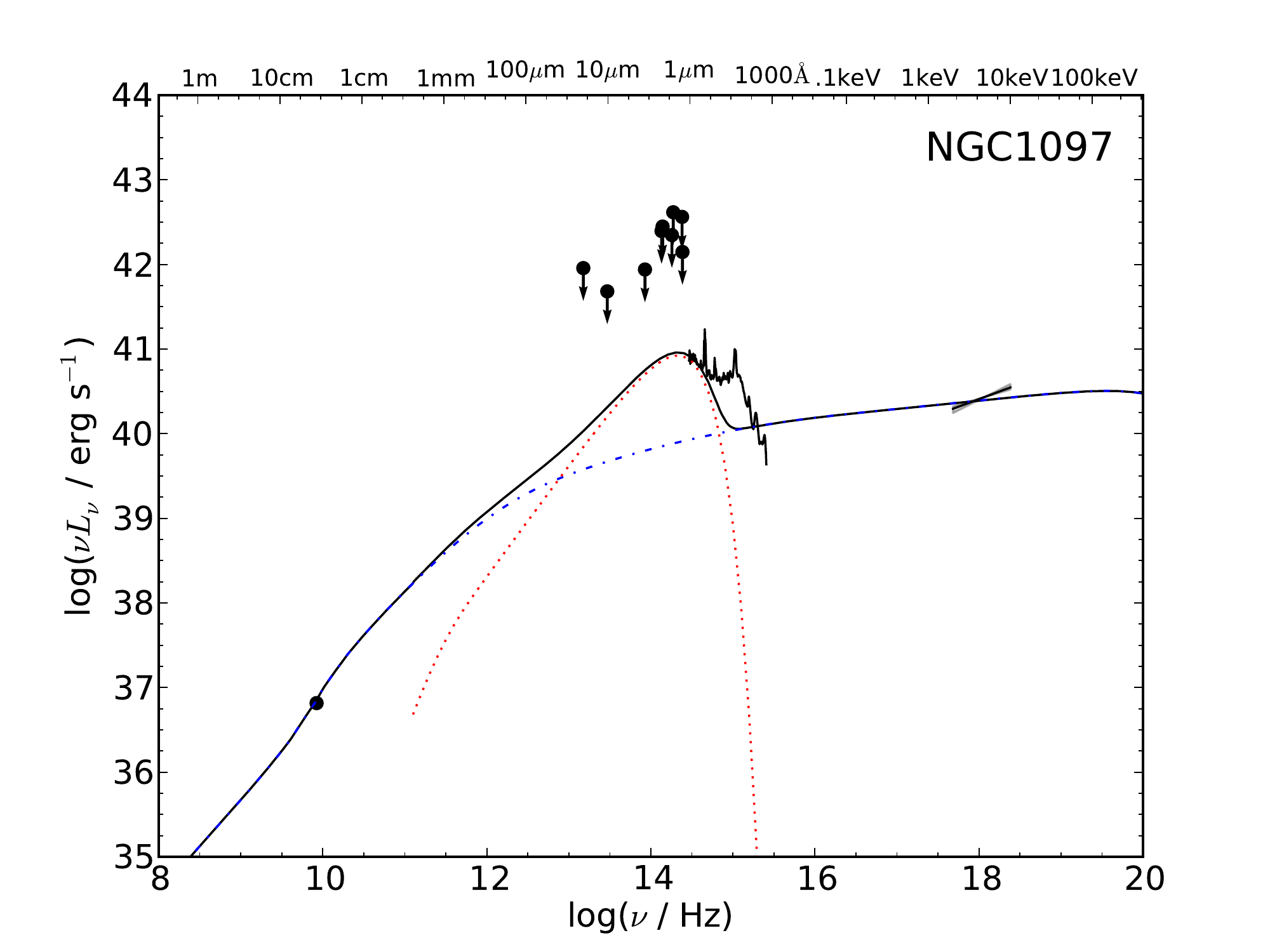}
}
\centerline{
\includegraphics[scale=0.5]{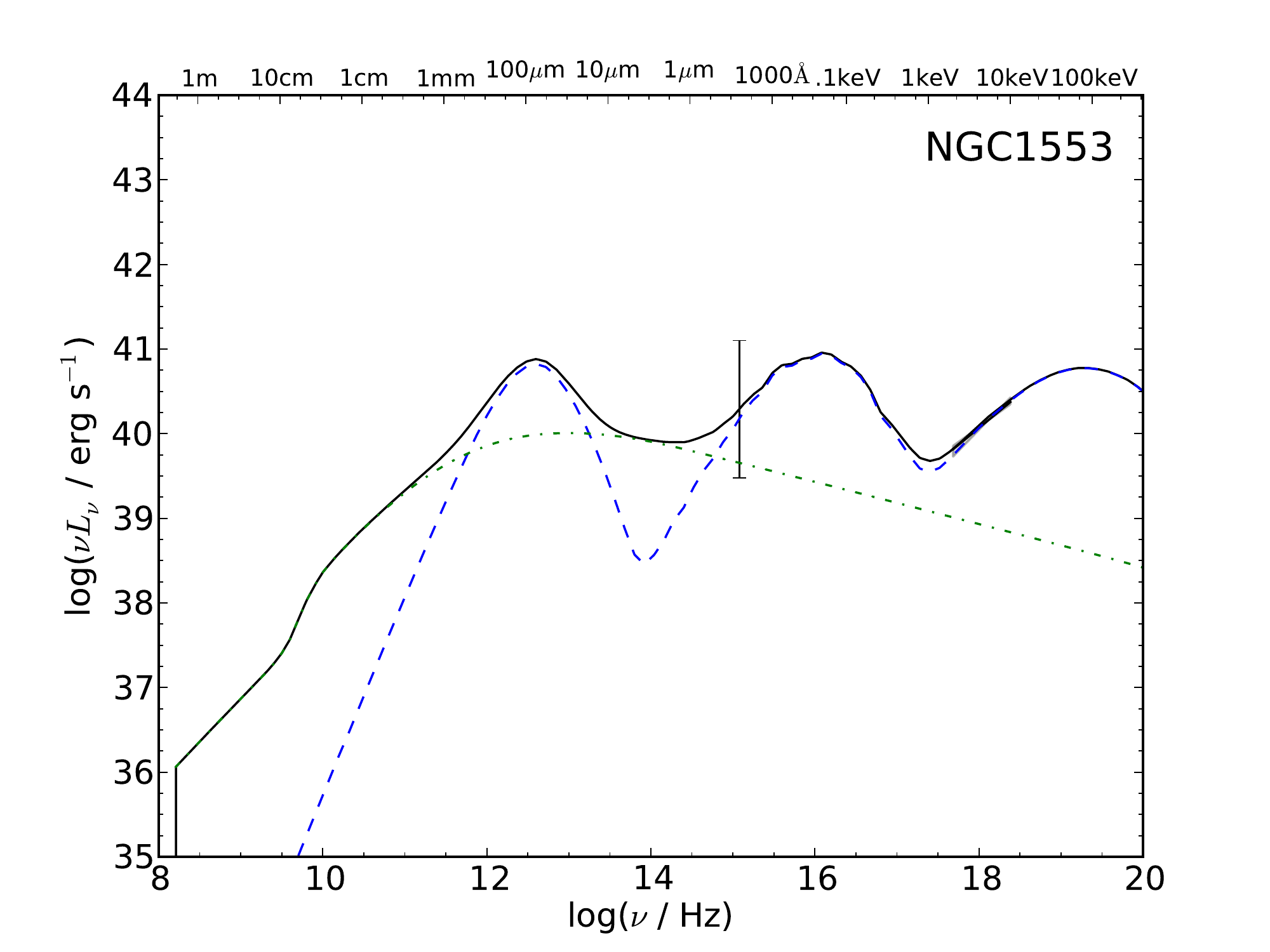}
\hskip -0.3truein
\includegraphics[scale=0.5]{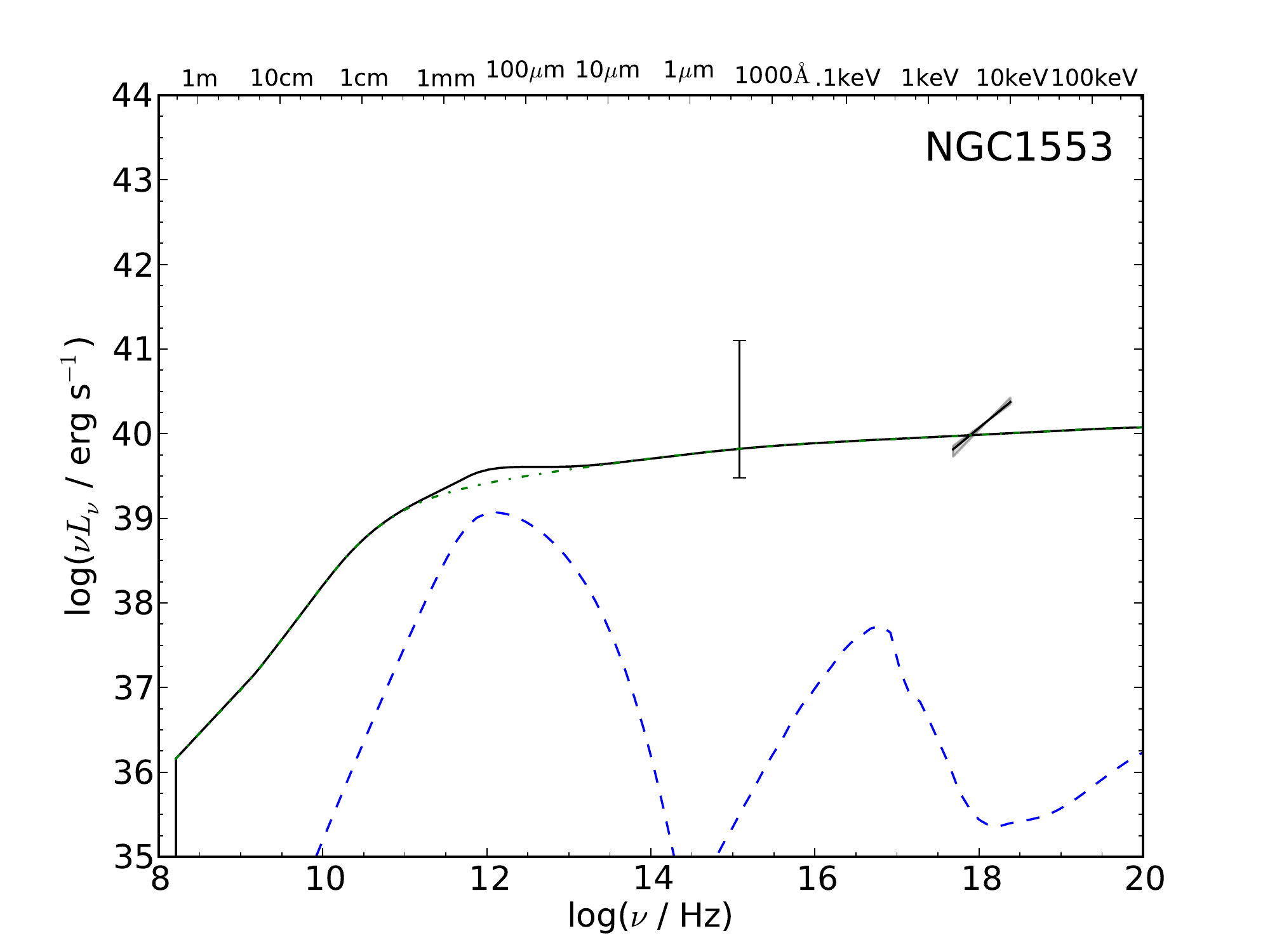}
}
\caption{SEDs and coupled accretion-jet models for NGC 0266, NGC 1097 and NGC 1553. The left panel shows the AD models for each object while the right panel displays the JD models. The dashed, dotted and dot-dashed lines correspond to the emission from the ADAF, truncated thin disk and jet, respectively. The solid line when present represents the sum of the emission from all components.}
\label{fig:seds01}
\end{figure*}

\begin{figure*}
\centerline{
\includegraphics[scale=0.5]{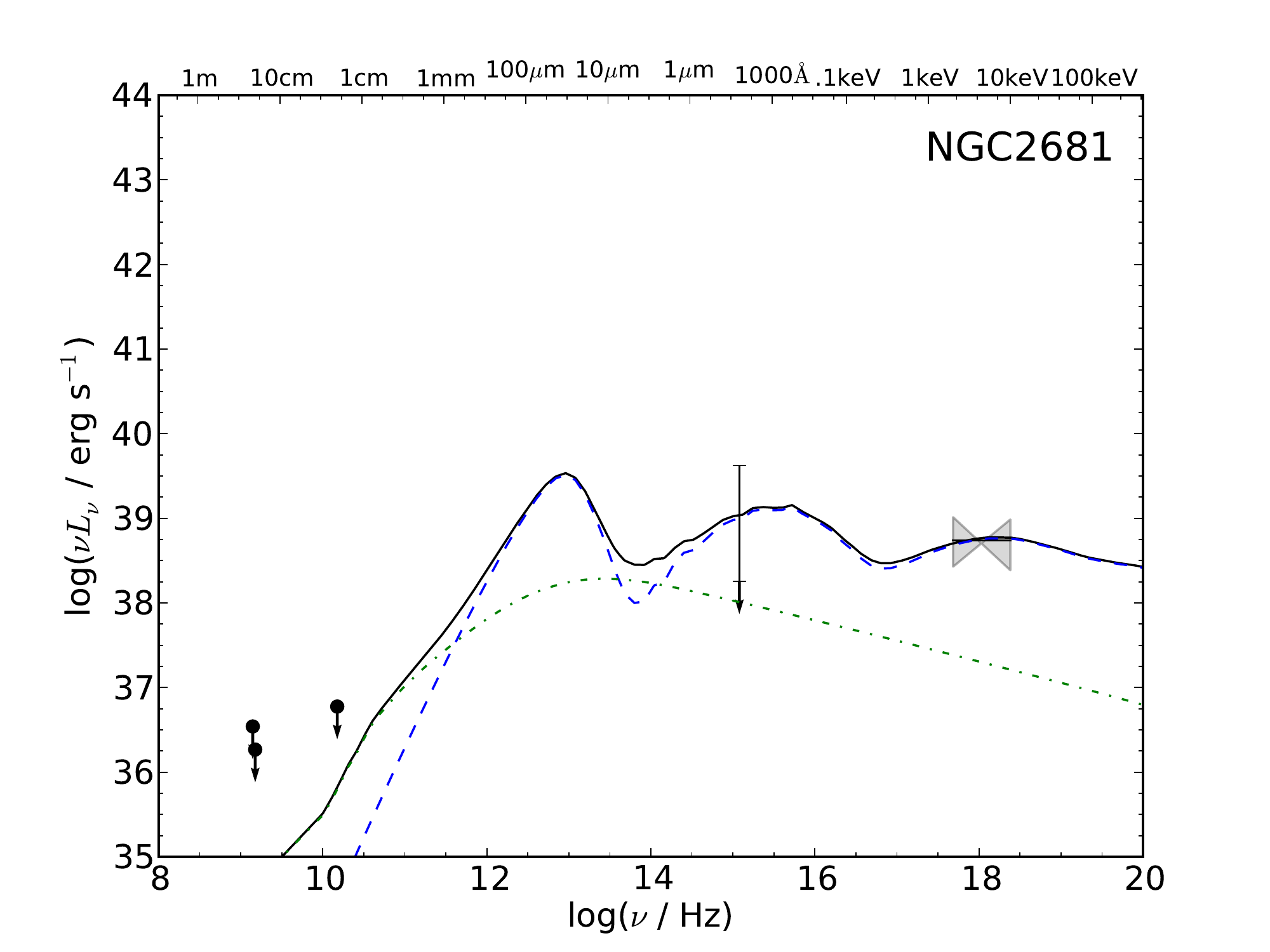}
\hskip -0.3truein
\includegraphics[scale=0.5]{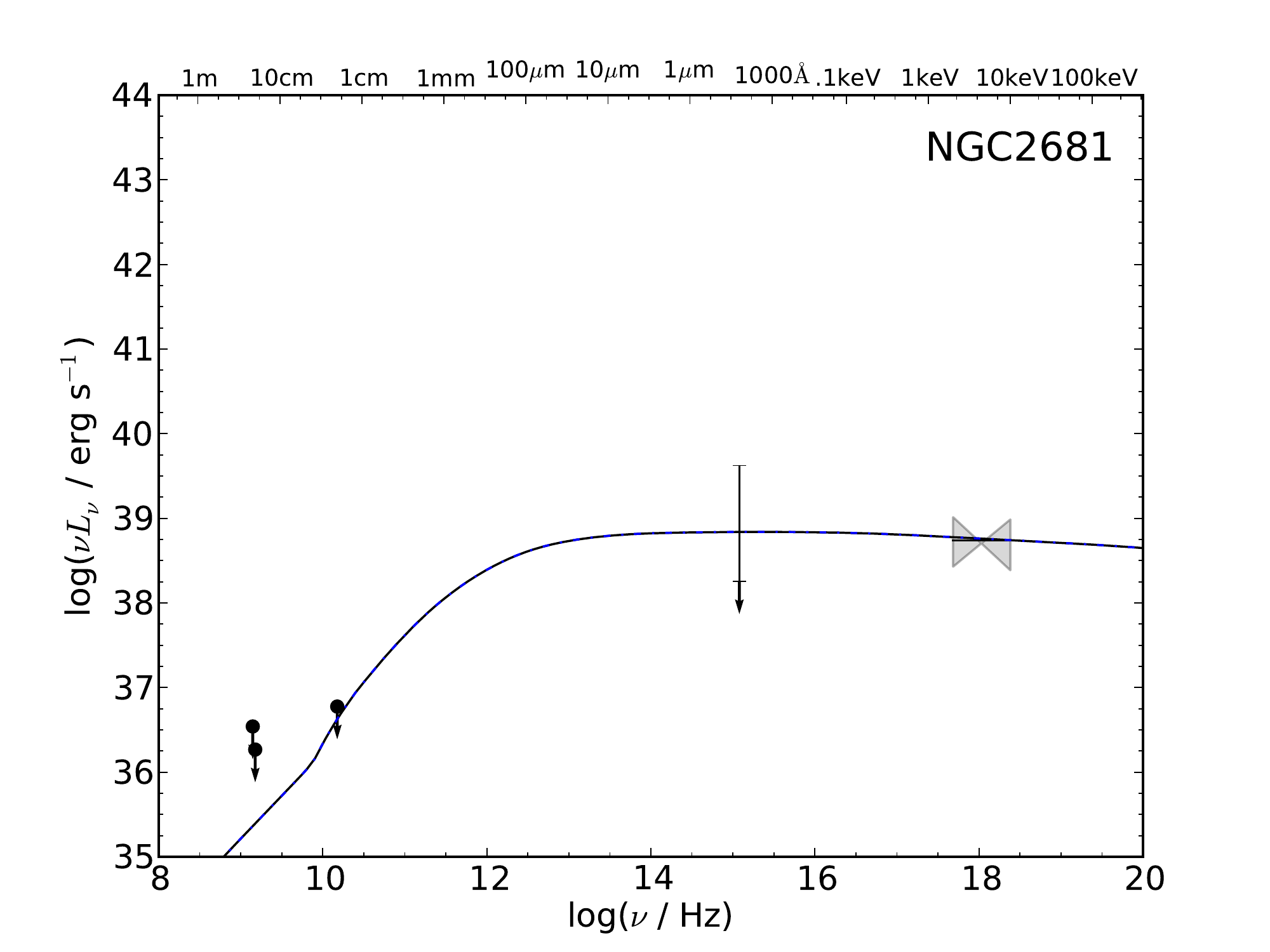}
}
\centerline{
\includegraphics[scale=0.5]{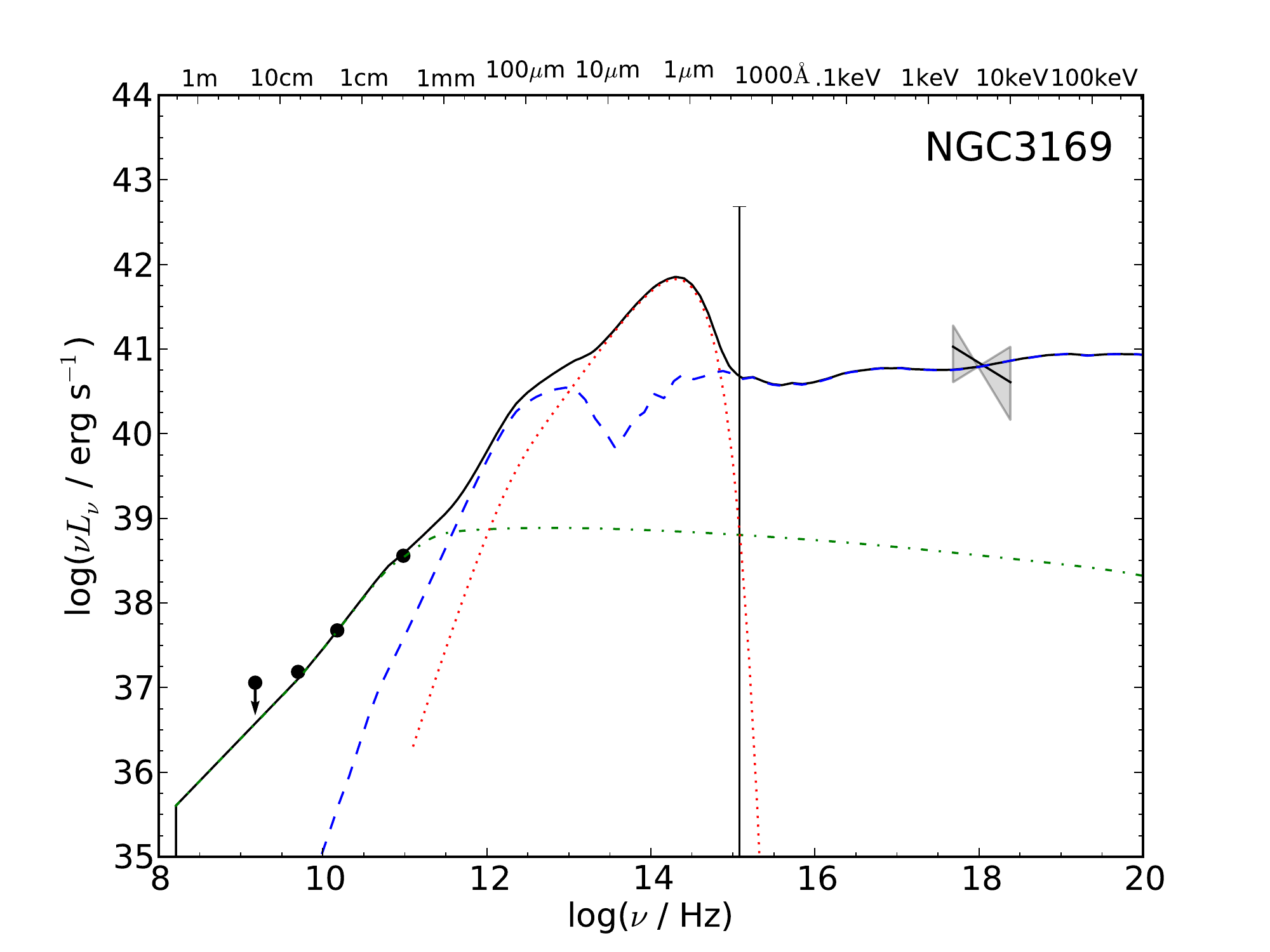}
\hskip -0.3truein
\includegraphics[scale=0.5]{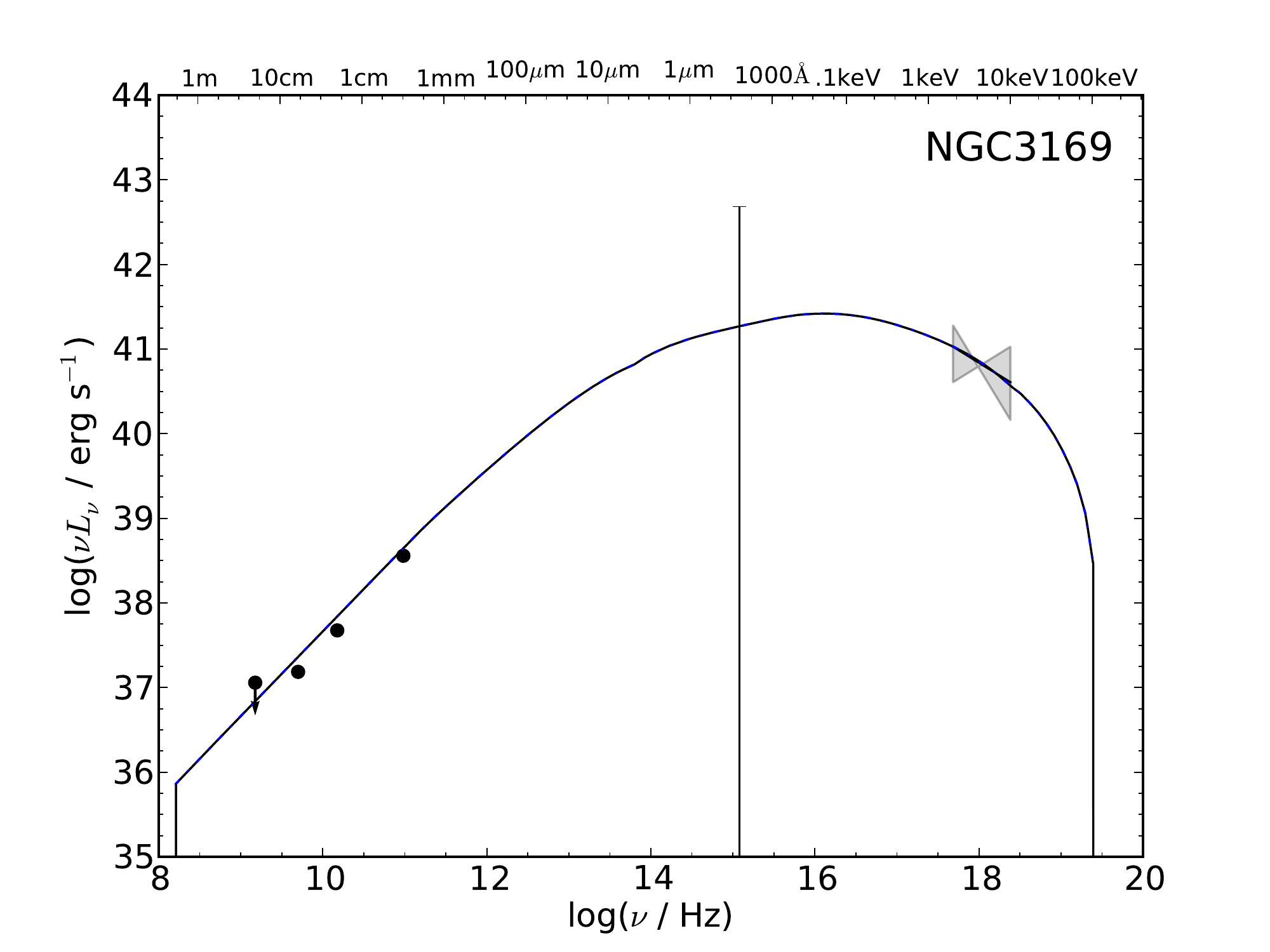}
}
\centerline{
\includegraphics[scale=0.5]{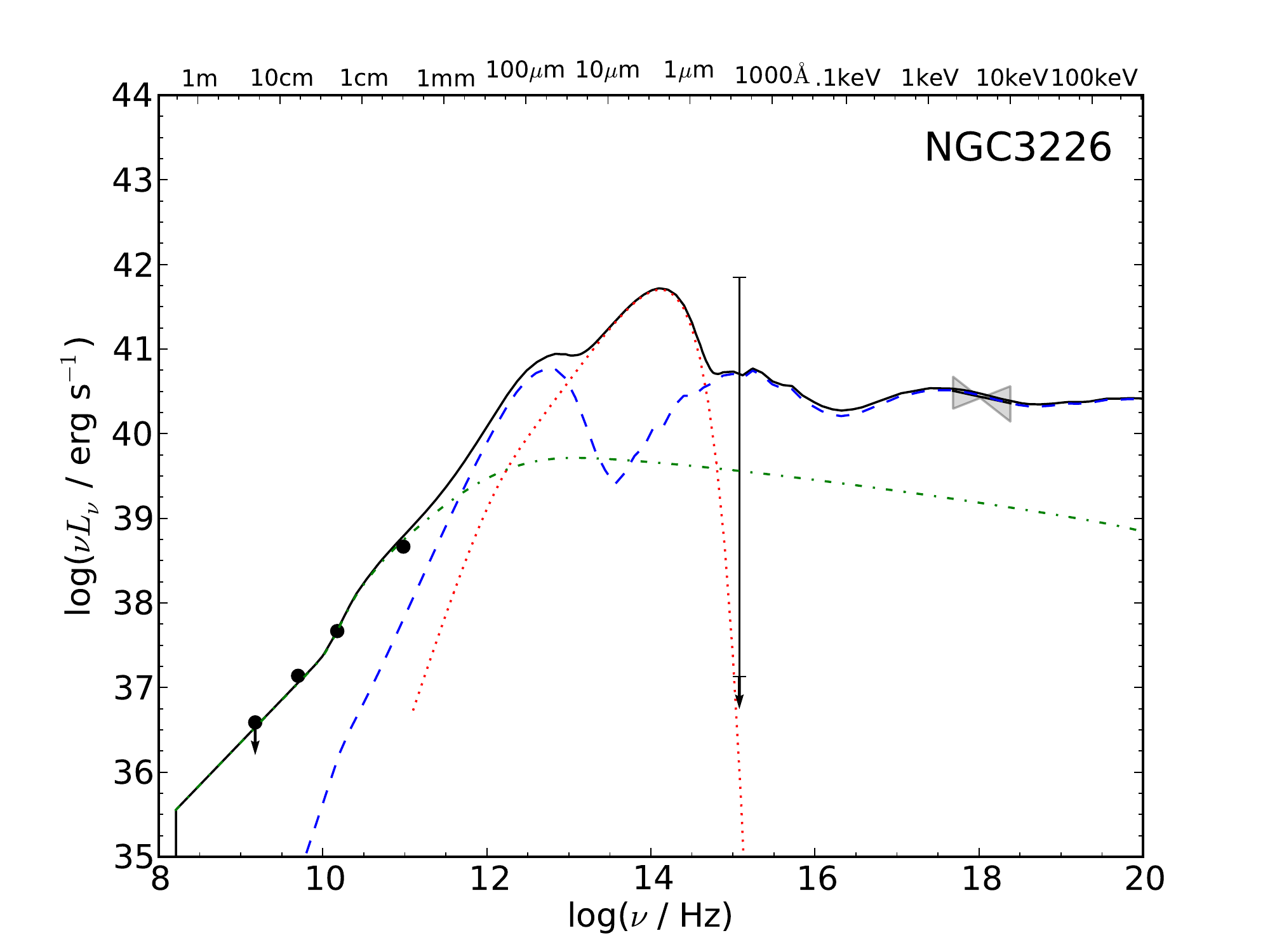}
\hskip -0.3truein
\includegraphics[scale=0.5]{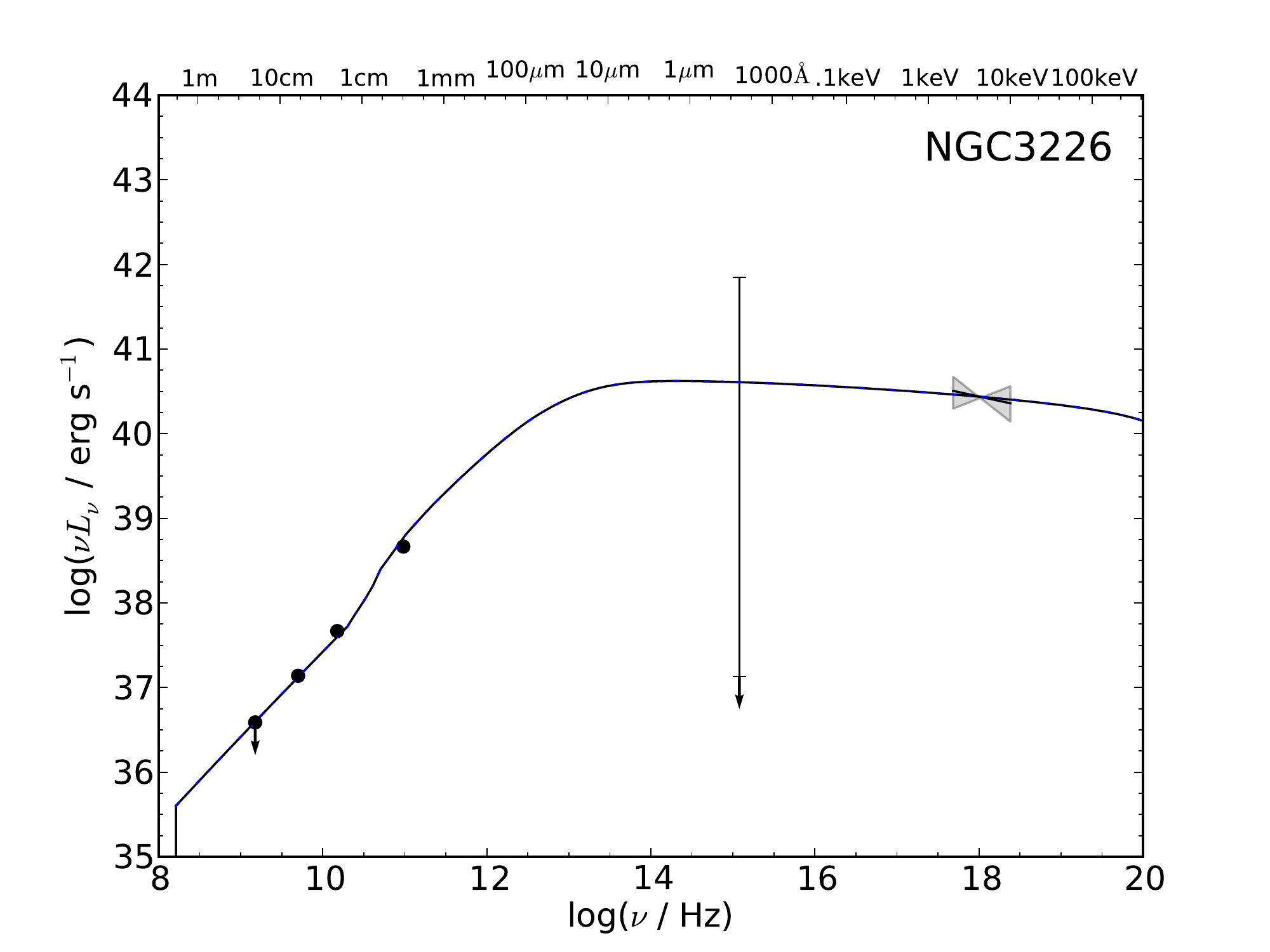}
}
\caption{Same as Figure \ref{fig:seds01} for NGC 2681, NGC 3169 and NGC 3226.}
\label{fig:seds02}
\end{figure*}

The AD model displayed in Figure \ref{fig:seds01} corresponds to the model obtained by \citet{nemmen06}. The JD model is also able to account for the whole SED, though it does not reproduce the slope of the X-ray spectrum as well as the AD model. Regarding the JD fit, given the available data there is no need to incorporate the contribution of the ADAF. Hence we do not include any ADAF model in the right panel of Fig. \ref{fig:seds01}.

\subsection{NGC 1553}	\label{sec:1553}

The Bondi accretion rate was estimated by \citet{pelle05}. 

Even though the JD model roughly accounts for the estimated $L_X$, it fails to fit the slope of the X-ray spectrum even with a small value of $p$. We take this as evidence that this source is unlikely to be JD.
The estimated $\dot{m}_{\rm out}$ is consistent with the lower limit on $\dot{m}_{\rm Bondi}$ obtained by \citet{pelle05}.

\subsection{NGC 2681}

For this LLAGN, there are not enough optical observations to fit the truncated thin disk model and estimate the transition radius.

\subsection{NGC 3169}

As was the case of NGC 2681, for this LLAGN there are not enough optical observations to fit the truncated thin disk model and estimate the transition radius. 
The JD model accounts well for the available data. The inferred extreme values of $\epsilon_e$ and $\epsilon_B$ imply that essentially all the energy in the post-shock region of the jet is carried by the particles.

\subsection{NGC 3226}

The SED of this LLAGN is well fitted by both an AD and JD type of models. As is the case of NGC 2681, there are no good optical band constraints on the emission of the truncated thin disk.

\subsection{NGC 3379}

This source has only two data points outside the X-ray band, one in the radio and the other in the optical, both of which correspond to upper limits. Therefore, there are few constraints for the accretion-jet model. The Bondi rate was estimated by \citet{david05}. 

Figure \ref{fig:seds03} shows that the AD model is able to reproduce the observed SED given the few constraints available, but there are no radio data to constrain the jet model in this case so we don't include it. The available data are not enough to constrain well the transition radius. 
The accretion rate required by the AD model is more than an order of magnitude higher than $\dot{m}_{\rm Bondi}$.  One possible explanation for this result is that the accretion rate is enhanced by gas released by stars. This is in line with the findings of \citet{soria06a, soria06b} for a sample of quiescent early-type galaxies.

\subsection{NGC 4143}

The contribution of the truncated thin disk is required in order to account for the optical emission. The required transition radius in particular is $r_{\rm tr} = 70$. 

The JD model requires a value of $p$ smaller than the usual values suggested by relativistic shock theory, as was the case of NGC 266.

\subsection{NGC 4261}

For this object, estimates of the Bondi rate, inclination angle and jet power are available \citep{gliozzi03, merloni07}. There are two independent estimates of the jet power. The first is based on energetic considerations and the VLBA observations of the pc-scale jet, which yield a lower limit to the jet power \citep{gliozzi03}. In the second method we use the 5 GHz specific luminosity and the \citet{merloni07} correlation obtaining a much higher jet power estimate.

\begin{figure*}
\centerline{
\includegraphics[scale=0.5]{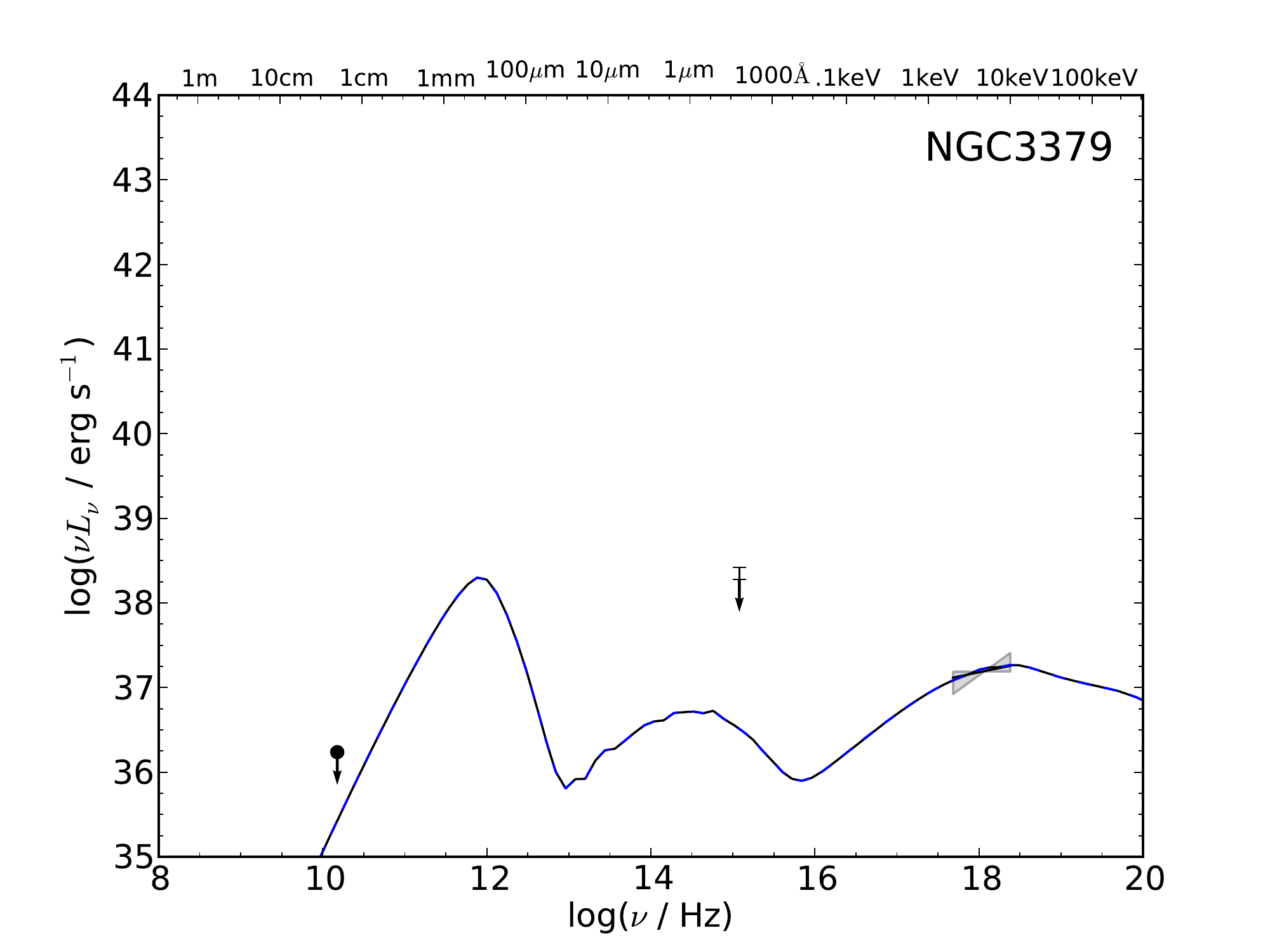}
\hskip -0.3truein
\includegraphics[scale=0.5]{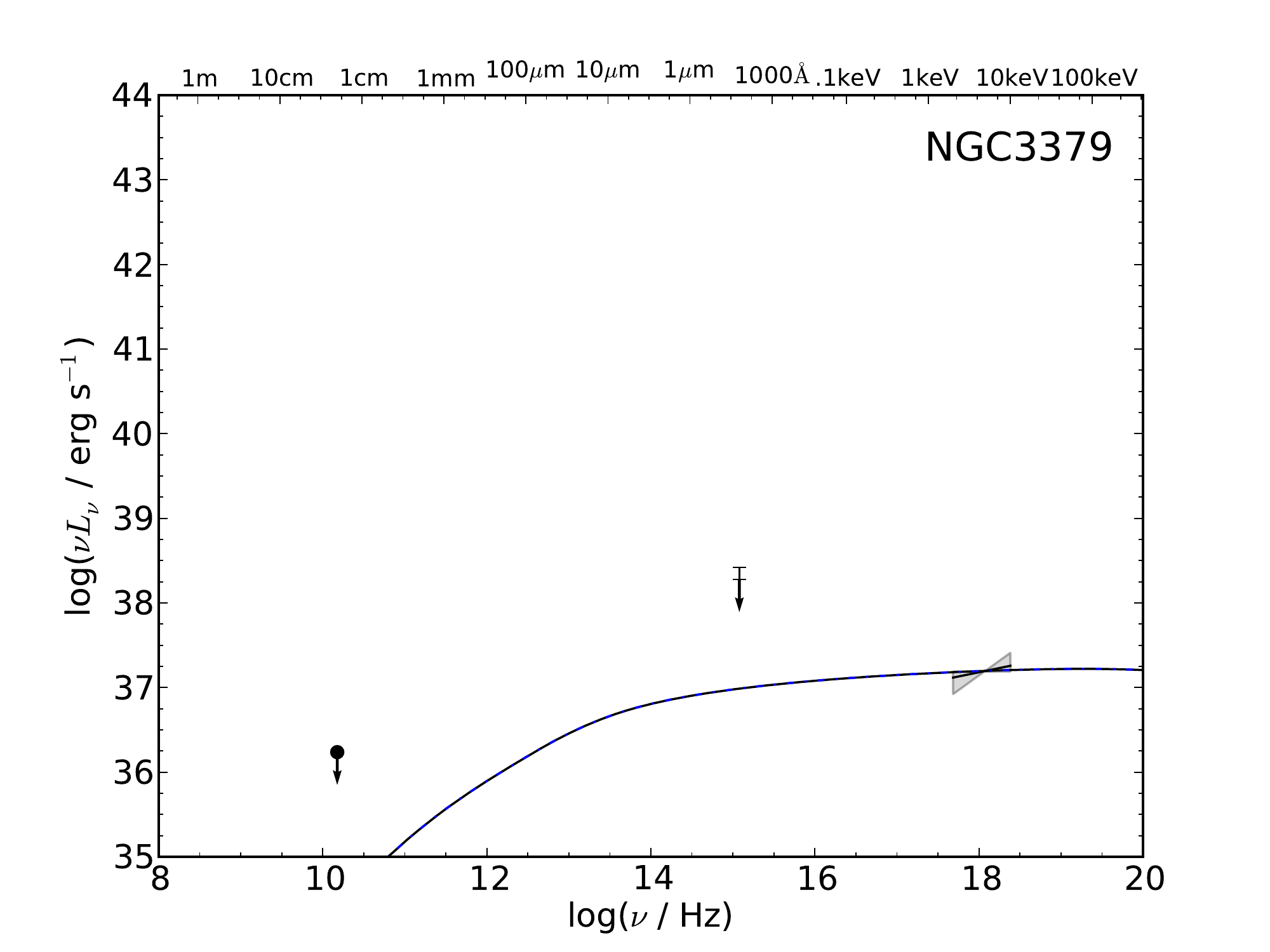}
}
\centerline{
\includegraphics[scale=0.5]{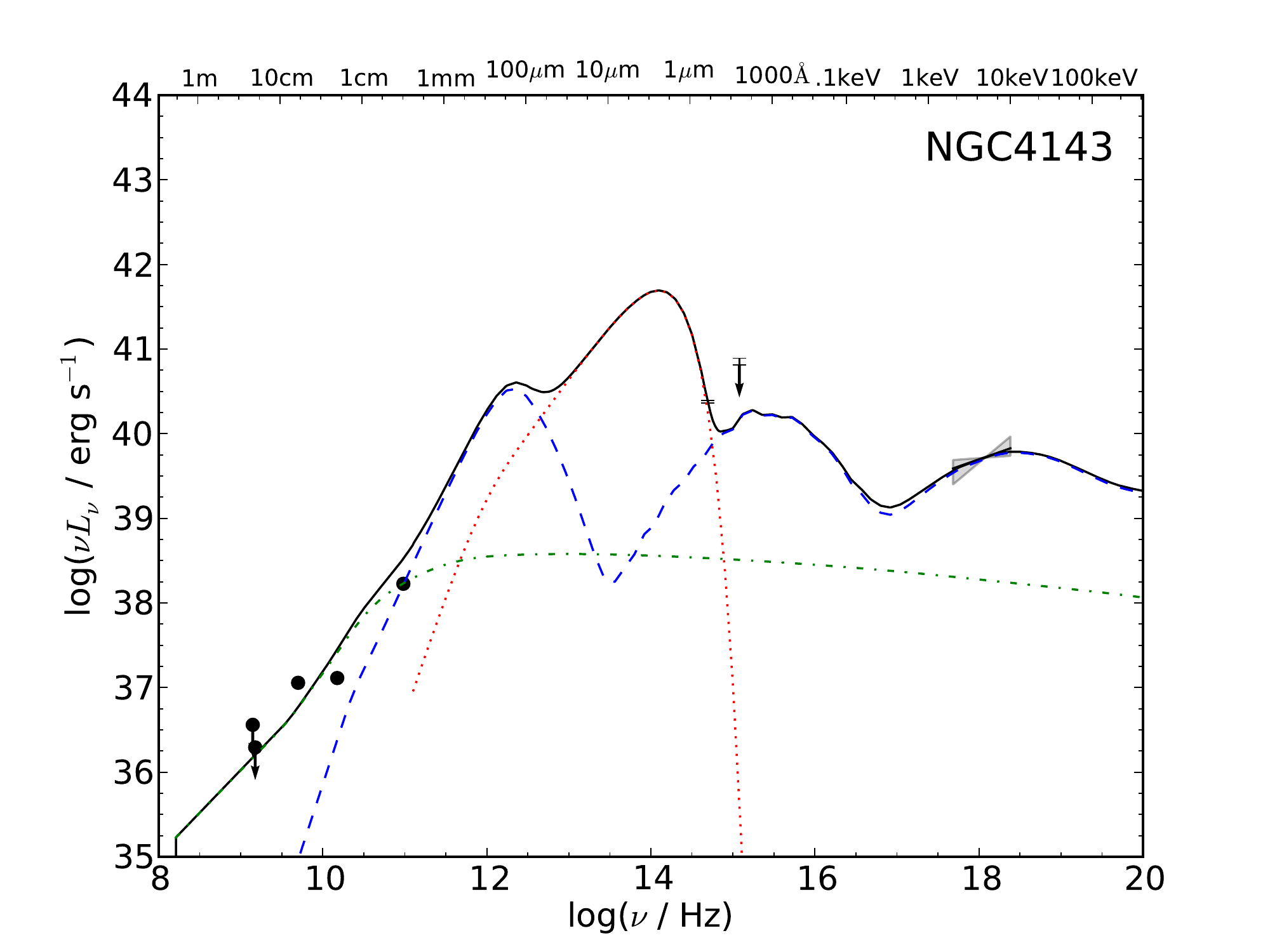}
\hskip -0.3truein
\includegraphics[scale=0.5]{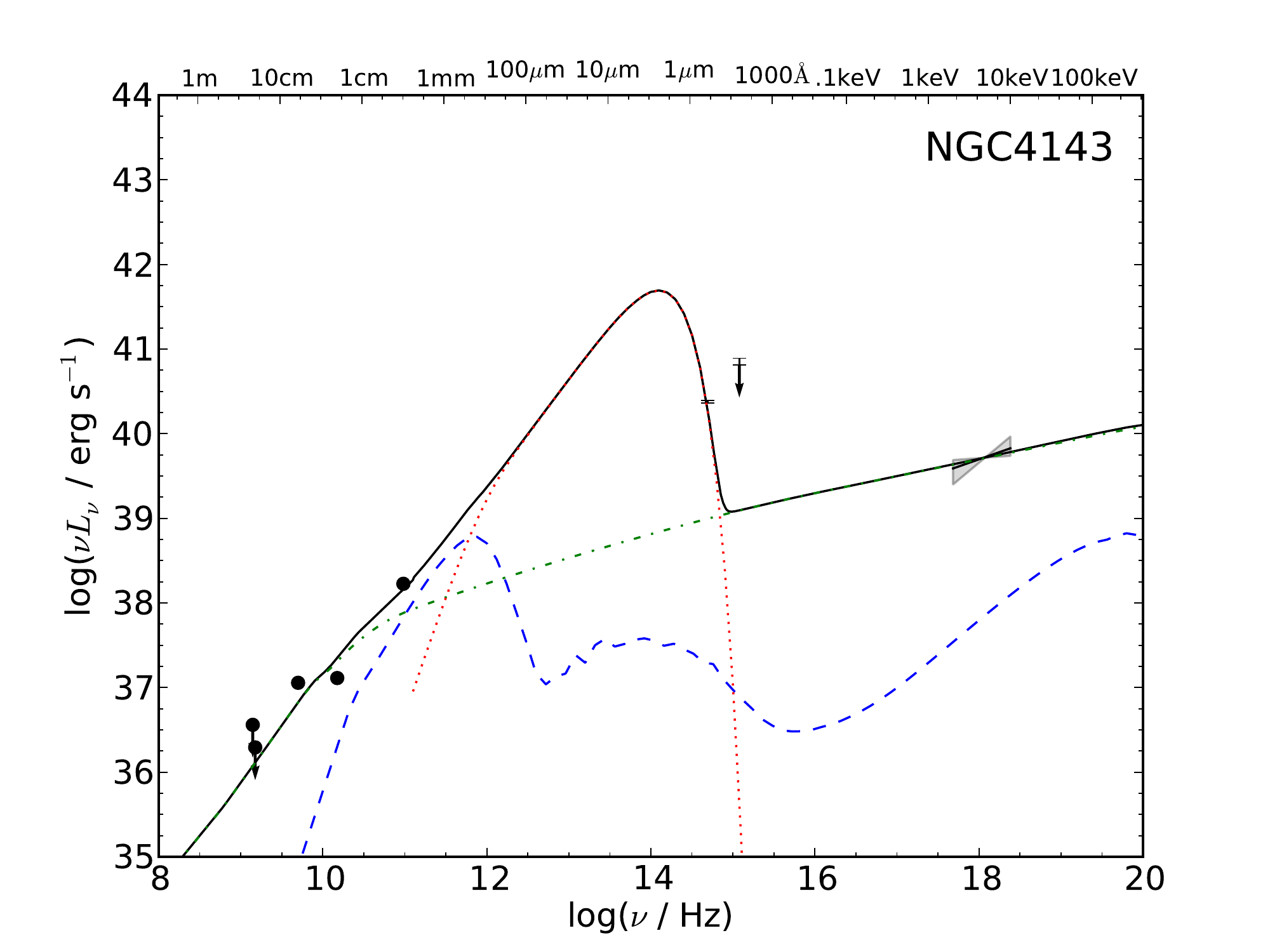}
}
\centerline{
\includegraphics[scale=0.5]{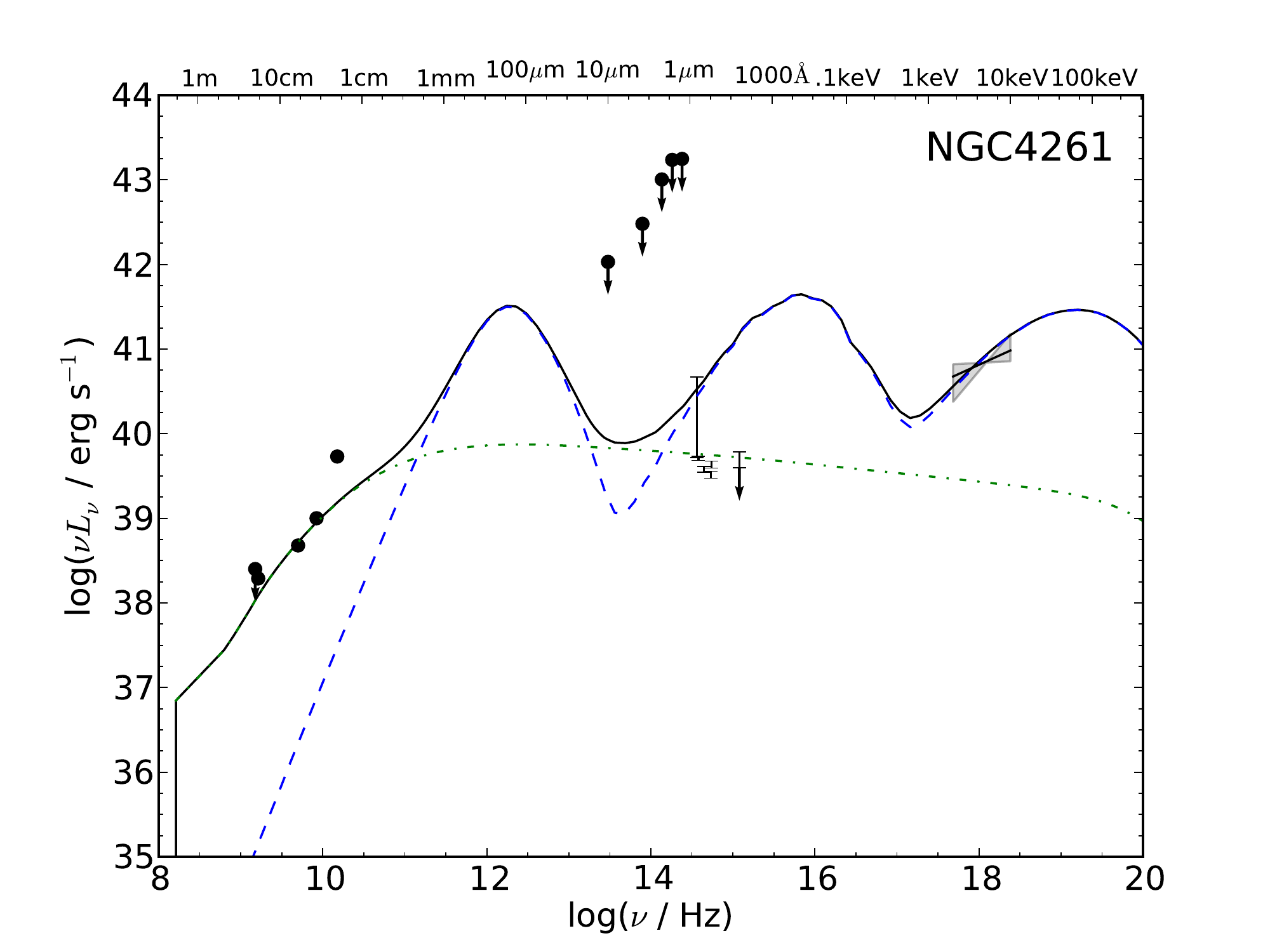}
\hskip -0.3truein
\includegraphics[scale=0.5]{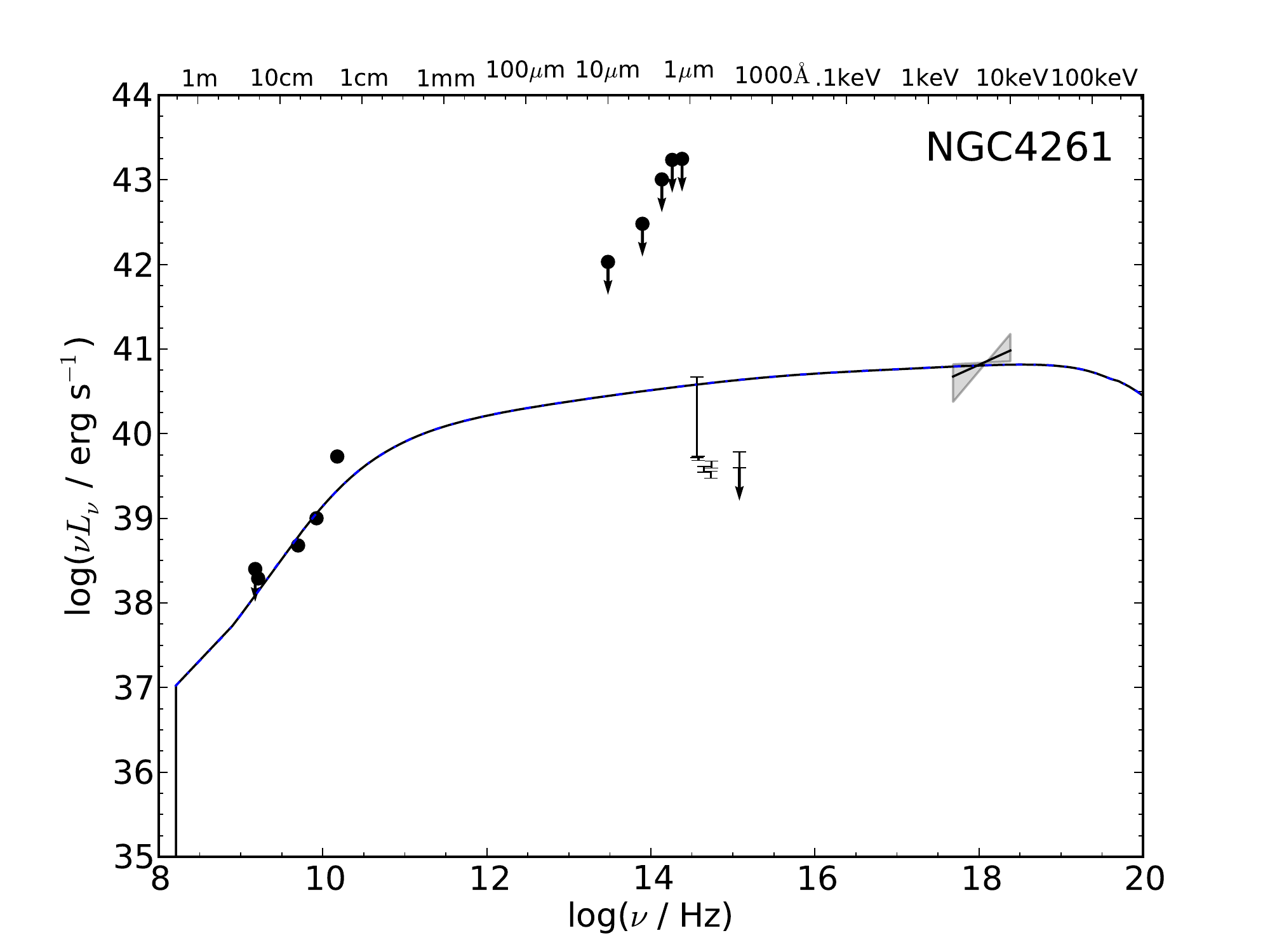}
}
\caption{Same as Figure \ref{fig:seds01} for NGC 3379, NGC 4143 and NGC 4261.}
\label{fig:seds03}
\end{figure*}

\begin{figure*}
\centerline{
\includegraphics[scale=0.5]{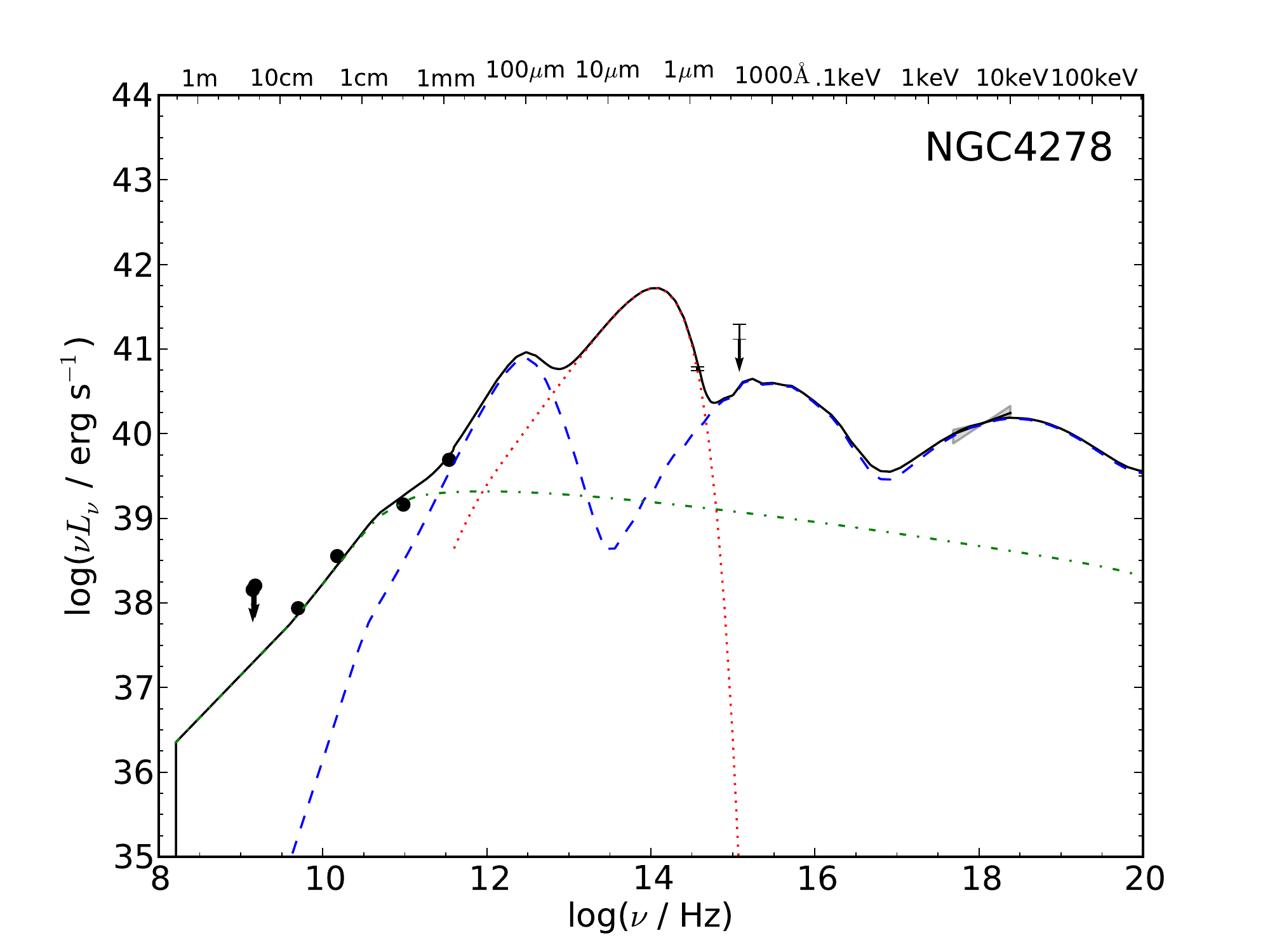}
\hskip -0.3truein
\includegraphics[scale=0.5]{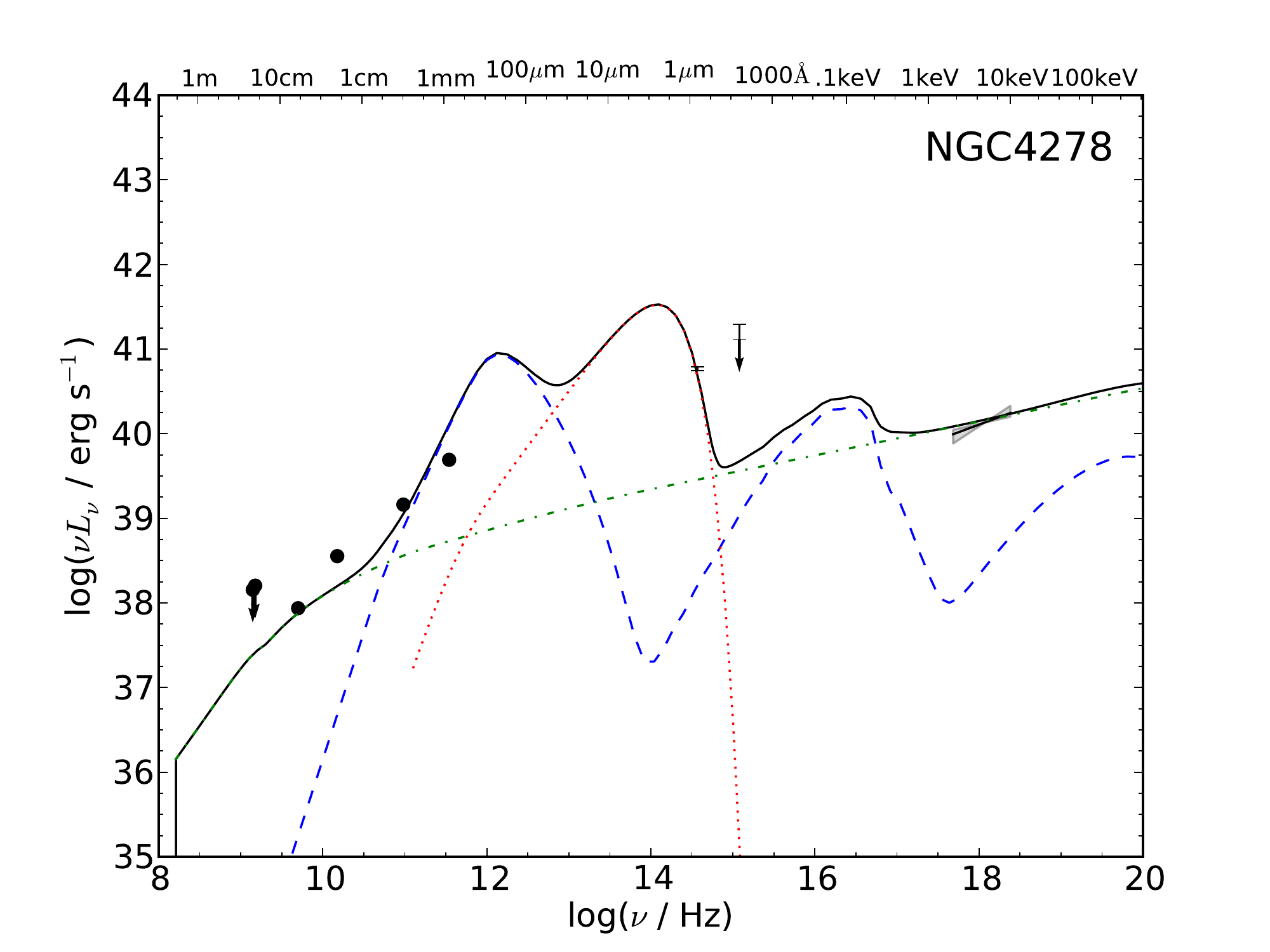}
}
\centerline{
\includegraphics[scale=0.5]{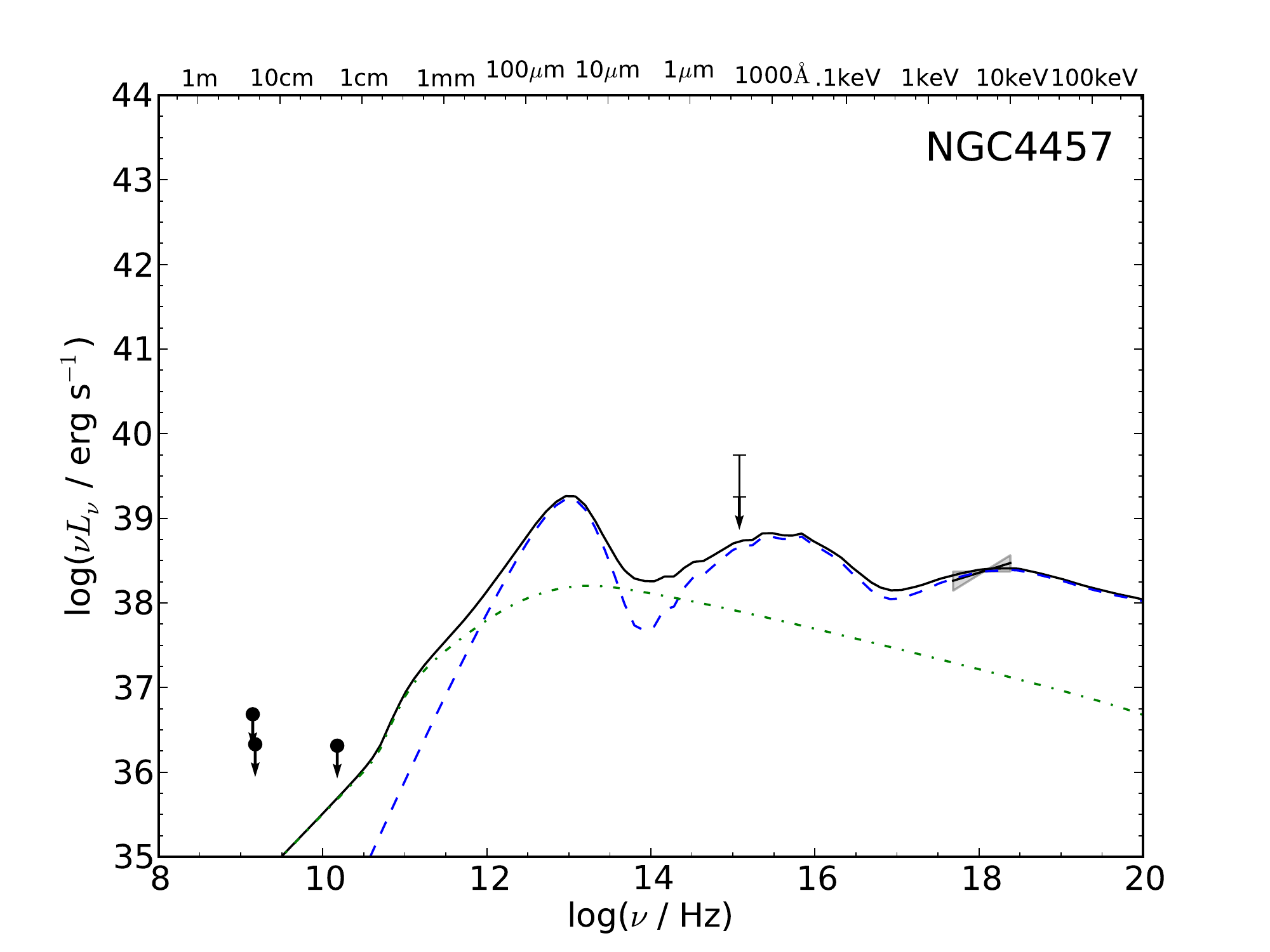}
\hskip -0.3truein
\includegraphics[scale=0.5]{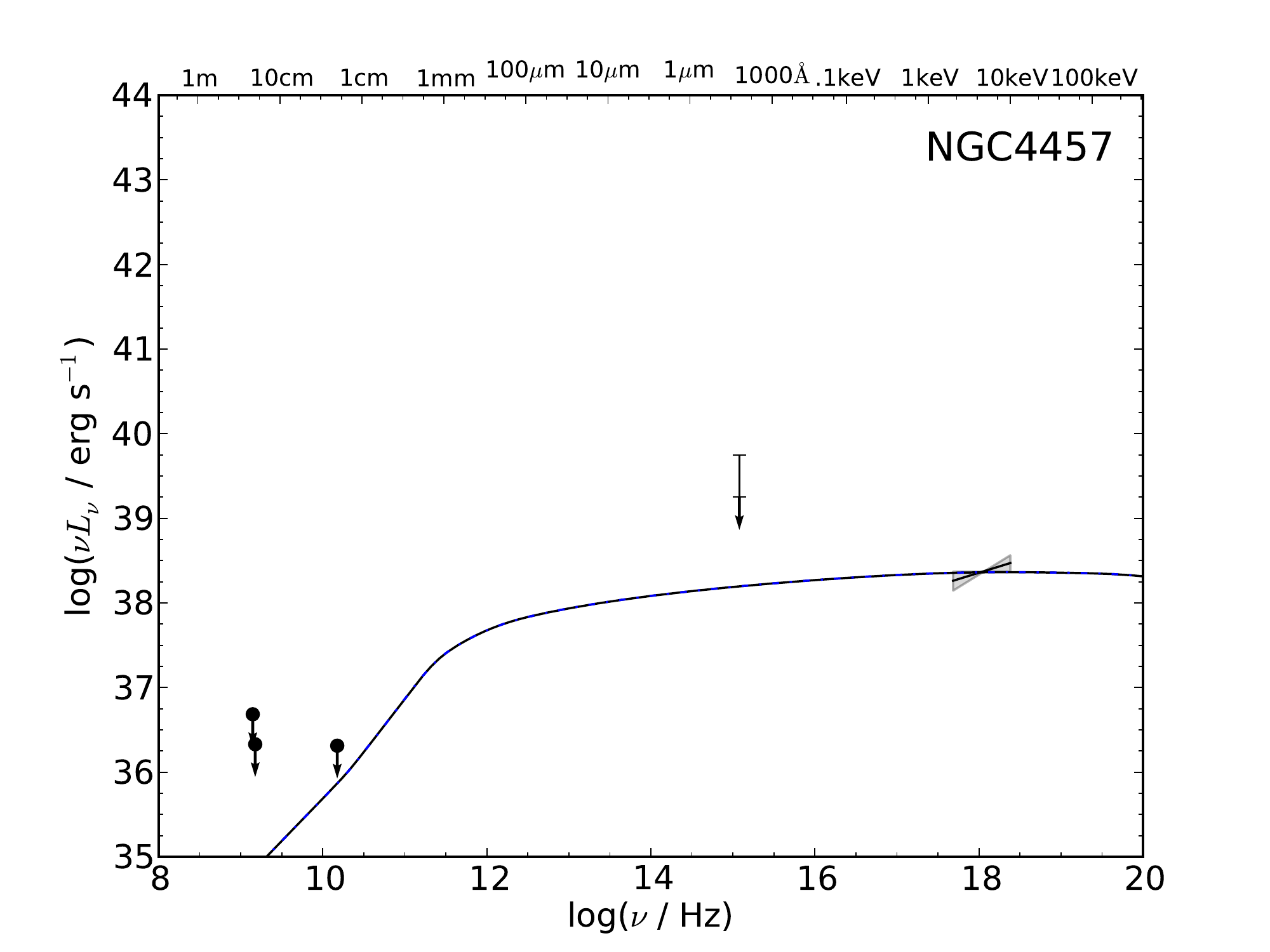}
}
\centerline{
\includegraphics[scale=0.5]{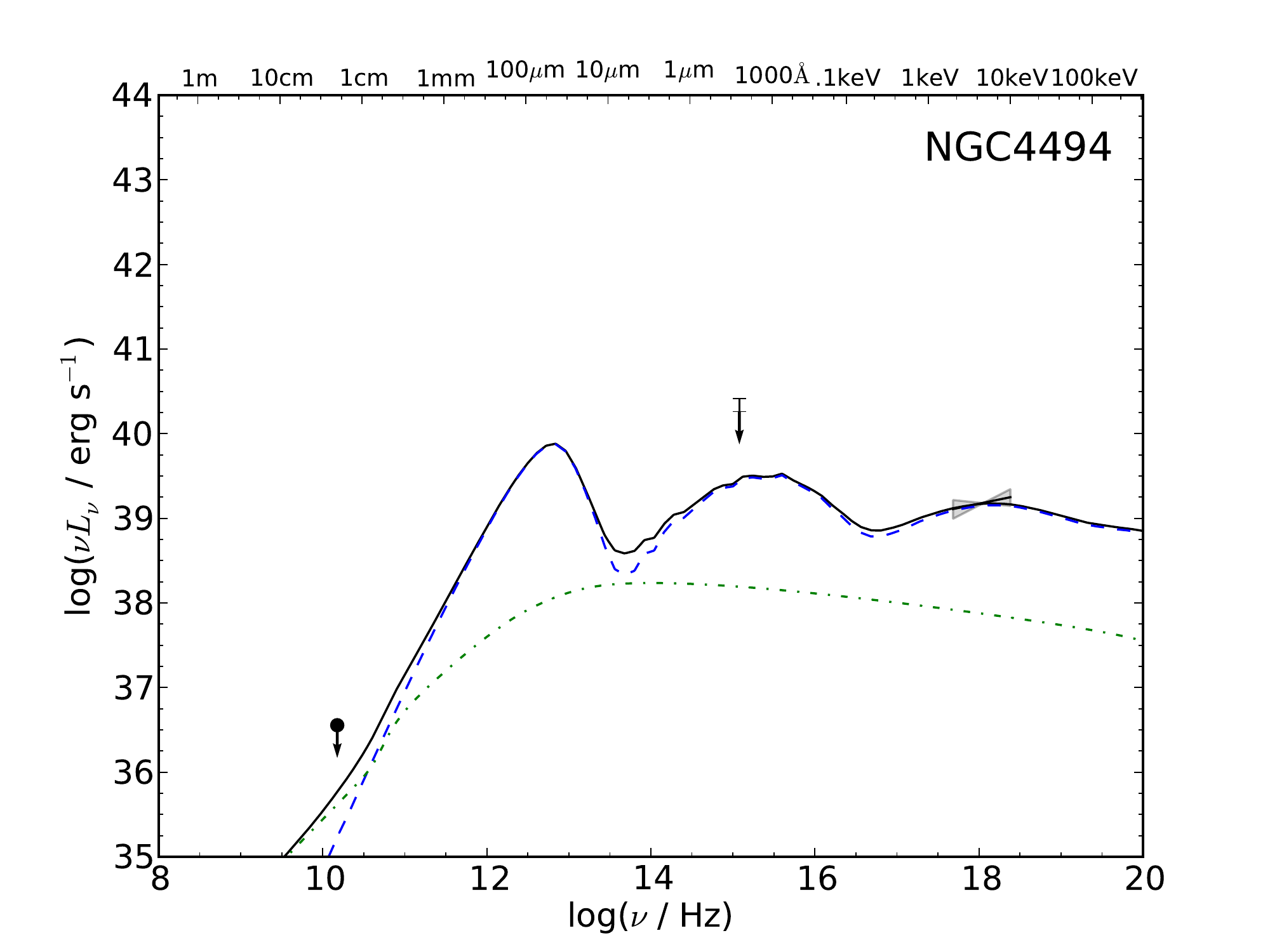}
\hskip -0.3truein
\includegraphics[scale=0.5]{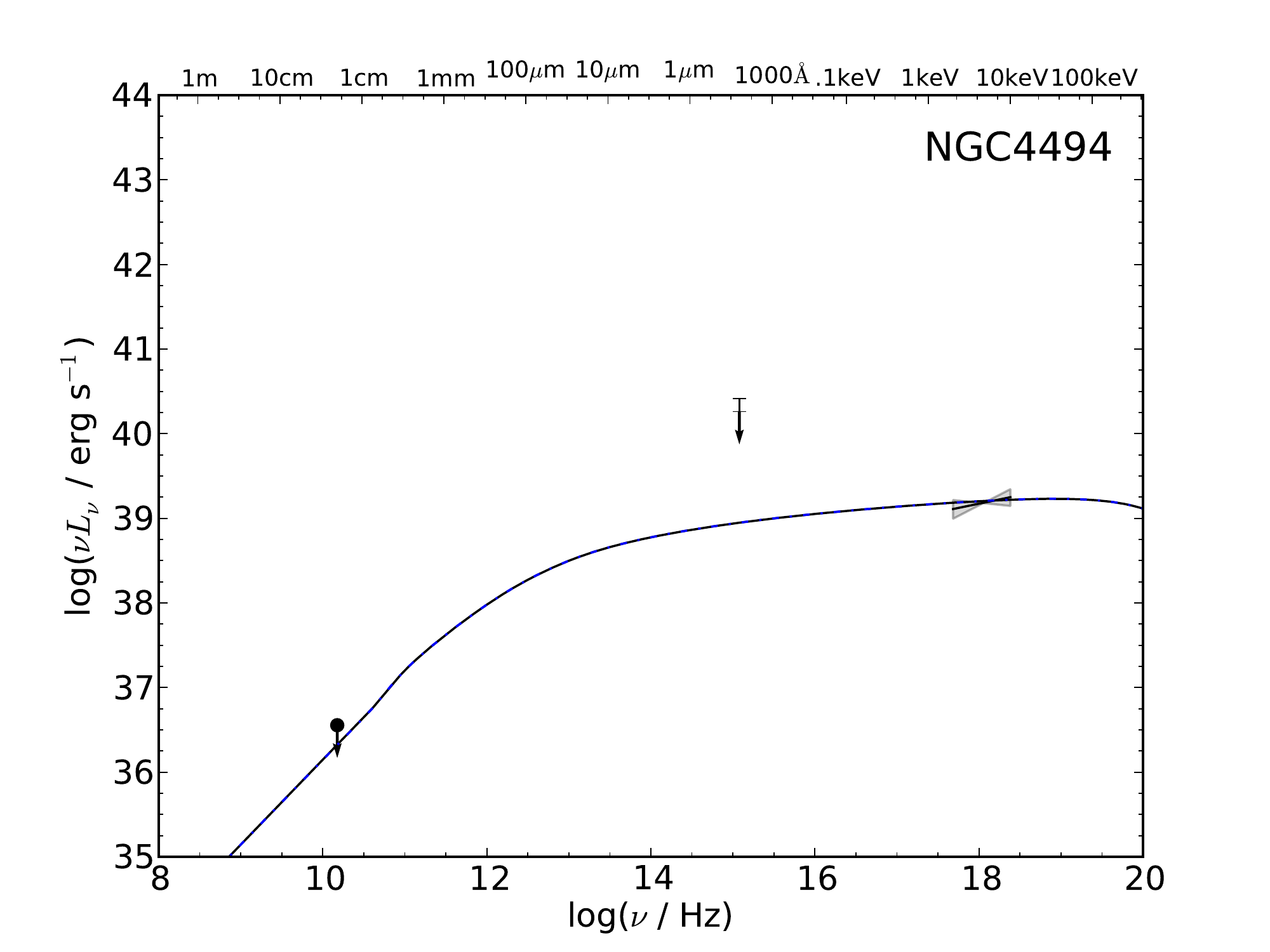}
}
\caption{Same as Figure \ref{fig:seds01} for NGC 4278, NGC 4457 and NGC 4494.}
\label{fig:seds04}
\end{figure*}

Note that the accretion rate that we obtain from the AD fit is somewhat smaller than $\dot{m}_{\rm Bondi}$. 
This AGN is likely to be strongly affected by extinction in the OUV band (EHF10). Hence, the OUV measurements probably do not capture the emission of the central engine. For this reason, it is not surprising that the models overpredict the OUV emission by almost one order of magnitude in Fig. \ref{fig:seds03}. 
 
\subsection{NGC 4278}

The radio emission of NGC 4278 was previously studied in the context of ADAF models by \citet{dmt01}, who found that $s > 0$ is required in order for the radio emission predicted by ADAF models not to overestimate the radio flux.
\citet{dmt01} also estimated that the Bondi rate is $\dot{m}_{\rm Bondi} \sim 0.001-0.01$. 
\citet{giro05} found a two-sided radio structure for the jet using VLA data, and estimated that the jet is oriented close to the line of sight ($2^\circ \lesssim i \lesssim 4^\circ$) and mildly relativistic ($\Gamma_j \sim 1.5$). In our jet models we therefore adopt $i=3^\circ$ and $\Gamma_j = 1.5$. 

Both the AD and JD models require a truncated thin disk with a small transition radius ($r_{\rm tr} \sim 30-40$) in order to account for the near-IR data. The values of the transition radius that we used in the models presented above ($r_{\rm tr}=30-40$) are not unique and models with larger radii can also reproduce the SED. For instance, an accretion model with $r_{\rm tr}=100$, $\dot{m}_{\rm out}=4 \times 10^{-3}$, $\delta=0.1$ and $s=0.77$ is able to explain the optical and X-ray observations. 

The jet emission by itself underpredicts the 1 mm radio observation, therefore the ADAF contribution is required even in the JD case. In the JD fit, a small value of $s$ is favored by the observations as is the case of the JD fit for NGC 4143.

\subsection{NGC 4457}

Since there are only upper limits in the radio band, the jet power for this object was estimated using the observation at $\nu = 1.5 \times 10^{10}$ Hz and the \citet{merloni07} correlation.

\subsection{NGC 4494}

Same as NGC 4457.

\subsection{NGC 4548}

Same as NGC 4457.
Given the large uncertainty in the photon index, the X-ray spectrum does not provide good constraints for the models. 

\subsection{NGC 4552}	\label{sec:4552}

The Bondi accretion rate was estimated by \citet{merloni07} (see also \citealt{allen06}) from the X-ray profiles of density and temperature. The kinetic power carried by the jet was estimated by \citet{merloni07} (see also \citealt{allen06}) from the energy deposited in the X-ray cavities.

Since the X-ray spectrum is quite soft, AD models are unable to account for the X-ray emission. 
The JD model in Fig. \ref{fig:seds05} is roughly consistent with the radio observations and explains quite well the X-ray data, but overpredicts the optical data. The resulting jet power is roughly consistent with the value estimated by \citet{allen06,merloni07}.



\subsection{NGC 4579}

NGC 4579 shares many characteristics with NGC 3031 (M81) and NGC 1097: its nucleus features broad double-peaked Balmer emission lines \citep{barth01} and a lack of the iron K$\alpha$ line emission \citep{erac02}.
\citet{quat99} previously modeled the SED of NGC 4579 compiled by \citet{ho99} using the ``old'' ADAF model \citep{ny95}. 

From the width of the broad H$\alpha$ line, \citet{barth01} obtained a rough estimate of the inner radius of the line-emitting portion of the accretion disk $r_{\rm tr} \sim 160$. Hence, we adopt in our SED models $r_{\rm tr} \lesssim 160$ and $i=45^\circ$.

\begin{figure*}
\centerline{
\includegraphics[scale=0.5]{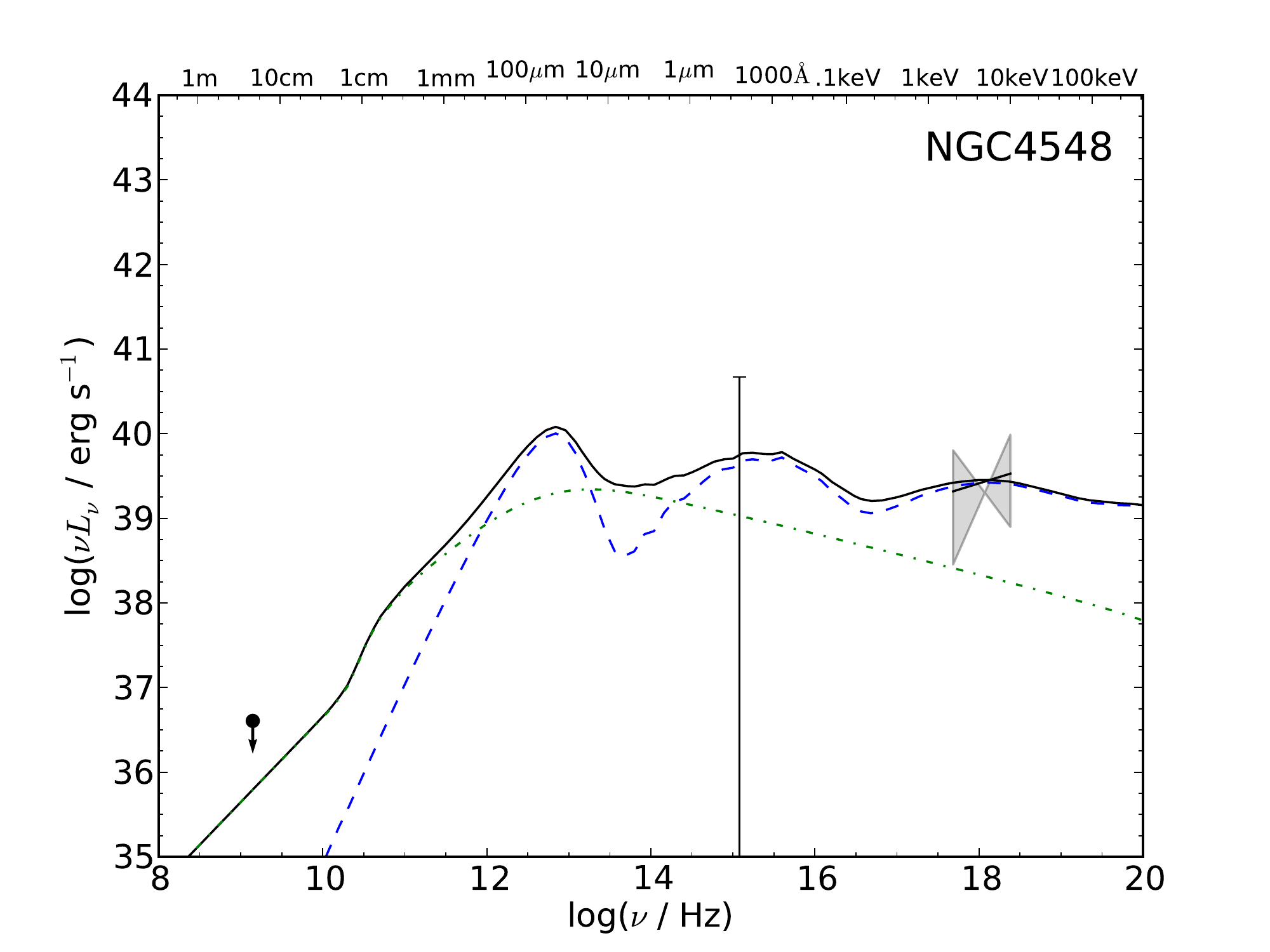}
\hskip -0.3truein
\includegraphics[scale=0.5]{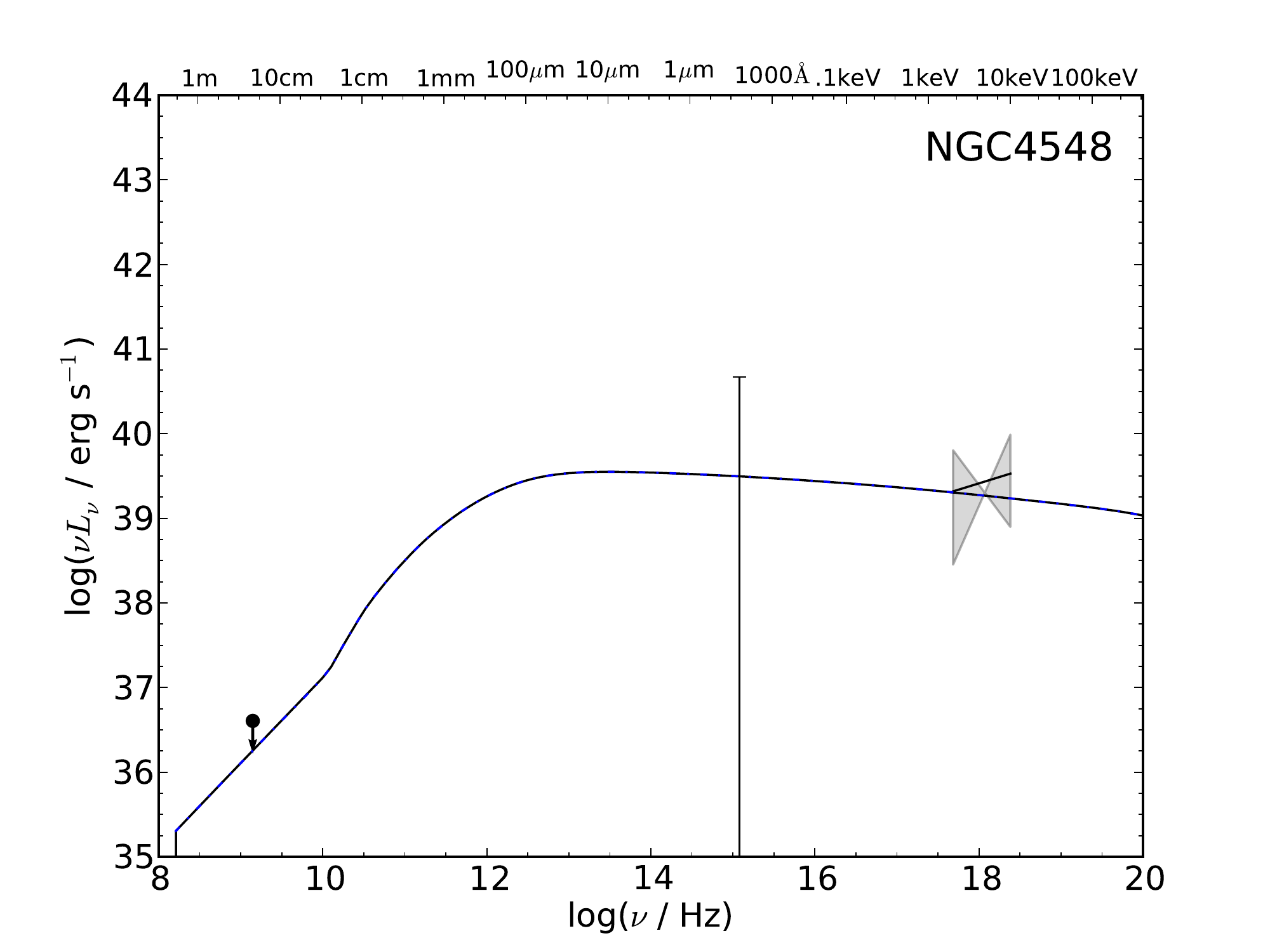}
}
\centerline{
\includegraphics[scale=0.5]{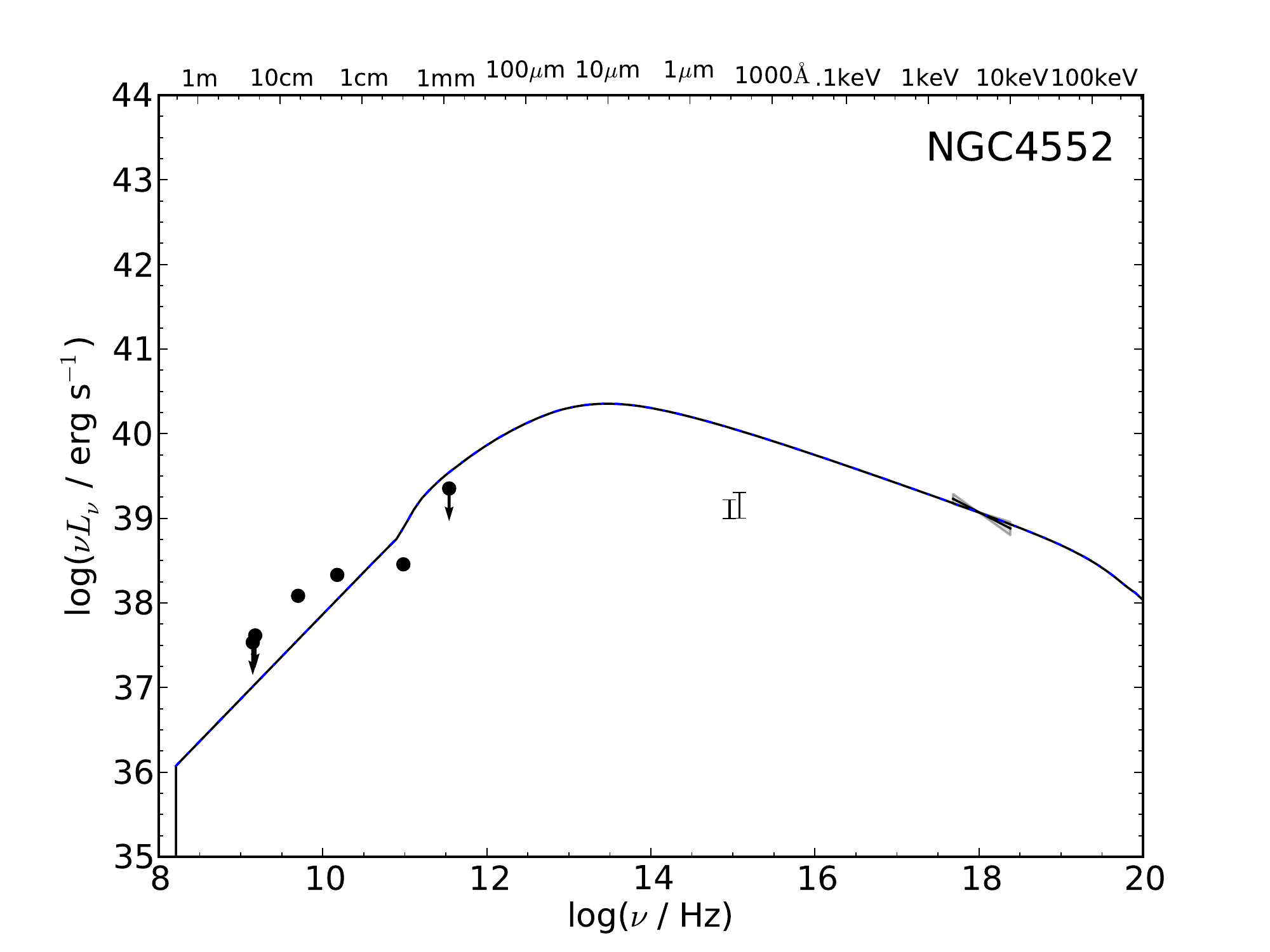}
\hskip -0.3truein
}
\centerline{
\includegraphics[scale=0.5]{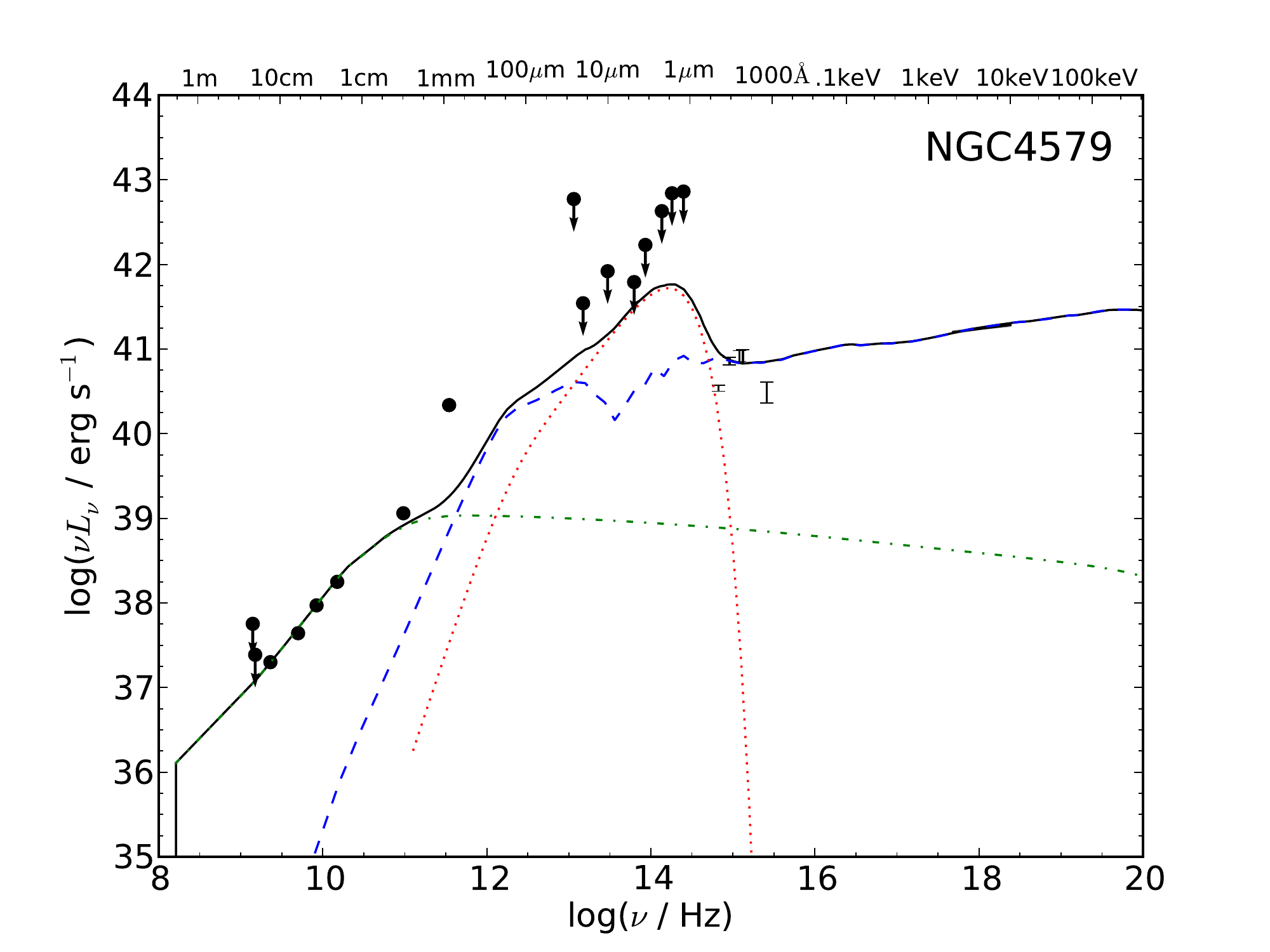}
\hskip -0.3truein
\includegraphics[scale=0.5]{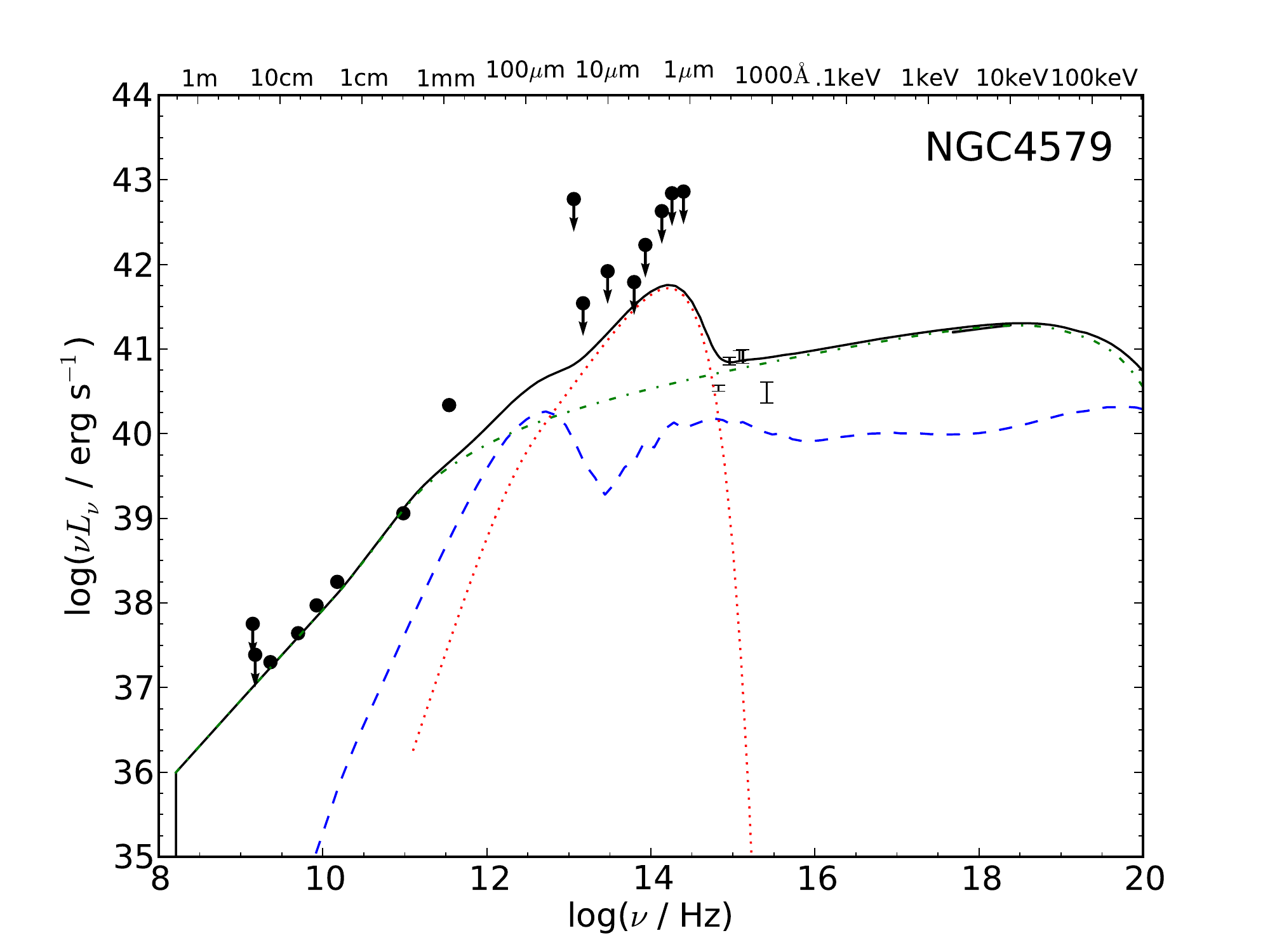}
}
\caption{Same as Figure \ref{fig:seds01} for NGC 4548, NGC 4552 and NGC 4579.}
\label{fig:seds05}
\end{figure*}

\begin{figure*}
\centerline{
\includegraphics[scale=0.5]{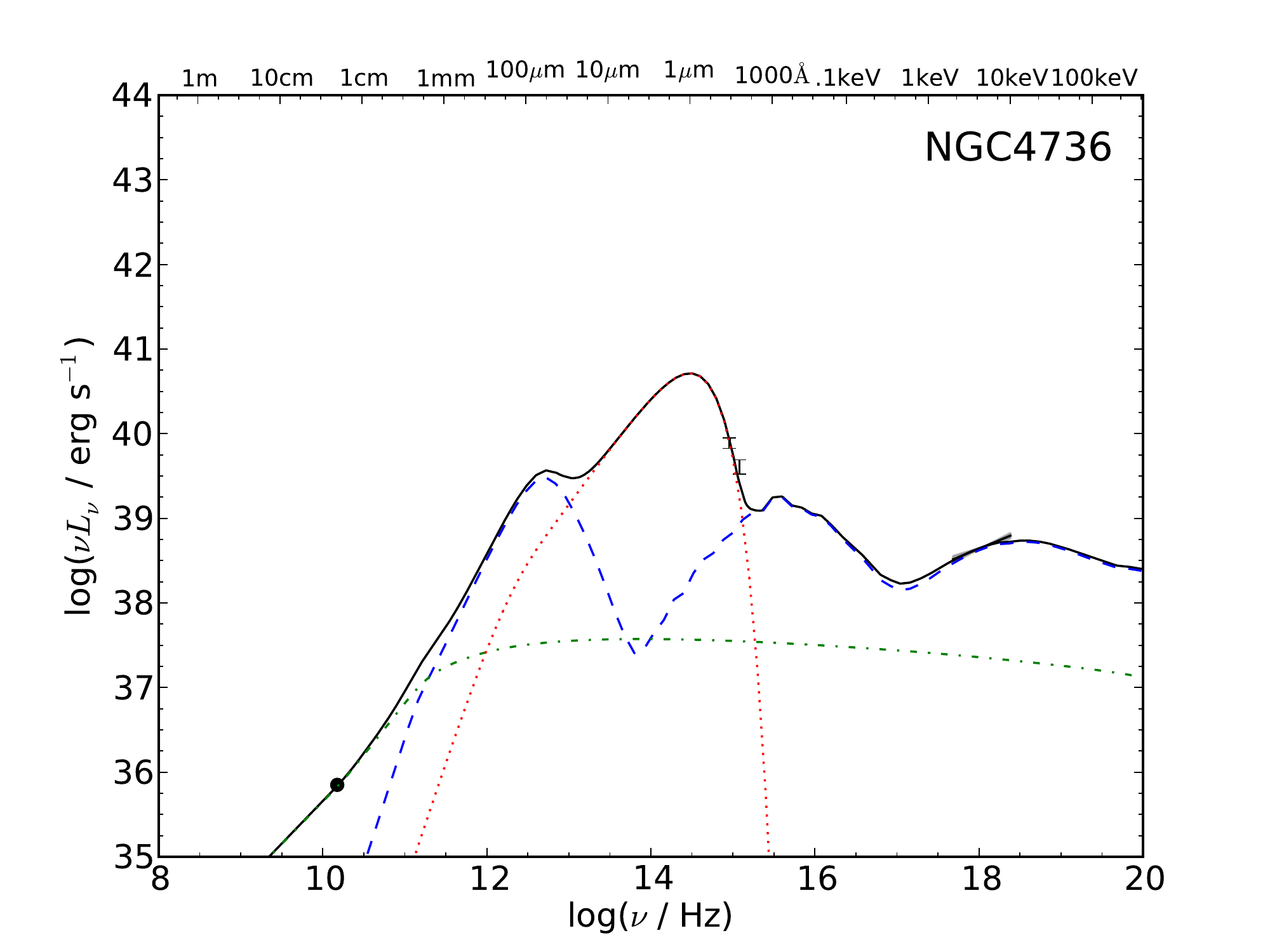}
\hskip -0.3truein
\includegraphics[scale=0.5]{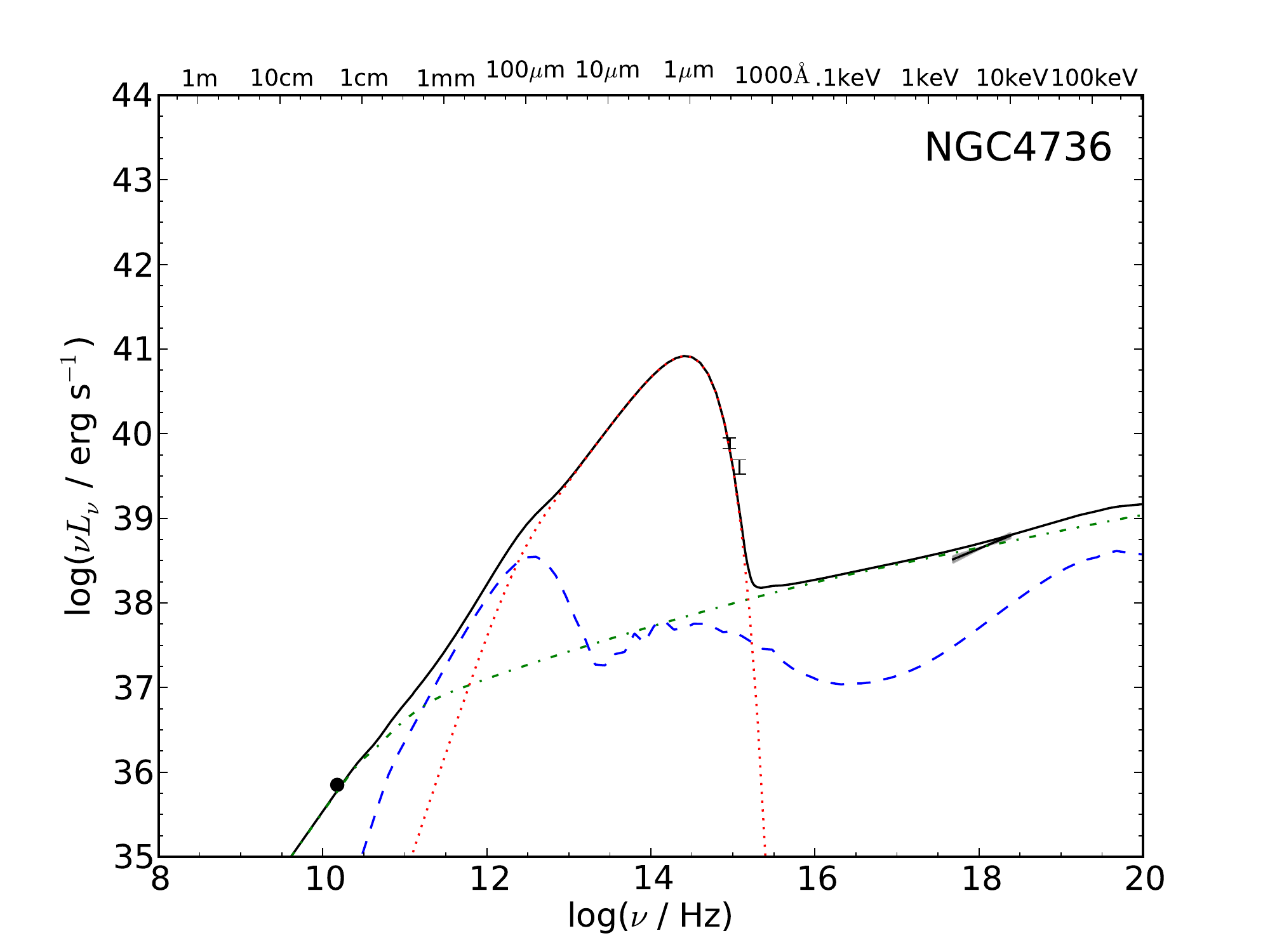}
}
\caption{Same as Figure \ref{fig:seds01} for NGC 4736.}
\label{fig:seds06}
\end{figure*}

As can be seen in Fig. \ref{fig:seds05}, the optical continuum data do not require a small transition radius since it can be accounted by either the jet (JD fit) or ADAF (AD fit) emission. The truncated thin disk in our fits is somewhat hotter than the thin disk modeled by \citet{quat99}, since their adopted black hole mass is outdated and is $\sim 10$ smaller than the value we use in this work. The transition radius can be chosen to be somewhat smaller if the accretion rate is lower and still be consistent with the data. 


We show together with the JD model an ADAF model which is the same as the one calculated for the AD fit except for the higher value of $s$.

\subsection{NGC 4736}

As was the case of NGC 1097, NGC 4143 and NGC 4278, this LINER requires a small transition radius in order to explain the OUV data.

\section{The distribution of the accretion-jet parameters for LLAGNs in LINERs} \label{sec:pars}

Our detailed SED modeling allows us to place independent constraints on the mass accretion rate onto the black hole, the jet mass-loss rate and jet kinetic power for the LLAGNs in our sample. These quantities are of direct relevance to studies of the feeding and feedback of the underfed supermassive black holes in nearby galaxies. It is therefore prudent to study the resulting distributions of the central engine parameters from our models.

\subsection{Accretion rates}

\begin{figure*}
\centering
\begin{minipage}[b]{0.49\linewidth}
\includegraphics[width=\linewidth]{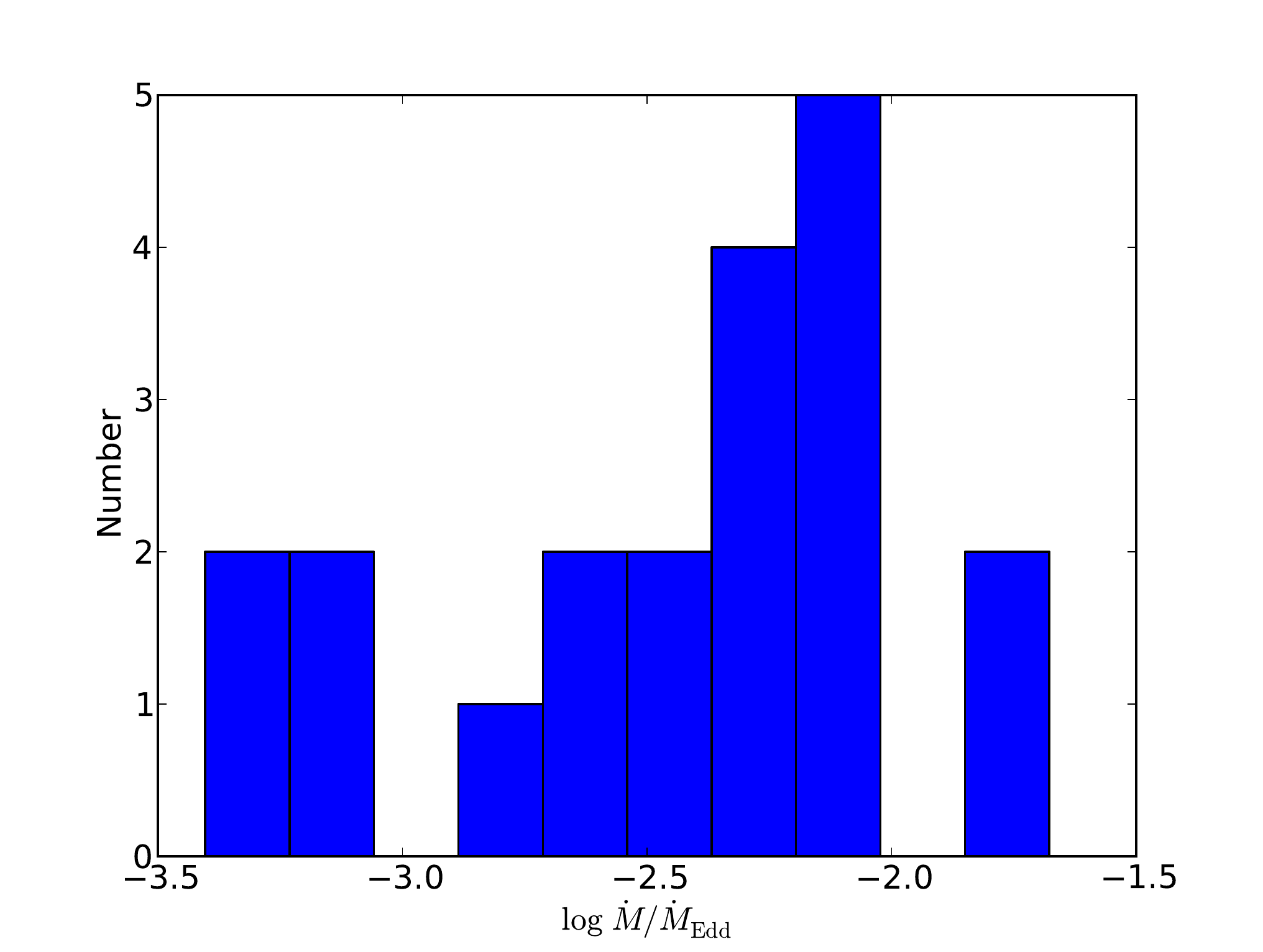}
\end{minipage}
\hfill
\begin{minipage}[b]{0.49\linewidth}
\includegraphics[width=\linewidth]{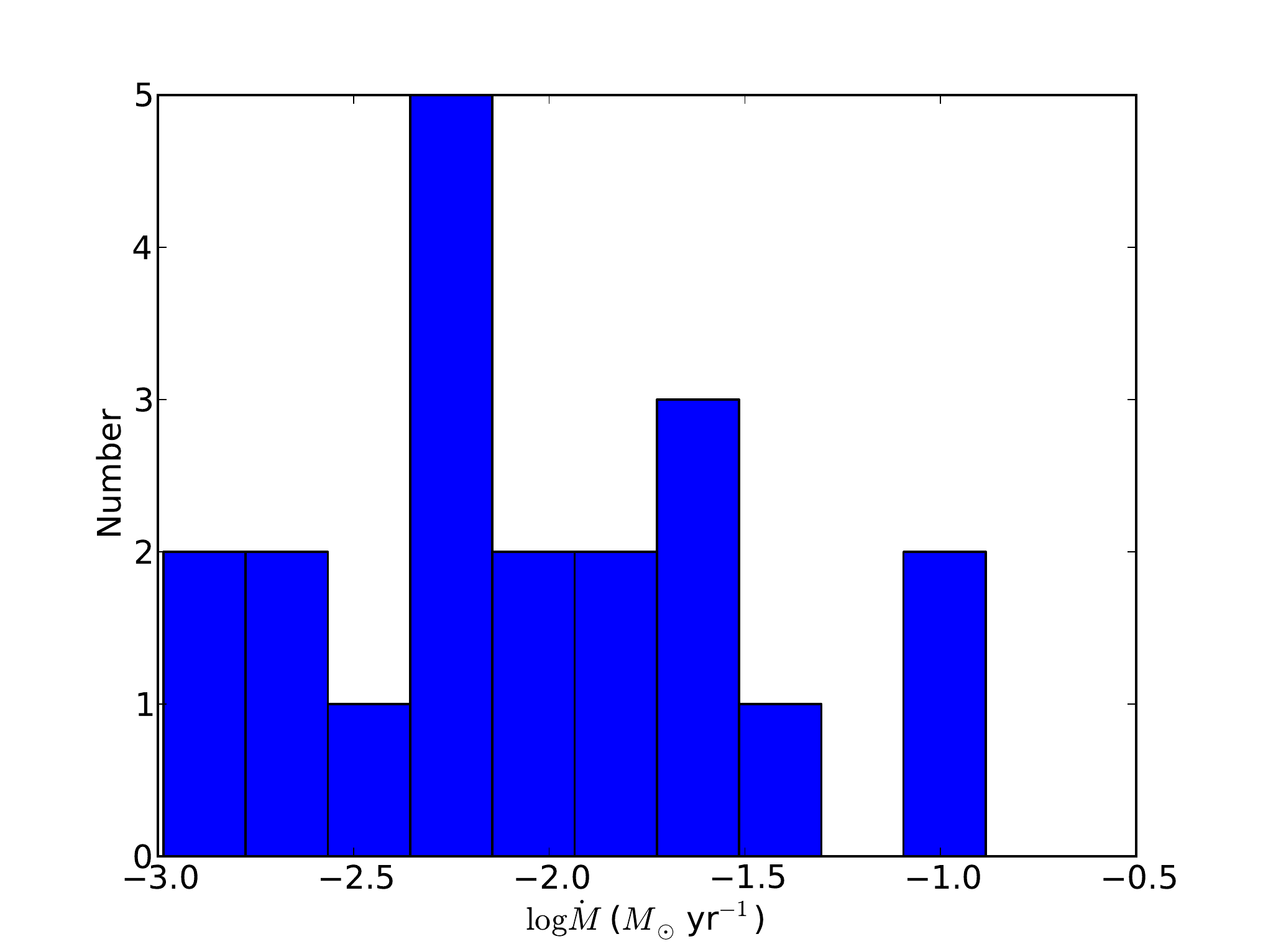}
\end{minipage}
\caption{Distribution of accretion rates at the outer radius of the ADAF obtained from our AD SED fits. \textbf{a, left:} Accretion rate in Eddington units. \textbf{b, right:} Accretion rate in physical units.}
\label{fig:mdot}
\end{figure*}

In the previous section we showed that essentially two types of models (AD and JD) are able to account for the SEDs, depending on the combination of parameters that we choose. We found that in most of the JD fits, the contribution of an underlying ADAF is not required at all (cf. the JD models for NGC 4594, M87 and NGC 4579). Since the estimate of the accretion rate relies on fitting the ADAF-thin disk model to the data, in those cases no robust $\dot{m}$ constraint is available. We therefore show in the left and right panels of Figure \ref{fig:mdot} the histograms displaying the distribution of values of $\dot{m}_{\rm out}$ (dimensionless) and $\dot{M}_{\rm out}$ (in units of $M_\odot$ year$^{-1}$) respectively, considering only the AD models. For one object (NGC 4552) we are unable to find a suitable AD model, therefore we do not include this object in the histogram.

Most of the objects have accretion rates in the range $\dot{m} \approx 0.003-0.01$. The mean accretion rate for the sample, in the case of the AD fits, is $\left \langle \log \dot{m} \right \rangle = -2.5$ or $\left \langle \log \left( \dot{M}/{\rm M_\odot \ year^{-1}} \right) \right \rangle = -2$ with standard deviations of 0.5 and 0.6, respectively. 

\begin{figure}
\centering
\includegraphics[scale=0.45]{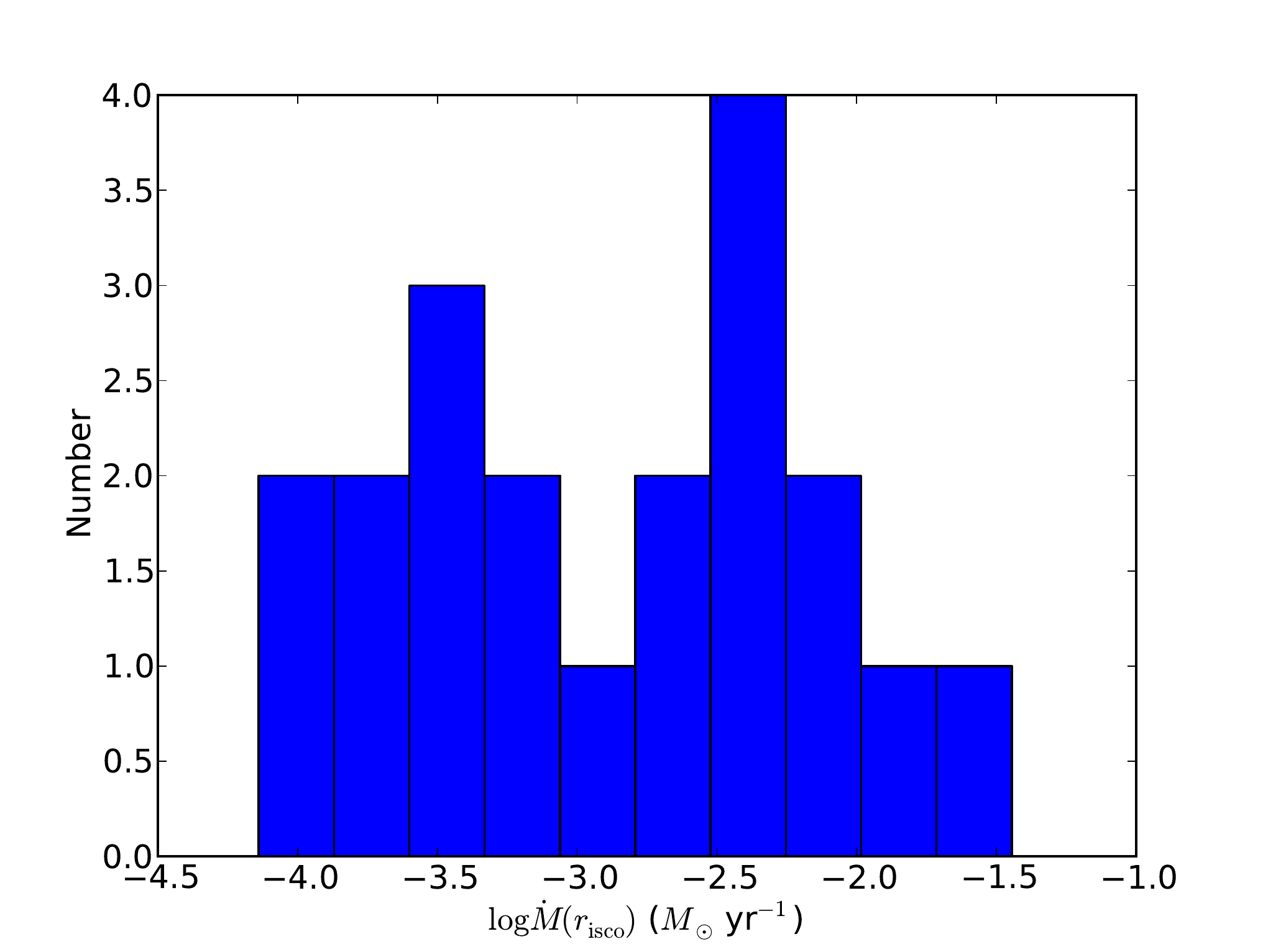}
\caption{Distribution of accretion rates at the radius corresponding to the ISCO of a Schwarzschild black hole, estimated from the AD fits.}
\label{fig:mdotisco}
\end{figure}

Not all gas supplied at the outer radius of the accretion flow ends up being accreted onto the black hole due to mass-loss via winds produced in the ADAF. 
From our derived parameters we can estimate the fraction of the gas that actually falls into the event horizon. Figure \ref{fig:mdotisco} shows the histogram of values of $\dot{M}(r_{\rm isco})$ taking into account the radial dependence of the density profile in the ADAF, where $r_{\rm isco} = 3$.
We obtain $\left \langle \log \left( \dot{M}(r_{\rm isco})/{\rm M_\odot \ year^{-1}} \right) \right \rangle = -3.3$ with a standard deviation of 0.5. In other words, typically only a fraction of 14\% of the available fuel supply reaches near the ISCO and presumably a similar fraction is accreted onto the black hole.

\subsection{Jet powers}

\begin{figure}
\centering
\includegraphics[scale=0.47]{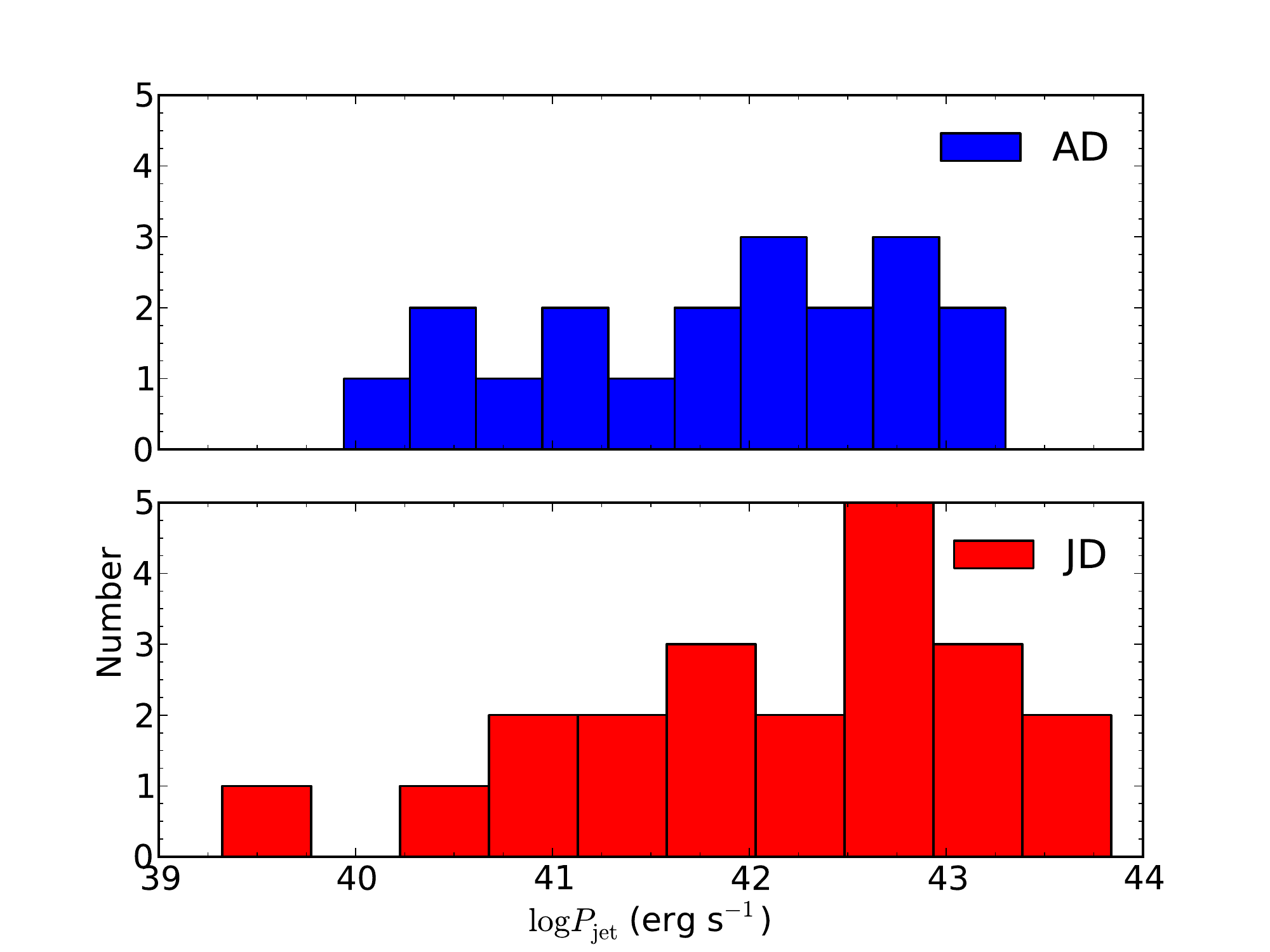}
\caption{Distribution of jet powers.}
\label{fig:jetpower}
\end{figure}

Especially relevant for studies of the feedback of massive black holes in the local universe is the amount of energy that the jets deposit in their environment. Figure \ref{fig:jetpower} shows the distribution of jet kinetic powers obtained via both the AD and JD type of SED fits. As we noted above, we were not able to estimate a jet power for the AD model fit to NGC 4552. The average jet power $\left \langle \log \left( P_{\rm jet}/{\rm erg \ s}^{-1} \right) \right \rangle$ is 41.8 and 42.2 respectively for the AD and JD scenarios, each with a standard deviation of $\approx 1$ dex. Therefore, we can say that the typical jet power is $\sim 10^{42 \pm 1} \ {\rm erg \ s}^{-1}$.
The distributions in Fig. \ref{fig:jetpower} are in qualitative agreement with the distribution of jet powers estimated by \citet{nagar05} (see their Fig. 6).

\begin{figure}
\centering
\includegraphics[scale=0.47]{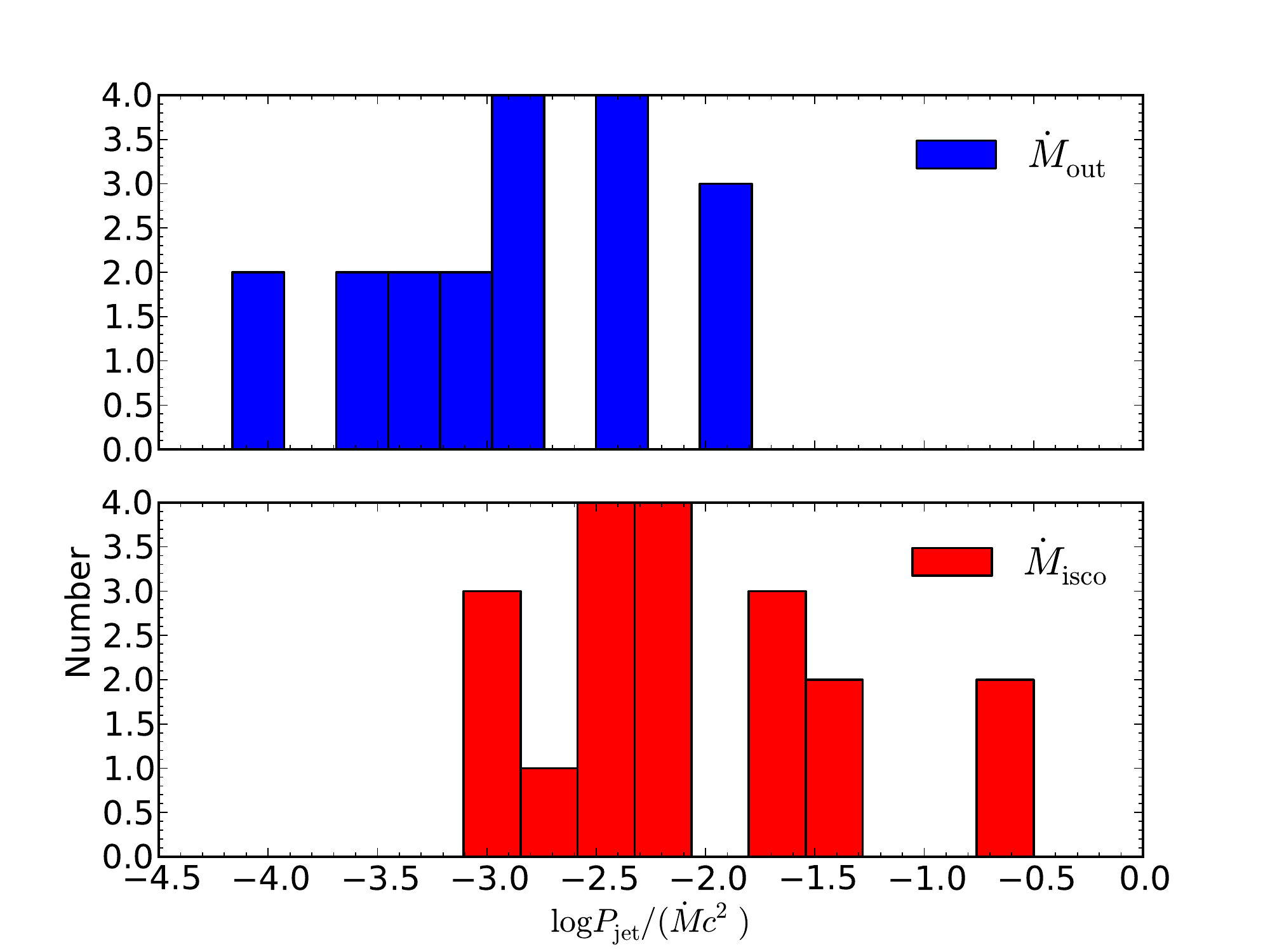}
\caption{Distribution of jet kinetic efficiencies $P_{\rm jet}/( \dot{M} c^2)$ considering $\dot{M}=\dot{M}_{\rm isco}$ and $\dot{M}=\dot{M}_{\rm out}$ for the AD models.}
\label{fig:effjet}
\end{figure}

Also relevant in studies of the kinetic AGN feedback is what fraction of the rest mass energy associated with the mass accretion rate is tapped by the central engine and converted into the jet kinetic power, which can be defined as $\eta_{\rm jet} \equiv P_{\rm jet}/( \dot{M} c^2)$. The choice of accretion rate in the definition of the ``jet kinetic efficiency'' is arbitrary and we could choose to use either the accretion rate $\dot{M}_{\rm out}$ at the outer radius of the ADAF or the one at the innermost stable circular orbit (ISCO), $\dot{M}_{\rm isco}$ (note that $\dot{M}_{\rm isco} < \dot{M}_{\rm out}$ given the mass-loss through winds in the ADAF). The ISCO radius in this case corresponds to $3 R_S$, appropriate for a Schwarzschild black hole. Figure \ref{fig:effjet} shows the distribution of the jet kinetic efficiency defined using $\dot{M}_{\rm out}$ (top panel) and $\dot{M}_{\rm isco}$ (bottom panel) for the case of AD SED fits, since in the JD fits there are no reliable estimates of the accretion rates. The average values of $\log \eta_{\rm jet}$ are -2.9 ($\dot{M} = \dot{M}_{\rm out}$) and  -2.1 ($\dot{M} = \dot{M}_{\rm isco}$) (both these averages have an uncertainty of $\approx 0.7$ dex). Clearly, the latter efficiency is higher in order to be consistent with the smaller ISCO accretion rates.

\begin{figure}
\centering
\includegraphics[scale=0.45]{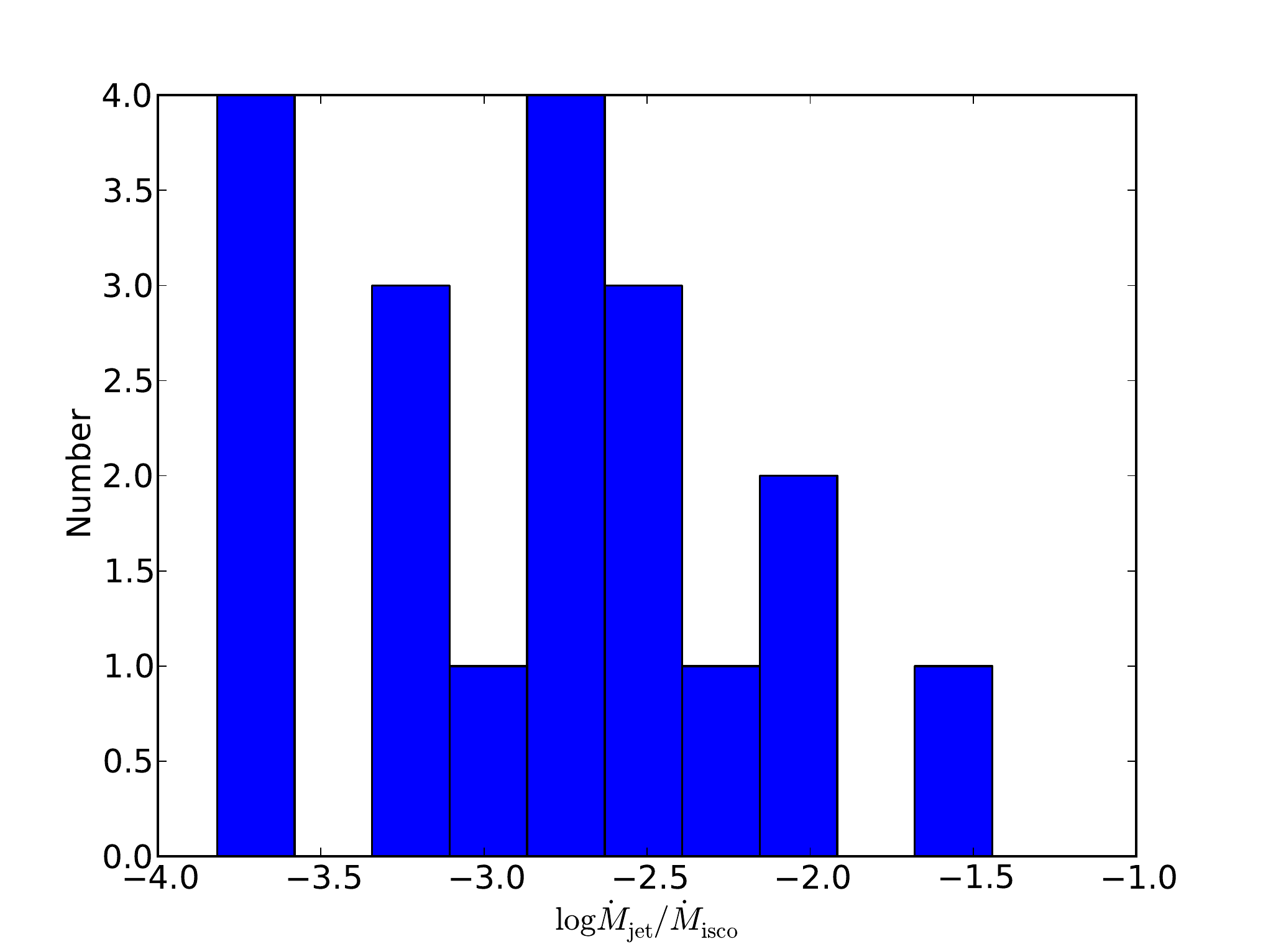}
\caption{Distribution of $\dot{M}_{\rm jet}/\dot{M}_{\rm isco}$.}
\label{fig:ratio}
\end{figure}

We are also able to quantify the fraction of the mass accretion rate that escapes the attraction of the black hole and is channeled into the jet. Figure \ref{fig:ratio} shows the distribution of values of $\dot{M}_{\rm jet}/\dot{M}_{\rm isco}$ for the AD fits. As it turns out, the average value of the ratio is $\left \langle \log \left( \dot{M}_{\rm jet}/\dot{M}_{\rm isco} \right) \right \rangle = -2.9$, i.e. on average $0.4\%$ of the material that reaches the ISCO is channeled in the jet. The corresponding uncertainty in this estimate is 0.6 dex.

\subsection{Bolometric luminosities and radiative efficiencies}


One possible way of estimating the bolometric luminosities for the LLAGNs using the observed SEDs is to integrate the SEDs directly, not taking into account upper limits (especially in the IR) and assuming a suitable interpolation. For instance, this procedure was carried out by EHF10 who also computed a bolometric correction for the X-ray luminosity.

\begin{figure}
\centering
\includegraphics[scale=0.47]{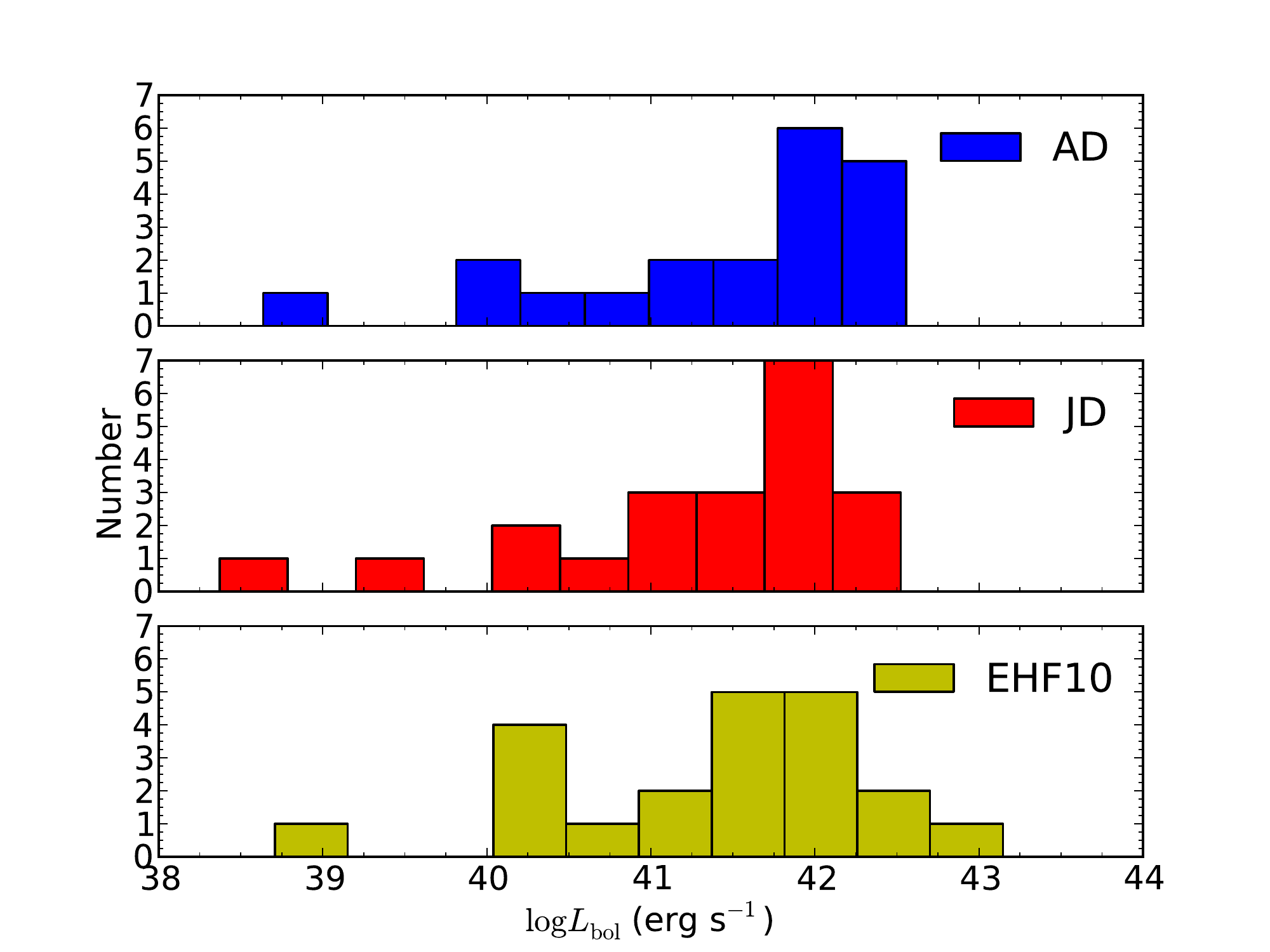}
\caption{Distribution of bolometric luminosities inferred from the models compared with the ones estimated by EHF10.}
\label{fig:lumbol}
\end{figure}

In our case, we have physically motivated fits for the 21 SEDs. 
Therefore, for each LINER in our sample, we calculated $L_{\rm bol}$ by integrating the total SED. Carrying this out using the SEDs obtained according to the AD scenario we obtain the average value of $\left \langle \log \left( L_{\rm bol} /  \ {\rm erg \ s}^{-1} \right) \right \rangle = 41.5$ with a standard deviation of $\approx 1$ dex. For the SED fits computed according to the JD scenario we find a corresponding average of 41.3 with a similar standard deviation, which gives the same average as the one computed using the bolometric luminosities estimated by EHF10. 
Figure \ref{fig:lumbol} displays the distribution of values of $L_{\rm bol}$ resulting from the models compared with the one estimated by EHF10.

Defining the X-ray bolometric correction as $\kappa_X \equiv L_{\rm bol}/L_X$, we calculated the average bolometric correction $\langle \log \kappa_X \rangle$ resulting from our models as 1.45 and 1.35 according to the AD and JD scenarios respectively, with standard deviations of 0.36 dex (AD) and 0.46 dex (JD). The corresponding geometric means of $\kappa_X$ are 28 (AD) and 22 (JD), the mean values are 43 (AD) and 47 (JD) and the medians are 22 (AD) and 14 (JD). Our models result in values of $\kappa_X$ which are somewhat lower than those estimated by EHF10.

Using our SED models we can calculate what is the typical fraction of the accreted matter that is converted into the combined radiation of the accretion flow and the jet, i.e. the radiative efficiency $\eta_{\rm rad} \equiv L_{\rm bol}/(\dot{M}_{\rm out} c^2)$ (note that the accretion rate in the definition of this efficiency is the one at the outer radius of the ADAF). We obtained the average value $\left \langle \log \eta_{\rm rad} \right \rangle = -3.25$ with a standard deviation of 0.75 dex. Hence, the typical radiative efficiency is $5 \times 10^{-4}$.

\begin{figure}
\centering
\includegraphics[scale=0.47]{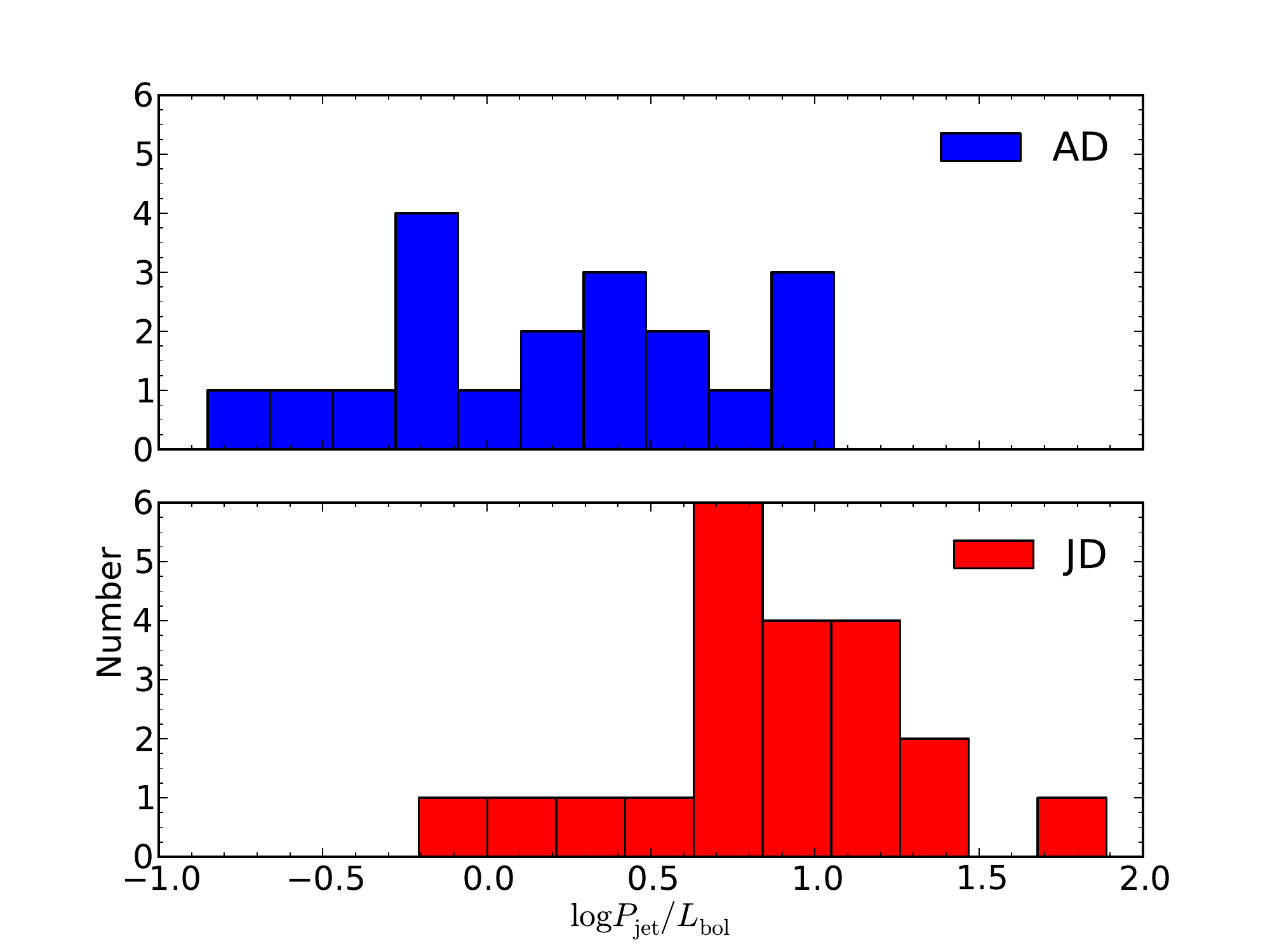}
\caption{Distribution of the ratio of jet power to bolometric luminosity.}
\label{fig:jetlbol}
\end{figure}

We estimated how the jet kinetic power compares to the total radiative output of the central engine, i.e. we calculate the ratio $P_{\rm jet}/L_{\rm bol}$ where $P_{\rm jet}$ is obtained from the jet model and $L_{\rm bol}$ from integrating the modeled SEDs. We obtain $\left \langle \log \left( P_{\rm jet}/L_{\rm bol} \right) \right \rangle \approx 0.2$ for the AD-type of fits and 0.9 for the JD models (there is a $\approx 0.5$ dex uncertainty in both values). 
The reason for the smaller jet powers in the AD fits compared to the JD ones is that the mass-loss rates required in the former fits are systematically smaller than in the ones in which the jet has a higher degree of contribution in the X-rays.
Based on our results and given the degeneracy in the fits, we can say that the typical kinetic to radiative power ratio of LLAGNs is $\approx 5$ (i.e. the average of the value from the AD and JD fits). This is in agreement with the findings of \citet{nagar05} that the accretion power output in LLAGNs is dominated by the jet power. Figure \ref{fig:jetlbol} shows the distribution of values of $P_{\rm jet}/L_{\rm bol}$ resulting from the models. 

\section{The average SED}	\label{sec:avgsed}


It is of interest to compute the average SEDs resulting separately from the AD and JD scenarios. These SEDs are useful for different purposes. First and foremost, since the observed SEDs are the main observables that we use to derive the central engine parameters via our fitting, as a consistency check the average model SEDs should reproduce the average SED obtained directly from the observed ones. 

Secondly, SEDs obtained from averaging observed data points are obviously limited by the observed bands. Outside the observed bands different authors adopt \textit{ad-hoc} interpolations between the data, usually a linear interpolation in log-log space (\citealt{ho99}; EHF10), which might not reflect the actual physical processes involved in the emission. The average from the AD and JD scenarios that we obtained can hence serve as guidelines -- with a physical justification -- to the typical shape of the SEDs in the bands which have not been constrained yet.

\begin{figure*}
\centerline{
\includegraphics[scale=0.5]{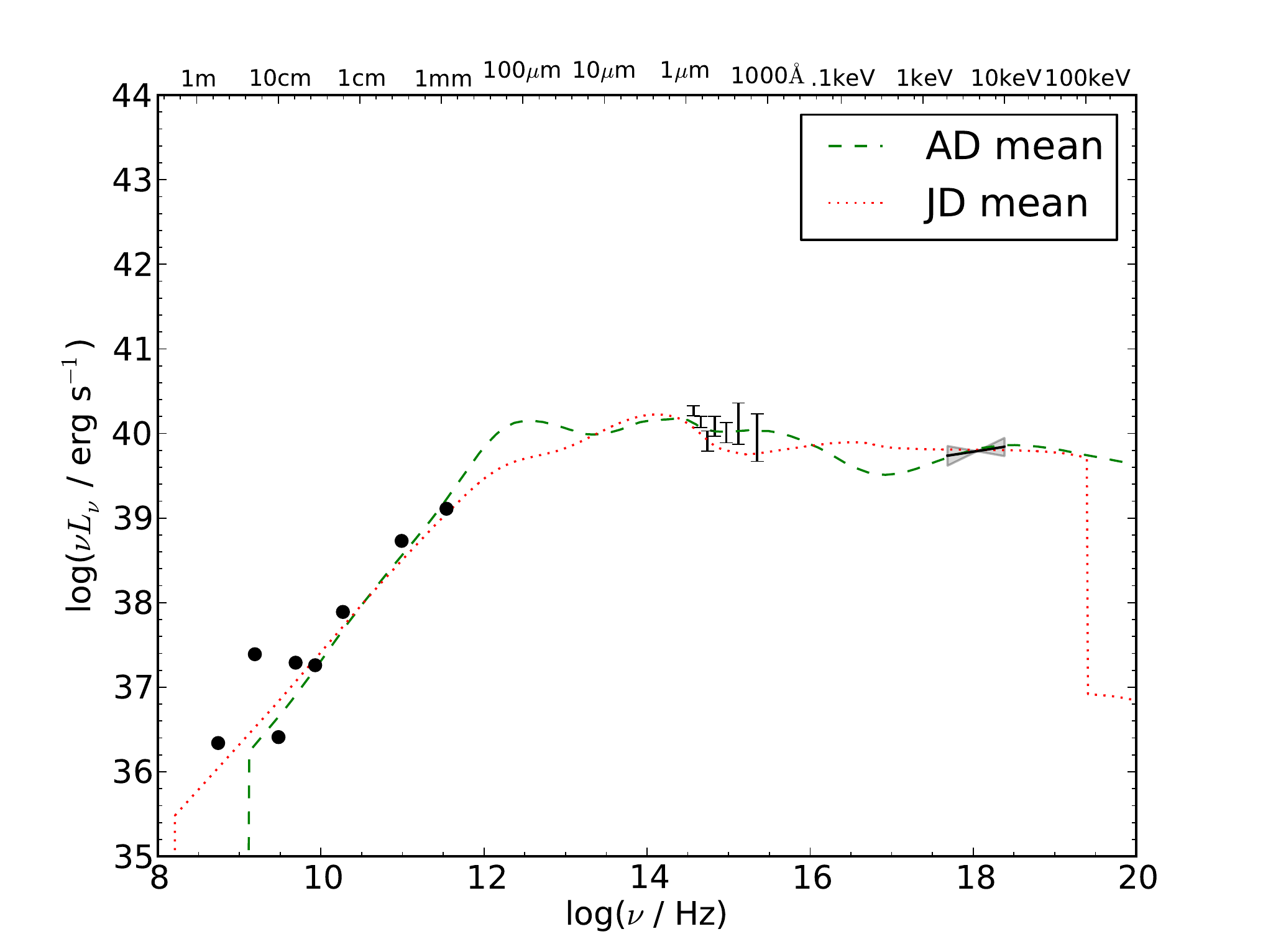}
\hskip -0.3truein
\includegraphics[scale=0.5]{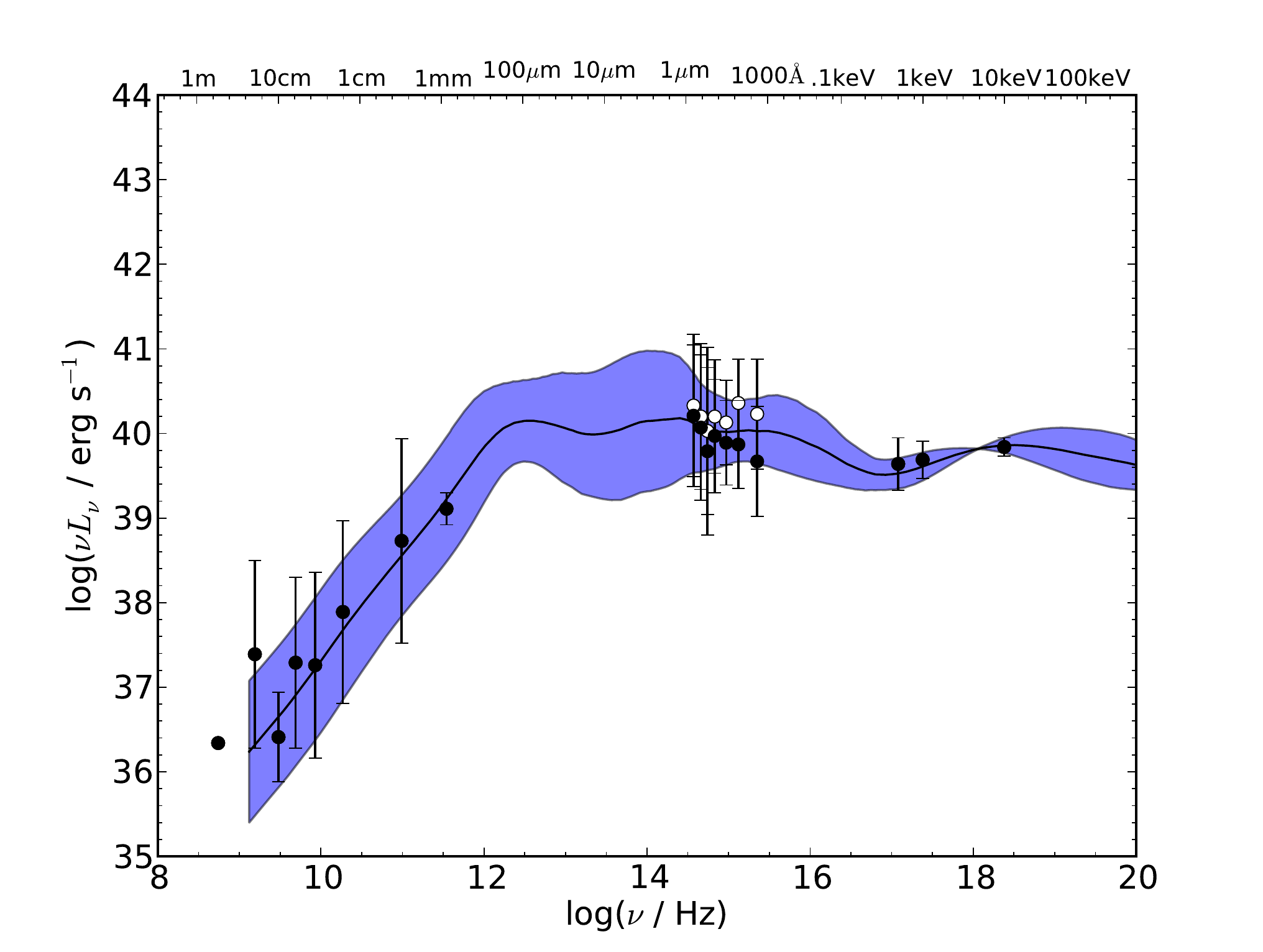}
}
\caption{\textbf{a, left:} The average SEDs (geometric mean) computed separately for the AD and JD models (dashed and dotted lines respectively). The data points correspond to the geometric mean computed by EHF10. \textbf{b, right:} $1\sigma$ scatter around the average (model and observed) SEDs illustrating the diversity of individual SEDs. The solid line shows the average AD SED and the shaded region corresponds to the standard deviation from the AD models. The points correspond to the mean computed by EHF10 and the error bars show the scatter in the measurements. In the OUV, the filled circles correspond to measurements without any reddening correction whereas the open circles correspond to the maximal extinction correction.}
\label{fig:avgsed}
\end{figure*}

Motivated by the reasons above, we show in Fig. \ref{fig:avgsed}a the average SEDs computed separately for the AD and JD scenarios. We first normalized the individual SEDs to the same X-ray luminosity of $10^{40} \ {\rm erg \ s}^{-1}$ in the band 2-10 keV. This value is approximately the average X-ray luminosity for the LINERs in our sample. After normalizing the SEDs, for each frequency bin we computed the geometric mean of the model SEDs corresponding to the particular model class as $\left\langle \log \nu L_\nu \right \rangle$. Following EHF10 we choose to compute the average of the logarithm of the luminosities instead of using the values of $\nu L_\nu$ themselves in order to reduce the effect of outliers in the resulting average. 

The different bumps in the average model SEDs reflect the different physical processes that operate in the flows. In the average AD model SED, the different bands are dominated by different radiative processes in the flow as described below:
\begin{description}
\item[1 m -- 1 mm] jet optically thick synchrotron emission.
\item[1 mm -- $\approx 100 \mu$m] the first bump and usually the strongest one in the average AD SED is due to the ADAF synchrotron emission.
\item[$\approx 10  \mu$m -- $\approx 1 \mu$m] the second bump is due to the thermal emission from the truncated thin disk.
\item[$\approx 1  \mu$m -- 0.1 keV] there is a weak UV bump due to the first inverse Compton scattering of synchrotron photons in the ADAF.
\item[$\approx 0.1$ keV -- 100 keV] the last bump in X-rays is dominated by the second inverse Compton scattering of synchrotron photons in the ADAF.
\end{description}

The average JD SED is more simple and has only one bump in the IR between a few $\times10  \mu$m -- a few $\times 1 \mu$m which is due to the truncated thin disk. The rest of the emission is due to synchrotron photons from the jet.
Overall, the spectral shapes of the average JD and AD SEDs are quite similar in the radio and X-rays above 1 keV. In between these bands the shapes are slightly dissimilar.
The bumps in the average SEDs are much less pronounced than the bumps in the individual SEDs. This occurs because the bumps in the individual SEDs do not peak at the same place and when the SEDs are averaged these bumps are smoothed out. For this reason, the average model SEDs will not resemble any one of the individual SEDs.

Figure \ref{fig:avgsed}a also displays the average data points computed from the observed SEDs in a similar way by EHF10 where the error bars represent the uncertainty in the emission due to the uncertainty affecting the amount of extinction correction involved. The shaded region around the average best-fit X-ray power-law represents the standard deviation in the value of the LINER photon index.
In order to illustrate the wide diversity of the individual SEDs, we show in Figure \ref{fig:avgsed}b the $1\sigma$ scatter affecting the model SEDs (where we show only the AD average SED for simplicity) and the observed ones. 

As expected, the average model SEDs agree well with the observed constraints. There are some details that are worth mentioning. 
The shape of the X-ray spectrum of the average JD SED is slightly softer than the corresponding shape of the average AD SED. Both are within the $1\sigma$ uncertainty in the photon index of the average observed SED.
In the OUV, even though the model SEDs agree with the observed constraints they are quite different from each other. For instance, the red bump is stronger in the average JD SED. The average JD SED predicts a lower level of UV flux.
In the radio band, the model SEDs are quite similar to each other.

\begin{figure}
\includegraphics[scale=0.5,trim=30 0 0 0,clip=true]{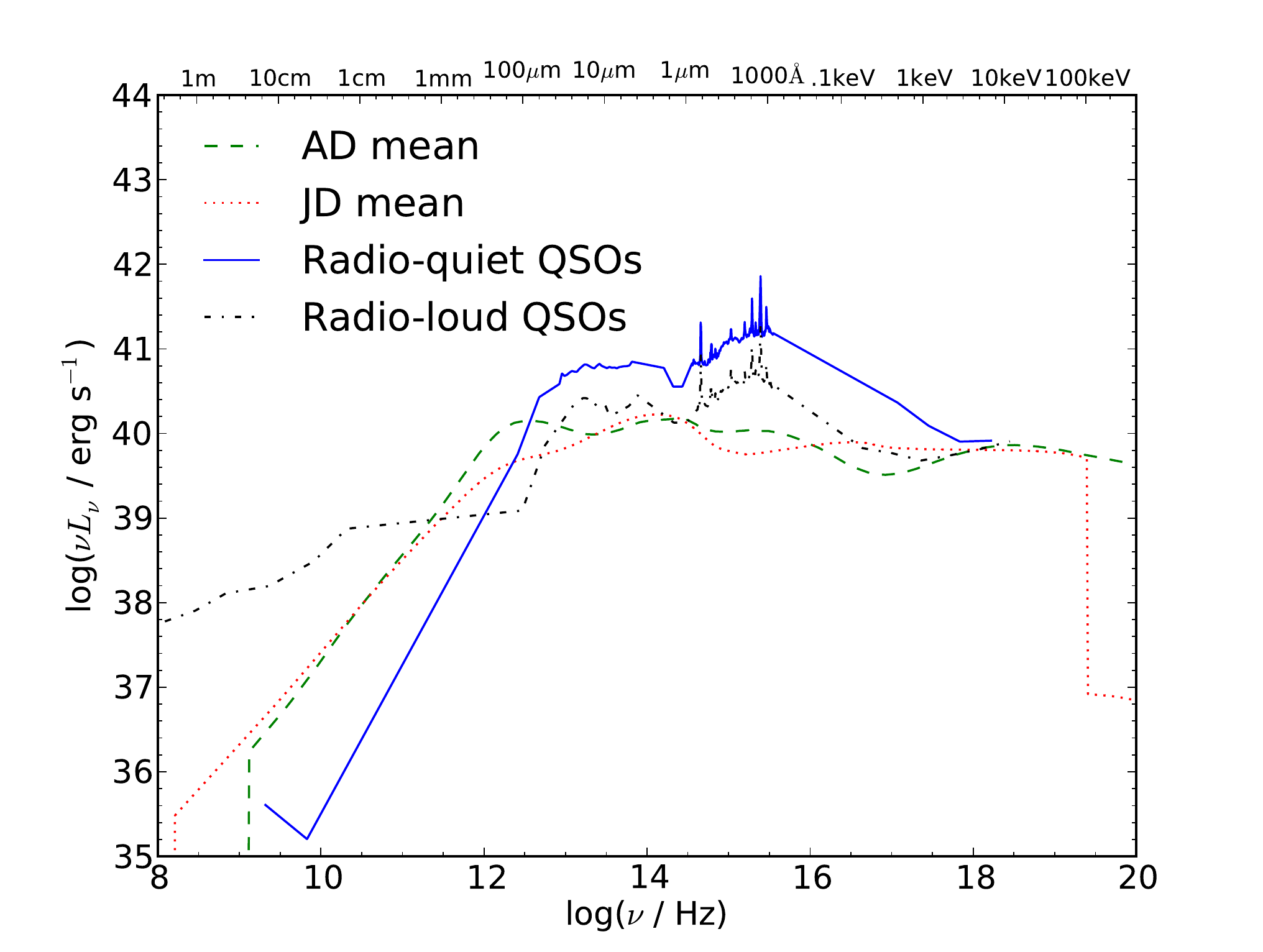}
\caption{The average JD and AD model SEDs compared to the average radio-loud and radio-quiet quasar SEDs computed by \citet{shang11}.}
\label{fig:qsoavg}
\end{figure}

Figure \ref{fig:qsoavg} shows the average AD and JD SEDs compared to the average ones of radio-loud and radio-quiet quasars computed by \citet{shang11}. The average quasar SEDs computed by \citet{shang11} are very similar to the ones of \citet{elvis94} but the former include more detailed features, and are based on more recent data obtained with improved instrumentation.
As in Fig. \ref{fig:avgsed}a, the SEDs were normalized such that they all have the same X-ray luminosity in the 2-10 keV band of $10^{40} \ {\rm erg \ s}^{-1}$. 
The UV excess in the quasar SEDs (the big blue bump) is clearly apparent in comparison to the LLAGN ones. It is also interesting that for $\nu > 10^{17}$ Hz the average AD, average JD and the radio-loud quasar SEDs are quite similar.

\section{Discussion: The nature of X-ray emission}	\label{sec:disc}

In our AD models, the X-rays are produced predominantly by inverse Compton scattering by the ADAF of seed synchrotron photons produced in the accretion flow itself; in contrast, in our JD models, the X-rays are dominated by the optically thin synchrotron emission in the base of the jet. In this regard, the \emph{Chandra} X-ray spectrum is one of the most important constraints for the origin of high-energy photons in LLAGNs.

There are only two objects for which the AD or the JD can be ruled out, within the observational uncertainties and the explored model parameter space: NGC 1553 (\textsection \ref{sec:1553}) and NGC 4552 (\textsection \ref{sec:4552}). For NGC 1553, the JD model does not reproduce the hardness of the X-ray spectrum whereas for NGC 4552 the AD model is not able to account for the corresponding X-ray softness.
For the remaining objects, by inspecting the SED fits we can see that both the AD and the JD models are able to reproduce with varying degrees of success the whole SED from radio to X-rays, in particular the X-ray portion.

In order to quantify which model systematically fits better the broadband SEDs we use as a goodness-of-fit parameter a modified $\chi^2$. We define this modified $\chi^2$ simply as 
\begin{equation}
\chi^2 \equiv \sum_i \left( y_{\rm obs} - y_{\rm model} \right)^2
\end{equation}
where $y=\log \nu L_\nu$ and the sum is carried over the observed data points. We sample the X-ray continuum using 3 data points from 0.5 keV to 10 keV and we do not take into account upper limit data points. The values of $\chi^2$ for each model are displayed in Table \ref{tab:models}.

Considering the 20 objects for which we were able to obtain both JD and AD fits, we find that $\chi^2 (JD) < \chi^2 (AD)$ for 9 objects and $\chi^2 (JD) > \chi^2 (AD)$ for the remaining 8. In other words, for 9 objects (10 including NGC 4552 for which we did not find an appropriate AD fit) the JD results in better SED fits as opposed to 8 for which the AD model is a better fit. The results from this goodness-of-fit test show that 52\% of the SEDs in our sample are better fitted by the JD model whereas 38\% are better reproduced by the AD scenario. For the 3 objects with $\chi^2 (JD)/\chi^2 (AD) \approx 1$ (NGC 1097, NGC 3998 and NGC 4374) it is hard to tell -- based on the broadband SED alone -- which model provides a better fit. 

We should point out a number of caveats involved when considering which scenario better accounts for the nature of the X-ray emission.
From an observational point of view, even though we are carrying the largest systematic analysis of the SEDs of LLAGNs in LINERs done up to date, the size of our sample is limited. We need to extend our modeling to a larger number of LLAGNs in order to increase the statistics and draw further conclusions. 
Secondly, for many sources we only have a few data points available to fit, so clearly for these objects we need to obtain more measurements to better constrain the models. 
In addition, in our modeling we find that one of the main observables that help to pinpoint the dominant source of X-rays is the hardness of the X-ray spectrum. ADAF models tend to predict harder X-ray spectra while jet models predict softer ones. In most of the LLAGNs in our sample there is a significant observational uncertainty in the X-ray hardness derived from the \emph{Chandra} observations which make it difficult to determine which whether the AD or JD is favored based on the X-ray spectrum.

From a theoretical perspective, there are considerable theoretical uncertainties involved in the the ADAF-jet models. For instance, the values of the model parameters that affect the microphysics and dynamics of the ADAF and jet are not sufficiently constrained. In particular, $\delta$ and $p$ have a large impact on the hardness of the X-ray spectrum predicted by ADAF/jet models but there is a considerable uncertainty in the range of possible values of these parameters despite recent progress (e.g., \citealt{sharma07,medvedev06}).
Considering the overall goodness-of-fit of the AD and JD models, it is important to point out that our jet model is basically a phenomenological one. Uncertainties affecting for example the jet model could translate in larger values of $\chi^2$. For instance, when we find $\chi^2 (JD) < \chi^2 (AD)$, that could be possibly because the JD model is less well constrained than the AD one, i.e, we have more freedom in the jet fitting compared to the ADAF model. 

Taking into account these caveats, it is still premature to favor one model over the other based only on the SED. 
The nature of the X-ray emission in LLAGNs has been debated in the last few years by several authors  favoring in some cases the AD or JD models based on the analysis of individual sources (e.g., \citealt{falcke00,yuan02sgr,yuan02_4258,yuan03,wu07,markoff08,miller10}) or in a statistical sense based on the fundamental plane of black hole activity \citep{merloni03,falcke04,yc05,yuan09,plotkin11}. We suggest different routes to clarify this issue in the future.

The AD and JD models should predict different characteristic radio and X-ray variability timescales (e.g., \citealt{ptak98}), hence we should be able to test which component is dominant in X-rays by carrying out a simultaneous monitoring of the variability of radio and X-rays. By comparing the variability pattern predicted by jet/ADAF models we could pinpoint the X-ray dominant component. This promising strategy is quite similar to what the observational campaigns carried out for M81 \citep{markoff08,miller10} and the radio-quiet Seyfert 1 NGC 4051 \citep{king11} and could be applied to many other LLAGNs. We note that in the case of M81, \cite{miller10} favor a scenario in which the X-rays originate in the transition region between the thin disk and the ADAF; NGC 4051 is a different case since it is a relatively bright AGN accreting at the $\sim 0.05 L_{\rm Edd}$ level \citep{peterson04} and its jet-disk coupling is quite distinct compared to LLAGNs.

The \emph{Fermi} Gamma-ray Space Telescope can be quite helpful in this regard. LLAGNs are potential sources of $\gamma$-rays \citep{takami11} and in fact one of the source in this sample (NGC 4486) has already been detected by the \emph{Fermi} LAT \citep{m87fermi}. By using the model parameters that we derived in our radio-to-X-rays fits we should be able to predict the $\gamma$-ray spectrum \citep{mahadevan97,takami11} and compare with future \emph{Fermi} detections. In this way we should be able to compare AD/JD predictions and through $\gamma$-ray observations place further constraints on the production site responsible for the high-energy emission and the jet-disk connection in LLAGNs.

Observations in the mm and sub-mm are also very helpful because they constrain the ADAF synchrotron emission and hence also the X-ray emission. This follows because the synchrotron photons in the ADAF are inverse-Compton-scattered to X-rays.

Finally, further refinements and a better understanding of the fundamental plane of black hole activity \citep{merloni03,falcke04,gultekin09,yuan09,plotkin11} should make the fundamental plane a better tool for constraining the radiative processes shaping the radio and X-ray emission in sub-Eddington black hole source and by consequence LLAGNs.

\section{Summary}	\label{sec:end}

We performed detailed modeling of the broadband -- radio-to-X-rays -- spectral energy distributions of a sample of 21 low-luminosity AGNs in LINERs selected from EHF10. With this exploratory modeling, our goal is to constrain the general properties of the central engines of the dominant population of AGNs at $z \approx 0$. Our coupled accretion-jet model consists of an accretion flow which is radiatively inefficient in the inner parts and becomes a thin disk outside a certain transition radius. The relativistic jet is modeled in the framework of the internal shock scenario. 

We demonstrated that there are two classes of models that can explain the majority of the observed SEDs. We call the first one AD which stands for ``ADAF-dominated'' since the ADAF dominates most of the broadband emission, particularly the X-rays. In the second class of models, the jet component dominates the majority of the continuum emission and for this reason we call this scenario JD as in ``jet-dominated''. The SEDs predicted by the AD models have a more complex shape than the JD ones, given the richer variety of radiative processes involved in ADAFs. For both scenarios though, the radio band is almost always dominated by the synchrotron emission from the jet and the near IR to optical band is dominated by the truncated thin disk thermal emission.


From our exploratory modeling of the SEDs we are able to constrain important parameters that characterize the central engines of LLAGNs, as summarized below.
\begin{description}
\item[Mass accretion rates] based on the AD models, the values of the accretion rates supplied to the accretion flow lie in the range $4 \times 10^{-4}-0.02$ in Eddington units with a mean $\left \langle \log \dot{m} \right \rangle = -2.5$ or alternatively $\left \langle \log \left( \dot{M}/{\rm M_\odot \ year^{-1}} \right) \right \rangle = -2$ (standard deviations of $\approx 0.5$ dex respectively). Of that gas supply, typically only $\sim 10\%$ is accreted onto the black hole with the remaining gas being lost due to outflows.
\item[Jet powers] the typical jet power resulting from the AD and JD scenarios is $\sim 10^{42} \ {\rm erg \ s}^{-1}$ ranging from $\sim 10^{40}$ to $\sim 10^{44} \ {\rm erg \ s}^{-1}$. 
\item[Jet production efficiencies] The typical efficiency with which the rest mass energy associated with gas supplied to the accretion flow is converted into jet kinetic power is estimated to be $\sim 0.001$ (range $\sim 10^{-4}-0.01$). The efficiency is ten times larger if we consider the gas that reaches a few gravitational radii only.
\item[Fraction of material ejected in the jet] On average, a fraction of $0.4\%$ of the gas that reaches a few gravitational radii is chanelled in the relativistic jet (range $~0.01-1 \%$).
\item[Radiative efficiencies] The typical efficiency of conversion of rest mass energy associated with gas supplied to the accretion flow into disk+jet radiation is $\approx 5 \times 10^{-4}$ (dispersion of 0.8 dex), producing an average bolometric luminosity of a few x$10^{41} \ {\rm erg \ s}^{-1}$.
\item[Kinetic to radiative output ratio] LLAGNs typically release $\approx 5$ times more kinetic power in the jets than the total energy radiated away.
\item[Transition radii between thin disk and ADAF] A truncated thin disk is required to fit the available mid and near-IR data for only 4 sources --  NGC 1097 (see also \citealt{nemmen06}), NGC 4143, NGC 4278 and NGC 4736 -- with the resulting transition radii in the range $30-225 R_S$.
\end{description}
The values that we derived from our SED models above provide useful indicators of the feeding and feedback properties of LLAGNs in LINERs, and by extension of the whole LLAGN population. 

Even though we used the physical scenario that is favored to account for the combined set of observational properties of LLAGNs \citep{ho08,nar08}, the inferred values of the parameters are model dependent. 
There are theoretical and observational sources of uncertainty in the values of the derived parameters. For instance, there are degeneracies between the effect of the model parameters on the resulting SEDs \citep{qn99b,nemmen06}. The uncertainty in the OUV -- due to uncertain degree of extinction correction -- and the X-ray data also contribute to uncertainty in the estimated parameters.
Furthermore, our sample is currently limited to 21 objects. In spite of these caveats, the estimates we obtained for the properties of central engines of LINER AGNs should be illustrative of the general behavior of the class.

We can draw a link between the supermassive black holes in LINERs and the quiescent black hole in our galaxy, Sgr A*. By modelling the SED of Sgr A* with ADAF models, an accretion rate of $\dot{m}_{\rm out} \sim 10^{-5}$ was estimated \citep{yuan03,yuan06}. This value is two orders of magnitude below the typical accretion rate of LINERs that we estimated.  Therefore, we could say that in order to Sgr A* ``light up'' and become a LINER-like LLAGN, it needs to accrete at a rate $\sim 100$ times higher than the present one.

The individual and average SED models can be useful for a different number of applications. They provide a library of templates for the compact emission of LLAGNs, which can be useful for studies of the emission of other astrophysical systems in the nucleus, i.e., stellar populations, dust and PAH features (e.g., \citealt{sb05, mason07}). Furthermore, the models provide the expected emission on bands that were not yet observed in many LLAGNs (e.g., mid and far IR, far UV).

\acknowledgments

We are grateful to Feng Yuan for his invaluable help with the SED models and for allowing us to use his codes as well as his comments which helped to improve the manuscript. We also thank Renyi Ma and Hui Zhang for their help with setting up the models, and Jo\~ao Steiner, Michael Brotherton, Francesco Tombezi, Thomas Nelson and Judith Racusin for productive discussions. RSN was supported by an appointment to the NASA Postdoctoral Program at Goddard Space Flight Center, administered by Oak Ridge Associated Universities through a contract with NASA. RSN and TSB acknowledge the financial support of the brazilian institutions CNPq and CAPES. This research has made use of the NASA/IPAC Extragalactic Database (NED) which is operated by the Jet Propulsion Laboratory, California Institute of Technology, under contract with the National Aeronautics and Space Administration.




\begin{sidewaystable}[!ht]
\centering
\scriptsize
\begin{tabular}{ccccccccccccccccc}
\hline
\hline
Galaxy & Model & $\dot{m}_{\rm out}$ & $r_{\rm tr}$ & $\delta$ & $s$ & $\dot{m}_{\rm jet}$ & $p$ & $\epsilon_e$ & $\epsilon_B$ & $i$ ($^\circ$) & $P_{\rm jet}^{\rm mod}$ & $\frac{\dot{m}_{\rm jet}}{\dot{m}(3 R_S)}$ & $P_{\rm jet}^{\rm obs}$ & $\dot{m}_{\rm Bondi}$ & $\chi^2$ & Refs. \\
& &  &   &  &  &  &  &  &  &  &  &  & &  & & and notes \\
\hline
NGC 0266 &  JD & - & - & - & - & $1.1 \times 10^{-5}$ & 1.8 & 0.1 & 0.01 & 30 & $3.1 \times 10^{42}$ & - & $3.2 \times 10^{42}$ & - & 0.4 & a,b  \\
NGC 0266 &  AD & $5.5 \times 10^{-3}$ & $10^4$ & 0.3 & 0.3 & $5 \times 10^{-6}$ & 2.5 & 0.1 & 0.01 & 30 & $1.1 \times 10^{42}$ & $8 \times 10^{-3}$ &  & - & 0.1 &  \\
\hline
NGC 1097 & JD & - & - & - & - & $5 \times 10^{-7}$ & 1.8 & 0.9 & 0.004 & 34 & $7.8 \times 10^{42}$ & - & $5.1 \times 10^{41}$ & - & 1.3 & b,c  \\
NGC 1097 & AD & $6.4 \times 10^{-3}$ & 225 & 0.1 & 0.8 & $7 \times 10^{-7}$ & 2.2 & 0.06 & 0.02 & 34 & $10^{43}$ & $3 \times 10^{-3}$ & - & - & 1.3 & \\
\hline
NGC 1553 &  JD & $10^{-4}$ & $10^{4}$ & 0.3 & 0.3 & $2 \times 10^{-6}$ & 1.9 & 0.1 & 0.1 & 30 & $10^{42}$ & 0.2 & $10^{42}$ & $>6.2 \times 10^{-5}$ & 0.8 & b,d, 1 \\ 
NGC 1553 &  AD & $2.5 \times 10^{-3}$ & $10^{4}$ & 0.3 & 0.3 & $7 \times 10^{-7}$ & 2.5 & 0.1 & 0.08 & 30 & $3.7 \times 10^{41}$ & 0.003 & & & 1 & \\
\hline
NGC 2681 &  JD & - & - & - & - & $5 \times 10^{-7}$ & 2.1 & 0.2 & 0.1 & 30 & $4.3 \times 10^{40}$ & - & $1.4 \times 10^{41}$ & - & 0.03 & b \\ 
NGC 2681 &  AD & $7.5 \times 10^{-3}$ & $10^{4}$ & 0.01 & 0.3 & $10^{-7}$ & 2.5 & 0.1 & 0.1 & 30 & $8.7 \times 10^{39}$ & $2 \times 10^{-4}$ & & - & 0.06 & \\
\hline
NGC 3031 & JD & $8 \times 10^{-4}$ & 50 & 0.3 & 0.6 & $1.2 \times 10^{-5}$ & 2.05 & 0.6 & $10^{-4}$ & 50 & $4.8 \times 10^{42}$ & 0.08 & $7.1 \times 10^{41}$ & - & 2.4 & b \\ 
NGC 3031 & AD & $3 \times 10^{-3}$ & 360 & 0.01 & 0.16 & $2 \times 10^{-6}$ & 2.2 & 0.1 & 0.01 & 50 & $8 \times 10^{41}$ & $1.4 \times 10^{-3}$ &  & - & 3.3 &  \\ 
\hline
NGC 3169 & JD & - & - & - & - & $3 \times 10^{-5}$ & 2.8 & 0.99 & $1.5 \times 10^{-7}$ & 30 & $1.3 \times 10^{43}$ & - & $10^{42}$ & - & 0.7 & b \\ 
NGC 3169  & AD & 0.015 & 100 & 0.01 & 0.3 & $2.5 \times 10^{-6}$ & 2.2 & 0.01 & 0.01 & 30 & $10^{42}$ & $4 \times 10^{-4}$ &  & - & 1.7 &  \\ 
\hline
NGC 3226 & JD & - & - & - & - & $10^{-6}$ & 2.2 & 0.5 & $2 \times 10^{-3}$ & 30 & $8.7 \times 10^{41}$ & - & $1.1 \times 10^{42}$ & - & 0.04 & b \\ 
NGC 3226  & AD & $5.5 \times 10^{-3}$ & 100 & 0.01 & 0.3 & $5 \times 10^{-7}$ & 2.3 & 0.1 & 0.01 & 30 & $4.3 \times 10^{41}$ & $3 \times 10^{-4}$ &  & - & 0.05 &  \\ 
\hline
NGC 3379 & JD & - & - & - & - & $2 \times 10^{-9}$ & 2.05 & 0.2 & 0.15 & 30 & $2.1 \times 10^{39}$ & - & - & $1.6 \times 10^{-5}$ & 0.04 & 2 \\ 
NGC 3379  & AD & $6.5 \times 10^{-4}$ & $10^{4}$ & 0.01 & 0.3 & - & - & - & - & 30 & - & - & - &  & 0.08 &  \\ 
\hline
NGC 3998 & JD & $4 \times 10^{-4}$ & 100 & 0.3 & 0.3 & $3.5 \times 10^{-6}$ & 2.01 & 0.75 & $3 \times 10^{-5}$ & 30 & $1.8 \times 10^{43}$ & 0.025 & $4.6 \times 10^{42}$ & - & 8.7 & b \\ 
NGC 3998  & AD & $7.2 \times 10^{-3}$ & $10^{4}$ & 0.1 & 0.4 & $1.7 \times 10^{-6}$ & 2.2 & 0.01 & $10^{-3}$ & 30 & $9 \times 10^{42}$ & $6 \times 10^{-3}$ &  & - & 8.7 & \\ 
\hline
NGC 4143 & JD & $2.5 \times 10^{-3}$ & 70 & 0.1 & 0.8 & $1.4 \times 10^{-6}$ & 1.6 & 0.03 & 0.1 & 30 & $1.9 \times 10^{42}$ & $7 \times 10^{-3}$ & $10^{42}$ & - & 0.2 & b \\ 
NGC 4143  & AD & $2.5 \times 10^{-3}$ & 70 & 0.1 & 0.76 & $1.2 \times 10^{-7}$ & 2.2 & 0.03 & 0.1 & 30 & $1.6 \times 10^{41}$ & $6 \times 10^{-4}$ & & - & 0.3 & \\ 
\hline
NGC 4261 & JD & - & - & - & - & $2 \times 10^{-5}$ & 1.9 & 0.3 & 0.002 & 63 & $6.9 \times 10^{43}$ & - & $10^{40}-10^{43}$ & $4 \times 10^{-3}$ & 4.6 & e,3 \\ 
NGC 4261  & AD & $1.9 \times 10^{-3}$ & $10^4$ & 0.3 & 0.3 & $6 \times 10^{-6}$ & 2.2 & 0.01 & 0.1 & 63 & $2 \times 10^{43}$ & 0.04 & & & 5.4 & \\ 
\hline
NGC 4278 & JD & $4 \times 10^{-4}$ & 30 & 0.3 & 0.7 & $9 \times 10^{-6}$ & 1.6 & 0.001 & 0.9 & 3 & $10^{43}$ & 0.1 & $2 \times 10^{42}$ & $0.001-0.1$ & 0.4 & b,4,5 \\ 
NGC 4278 & AD & $7 \times 10^{-4}$ & 40 & 0.01 & 0.18 & $3.5 \times 10^{-6}$ & 2.3 & 0.001 & 0.01 & 3 & $4 \times 10^{42}$ & $8 \times 10^{-3}$ & & & 0.1 & \\ 
\hline
NGC 4374 & JD & $1.5 \times 10^{-4}$ & 30 & 0.3 & 0.3 & $1.6 \times 10^{-6}$ & 2.2 & 0.009 & 0.008 & 30 & $8.4 \times 10^{42}$ & 0.02 & $3.9 \times 10^{42}$ & $4 \times 10^{-4}$ & 2.3 & 1,6,7 \\ 
NGC 4374 & AD & $4 \times 10^{-4}$ & 150 & 0.01 & 0.1 & $4 \times 10^{-7}$ & 2.4 & 0.01 & 0.1 & 30 & $2 \times 10^{42}$ & 0.001 & & & 2 & \\ 
\hline
NGC 4457 & JD & - & - & - & - & $1.1 \times 10^{-6}$ & 2.01 & 0.3 & 0.01 & 30 & $6 \times 10^{40}$ & - & $<4 \times 10^{41}$ & - & 0.1 & b \\ 
NGC 4457 & AD & $2.2 \times 10^{-3}$ & $10^{4}$ & 0.1 & 0.3 & $4 \times 10^{-7}$ & 2.5 & 0.05 & 0.01 & 30 & $2 \times 10^{40}$ & 0.002 & & - & 0.02 & \\ 
\hline
NGC 4486 & JD & - & - & - & - & $6 \times 10^{-8}$ & 2.3 & 0.001 & 0.008 & 10 & $8.2 \times 10^{42}$ & - & $10^{43}-10^{44}$ & $7 \times 10^{-4}$ & 3.4 & 6,7,8,9 \\ 
NGC 4486 & AD & $5.5 \times 10^{-4}$ & $10^{4}$ & 0.01 & 0.1 & $5 \times 10^{-8}$ & 2.6 & 0.001 & 0.001 & 10 & $6.8 \times 10^{42}$ & $2 \times 10^{-4}$ & & & 1.3 & \\ 
\hline
NGC 4494 & JD & - & - & - & - & $10^{-6}$ & 2.01 & 0.6 & 0.001 & 30 & $2.3 \times 10^{41}$ & - & $<1.4 \times 10^{41}$ & - & 0.04 &  \\ 
NGC 4494 & AD & $6.5 \times 10^{-3}$ & $10^{4}$ & 0.01 & 0.3 & $10^{-7}$ & 2.3 & 0.1 & 0.01 & 30 & $2.3 \times 10^{40}$ & $2 \times 10^{-4}$ &  & - & 0.01 & \\ 
\hline
\end{tabular}
\begin{flushleft}
\footnotesize 
\textbf{Notes:} \\
(a) Black hole mass estimated from the fundamental plane of black holes \citep{merloni03}, using the X-ray and radio luminosities. \\
(b) Observed jet power estimated using the radio data and the \citet{merloni07} correlation. \\
(c) For NGC 1097, $\Gamma=10$ (for consistency with the work of \citealt{nemmen06}). \\
(d) 5 GHz luminosity estimated using the \citet{merloni03} correlation. \\
(e) The lower limit on $P_{\rm jet}^{\rm obs}$ was derived by \citet{gliozzi03} and the upper limit results from using the radio data and the \citet{merloni07} correlation. The Bondi rate and inclination angle is from \citet{gliozzi03}. \\
\textbf{References:} 1. \citet{pelle05}, 2. \citet{david05}, 3. \citet{gliozzi03}, 4. \citet{dmt01}, 5. \citet{giro05}, 6. \citet{allen06}, 7. \citet{merloni07}, 8. \citet{biretta99}, 9. \citet{dmt03}, 10. \citet{barth01}
\\
\end{flushleft}
\caption{Model parameters resulting from the SED fits. The meaning of the parameters is described in Section \ref{sec:models}.}
\label{tab:models}
\end{sidewaystable}

\newpage

\begin{sidewaystable}[!ht]
\centering
\scriptsize
\begin{tabular}{ccccccccccccccccc}
\hline
\hline
Galaxy & Model & $\dot{m}_{\rm out}$ & $r_{\rm tr}$ & $\delta$ & $s$ & $\dot{m}_{\rm jet}$ & $p$ & $\epsilon_e$ & $\epsilon_B$ & $i$ ($^\circ$) & $P_{\rm jet}^{\rm mod}$ & $\frac{\dot{m}_{\rm jet}}{\dot{m}(3 R_S)}$ & $P_{\rm jet}^{\rm obs}$ & $\dot{m}_{\rm Bondi}$ & $\chi^2$ & Refs. \\
& &  &   &  &  &  &  &  &  &  &  &  & &  & & and notes \\
\hline
NGC 4548 &  JD & - & - & - & - & $1.5 \times 10^{-6}$ & 2.2 & 0.1 & 0.01 & 30 & $3.4 \times 10^{41}$ & - & $4.2 \times 10^{41}$ & - & 0.04 & b  \\
NGC 4548 &  AD & $8 \times 10^{-3}$ & $10^4$ & 0.01 & 0.3 & $5 \times 10^{-7}$ & 2.5 & 0.1 & 0.01 & 30 & $1.1 \times 10^{41}$ & $7 \times 10^{-4}$ &  & - & 0.2 &  \\
\hline
NGC 4552 &  JD & - & - & - & - & $1.8 \times 10^{-6}$ & 2.1 & 0.01 & 0.01 & 30 & $2.2 \times 10^{42}$ & - & $1.6 \times 10^{42}$ & 0.001 & 2.8 & 7  \\
\hline
NGC 4579 & JD & 0.02 & 150 & 0.01 & 0.6 & $1.3 \times 10^{-4}$ & 1.8 & 0.12 & $2 \times 10^{-5}$ & 45 & $5.5 \times 10^{43}$ & 0.01 & $10^{42}$ & - & 1 & b,10  \\
NGC 4579 &  AD & 0.02 & 150 & 0.01 & 0.2 & $1.4 \times 10^{-5}$ & 2.2 & 0.011 & 0.008 & 45 & $5.9 \times 10^{42}$ & $1.4 \times 10^{-3}$ &  & - & 1.7 &  \\
\hline
NGC 4594 & JD & $2 \times 10^{-3}$ & $10^4$ & 0.01 & 0.3 & $4.5 \times 10^{-7}$ & 2 & 0.8 & 0.03 & 30 & $10^{42}$ & 0.009 & $3 \times 10^{42}$ & 0.002 & 1.3 & b,1  \\
NGC 4594 &  AD & 0.006 & $10^4$ & 0.01 & 0.3 & $9 \times 10^{-7}$ & 2.3 & 0.005 & 0.003 & 30 & $2 \times 10^{42}$ & 0.002 & & & 2.2 &  \\
\hline
NGC 4736 & JD & 0.009 & 100 & 0.01 & 0.8 & $1.3 \times 10^{-6}$ & 1.6 & 0.045 & 0.1 & 30 & $10^{41}$ & 0.002 & $7 \times 10^{40}$ & - & 0.3 & b  \\
NGC 4736 &  AD & 0.0035 & 60 & 0.1 & 0.89 & $6 \times 10^{-7}$ & 2.2 & 0.01 & 0.01 & 30 & $6.5 \times 10^{40}$ & 0.002 & & & 0.03 &  \\
\hline
\end{tabular}
\caption{Model parameters resulting from the SED fits (continued).}
\label{tab:models02}
\end{sidewaystable}

\end{document}